\documentclass[11pt]{article}
\usepackage[margin=1in]{geometry}
\usepackage{amsmath,amssymb,amsthm}
\usepackage{mathtools}
\usepackage{bm}
\usepackage[round,authoryear]{natbib}
\usepackage[hidelinks]{hyperref}
\makeatletter
\renewcommand\hyper@natlinkstart[1]{}
\renewcommand\hyper@natlinkend{}
\renewcommand\hyper@natlinkbreak[2]{#1}
\makeatother
\usepackage{booktabs}
\usepackage{float}
\usepackage{graphicx}
\usepackage{tikz}
\usetikzlibrary{calc}
\usepackage{algorithm}
\usepackage{algpseudocode}
\graphicspath{{figs/}}

\newtheorem{theorem}{Theorem}[section]

\newtheorem{proposition}[theorem]{Proposition}

\theoremstyle{definition}
\newtheorem{remark}[theorem]{Remark}
\theoremstyle{plain}

\DeclareMathOperator{\St}{St}
\DeclareMathOperator{\E}{\mathbb{E}}

\newcommand{\R}{\mathbb{R}}
\newcommand{\N}{\mathcal{N}}
\newcommand{\MN}{\mathcal{MN}}
\newcommand{\one}{\mathbf{1}}
\newcommand{\Prob}{\mathbb{P}}

\usepackage[dvipsnames]{xcolor}

\title{Selective Inference for Clustering Under Unknown Covariance}
\author{%
  \begin{minipage}[t]{0.48\textwidth}\centering
    Yiqun T. Chen \\
    Departments of Biostatistics \\ and Computer Science \\
    Johns Hopkins University \\
    \texttt{yiqunc@jhu.edu}
  \end{minipage}%
  \hfill
  \begin{minipage}[t]{0.48\textwidth}\centering
    Lucy L. Gao \\
    Department of Statistics \\ and Actuarial Science \\
    University of Waterloo \\
    \texttt{lucy.gao@uwaterloo.ca}
  \end{minipage}
}
\date{\today}

\begin{document}
\maketitle

\begin{abstract}
Unsupervised clustering is widely used to discover underlying group structure, after which researchers often seek to validate these groups via post-hoc hypothesis testing on the population means. However, using the same data both to form clusters and to test for a difference in means inflates the Type I error rate. Recent work in selective inference has proposed solutions when the underlying data are drawn from a multivariate Gaussian distribution with a \emph{known} covariance matrix or an \emph{unknown isotropic} covariance matrix. We propose a new selective test for the difference in means between two clusters that \emph{provably} controls the Type I error rate under an \emph{arbitrary unknown covariance matrix}. Our test statistic does not require estimating the feature covariance matrix; instead, it leverages a Monte Carlo sampling procedure over the set of orthogonal matrices. We show that this new selective test can be computed efficiently, and we derive closed-form solutions when clusters are formed using $k$-means or hierarchical clustering. In simulations, we demonstrate that our proposed $p$-value controls the selective Type I error rate and achieves substantial power, and we illustrate its application on gene expression and drug review data sets.
\end{abstract}

\section{Introduction}
\label{sec:intro}

In many analyses, the groups that are compared are not fixed in advance but are
estimated from the data. Clinical researchers cluster patients on routine
measurements to find subgroups that differ in disease course and in who is most
likely to benefit from a treatment \citep{AhlqvistEtAl2018}. Marketers segment
customers by their purchases to decide whom to target \citep{WedelKamakura2000}.
It is then tempting to apply a classical two-sample test to ask whether the
estimated subgroups, or clusters, truly differ. But the hypothesis is generated
from the same data that are used to test it, so classical tests no longer control
the Type I error rate. This problem has been studied in a recent line of work known
as \emph{selective inference} or \emph{post-selection inference}
\citep[see][for a review]{NeufeldPerryWitten2026}.

\begin{figure}[t]
\centering
\includegraphics[width=\linewidth]{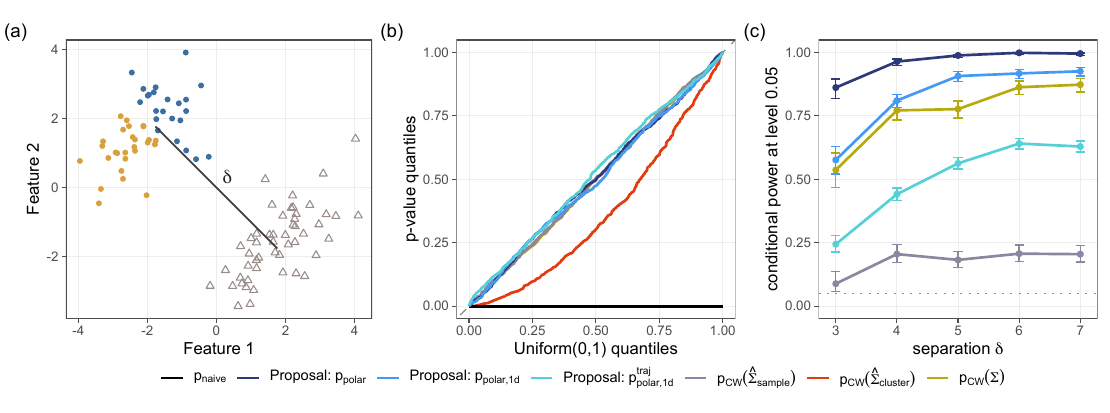}
\vspace{-8pt}
\caption{Estimating the covariance trades Type I error control against power.
\textbf{(a)} One data set from \eqref{eq:intro_model} with two true groups (shapes), split into
$K = 3$ clusters (colors) by $k$-means, which divides one true group in two.
\textbf{(b)} QQ plots of the Uniform$(0,1)$ distribution against the $p$-values when all
observations come from one Gaussian group: the naive test, the test of \citet{ChenWitten2023}
with three choices of covariance ($p_{\mathrm{CW}}$, Section~\ref{sec:sim}), and our
proposals $p_{\mathrm{polar}}$ in \eqref{eq:pstar}, $p_{\mathrm{polar,1d}}$ in
\eqref{eq:onedim_def}, and $p_{\mathrm{polar,1d}}^{\mathrm{traj}}$ of Section~\ref{sec:kmeans}. \textbf{(c)} Conditional power against the separation $\delta$ between
the true groups, for the tests that control the Type I error in (b).}
\label{fig:motivation}
\end{figure}

We illustrate the problem with a toy example. Let $\mathbf{X} \in \R^{n \times p}$ follow
the matrix normal model
\begin{equation}
\label{eq:intro_model}
\mathbf{X} \sim \MN_{n \times p}(\boldsymbol\mu, \mathbf{I}_n, \boldsymbol\Sigma),
\end{equation}
that is, the rows $\mathbf{X}_i \sim \N_p(\boldsymbol\mu_i, \boldsymbol\Sigma)$,
$i = 1, \ldots, n$, are independent, where $\boldsymbol\mu_i$ is the $i$th row of
$\boldsymbol\mu \in \R^{n \times p}$. In Figure~\ref{fig:motivation}(a), half of the
observations share one mean, the other half share another, and the two features are
correlated. We cluster the data into three groups with $k$-means, so the algorithm
must split one of the two true groups in two. It does so cleanly by construction,
because it chooses the partition that maximizes the between-cluster variation. The
two halves of the split group therefore have very different sample means, although
they have the same population mean.
Let $\hat{\mathcal C}_1$ and $\hat{\mathcal C}_2$ be two of the estimated clusters and write
$\bar{\boldsymbol\mu}_{G} = |G|^{-1} \sum_{i \in G} \boldsymbol\mu_i$ for the average mean of a
set $G$ of observations. The question is whether
\begin{equation}
\label{eq:intro_null}
H_0 : \bar{\boldsymbol\mu}_{\hat{\mathcal C}_1} = \bar{\boldsymbol\mu}_{\hat{\mathcal C}_2}
\quad \text{versus} \quad
H_1 : \bar{\boldsymbol\mu}_{\hat{\mathcal C}_1} \neq \bar{\boldsymbol\mu}_{\hat{\mathcal C}_2}.
\end{equation}
For the two halves of a split group this null is true. A textbook two-sample test,
such as Hotelling's $T^2$ referred to its $F$ distribution, nevertheless rejects it
almost always (the red line in Figure~\ref{fig:motivation}(b)). The $F$ distribution describes the distance
between groups that were fixed before the data were seen, not between groups that
the data selected. Across the simulation study of Section~\ref{sec:sim}, the naive
test rejects the true null in $85\%$ to $100\%$ of replicates
(Section~\ref{sec:typeI}).

Conditional selective inference repairs the reference distribution by
\emph{conditioning on the selection event}. The goal is a $p$-value $p$ that controls the
\emph{selective Type I error} \citep{FithianSunTaylor2017},
\begin{equation}
\label{eq:sel_typeI}
\Prob_{H_0}\big(p \le \alpha \;\big|\; \hat{\mathcal C}_1, \hat{\mathcal C}_2 \in \mathcal C(\mathbf{X})\big) \le \alpha
\quad \text{for all } \alpha \in (0, 1).
\end{equation}
That is, among the data sets on which the clustering returns the tested clusters, a true null hypothesis is
rejected at level $\alpha$ with probability at most $\alpha$. We call such a test valid. Let $T(\mathbf{X})$ be a test statistic,
such as Hotelling's $T^2$ or the distance $\|\bar{\mathbf{X}}_{\hat{\mathcal C}_1} - \bar{\mathbf{X}}_{\hat{\mathcal C}_2}\|_2$
between the two cluster means. Instead of asking whether $T(\mathbf{X})$ is extreme
under its unconditional null law, one asks whether it is extreme given that the
clustering procedure returned the observed clusters. \citet{GaoBienWitten2024}
formalized this idea. For a known $\boldsymbol\Sigma$ in \eqref{eq:intro_model}, the selective
$p$-value is
\begin{equation}
\label{eq:intro_psel}
p_{\mathrm{sel}}
= \Prob_{H_0}\Big( \|\boldsymbol\Sigma^{-1/2}(\bar{\mathbf{X}}_{\hat{\mathcal C}_1} - \bar{\mathbf{X}}_{\hat{\mathcal C}_2})\|_2
\ge \|\boldsymbol\Sigma^{-1/2}(\bar{\mathbf{x}}_{\hat{\mathcal C}_1} - \bar{\mathbf{x}}_{\hat{\mathcal C}_2})\|_2
\;\Big|\; \hat{\mathcal C}_1, \hat{\mathcal C}_2 \in \mathcal C(\mathbf{X}) \Big),
\end{equation}
where $\mathbf{x}$ is the observed data, $\mathcal C(\mathbf{X})$ is the set of clusters
obtained from $\mathbf{X}$, and $\boldsymbol\nu$ is the contrast vector with entries
$\one\{i \in \hat{\mathcal C}_1\}/|\hat{\mathcal C}_1| - \one\{i \in \hat{\mathcal C}_2\}/|\hat{\mathcal C}_2|$, so that
$\bar{\mathbf{X}}_{\hat{\mathcal C}_1} - \bar{\mathbf{X}}_{\hat{\mathcal C}_2} = \mathbf{X}^\top\boldsymbol\nu$. To compute
\eqref{eq:intro_psel}, \citet{GaoBienWitten2024} also condition on statistics that remove the
nuisance parameters and on one further direction, which leaves a single scalar with a truncated
$\chi$ distribution (Section~\ref{sec:gao_review}), and they show that the truncation set can be
computed efficiently for hierarchical clustering with commonly used linkages.
\citet{ChenWitten2023} extended this approach to $k$-means by conditioning on
the full $k$-means clustering trajectory of Lloyd's algorithm. That is, they replace the event
$\hat{\mathcal C}_1, \hat{\mathcal C}_2 \in \mathcal C(\mathbf{X})$ in \eqref{eq:intro_psel} with
the event that Lloyd's algorithm repeats every observed assignment.

Both approaches require $\boldsymbol\Sigma$ in \eqref{eq:intro_model} to be known.
Section~\ref{sec:setup} shows that plugging in an estimate of $\boldsymbol\Sigma$ trades
Type I error control against power (Figure~\ref{fig:motivation}(b) and (c)).

In this work, we give an exactly valid selective test after clustering under model
\eqref{eq:intro_model} with unknown means and an unknown, unstructured covariance
$\boldsymbol\Sigma \succ 0$. To avoid estimating $\boldsymbol\Sigma$, we test a stricter null
hypothesis than \eqref{eq:intro_null},
\begin{equation}
\label{eq:intro_null_strong}
H_0' : \boldsymbol\mu_i = \boldsymbol\mu_{i'} \quad \text{for all } i, i' \in \hat{\mathcal C}_1 \cup \hat{\mathcal C}_2,
\end{equation}
which postulates that \emph{all observations in the two clusters share a common mean}
\citep{YunFoygelBarber2023} (Remark~\ref{rem:which_null} explains why the weaker null is
not enough without a known covariance).

Our first insight is that if we stack the $r = |\hat{\mathcal C}_1| + |\hat{\mathcal C}_2|$ rows from
$\hat{\mathcal C}_1$ and $\hat{\mathcal C}_2$ and center them into a matrix
$\mathbf{Z} \in \R^{(r-1) \times p}$, then under \eqref{eq:intro_null_strong}, $\mathbf{Z}$ has
independent $\N_p(\mathbf 0, \boldsymbol\Sigma)$ rows. We then take the polar decomposition
$\mathbf{Z} = \mathbf{Q}\,\mathbf{H}$, where $\mathbf{Q}$ is uniformly distributed on the set
$\St(r-1, p)$, i.e., all $(r-1) \times p$ matrices with orthonormal columns.
Importantly, the distribution of $\mathbf{Q}$ does not involve $\boldsymbol\Sigma$ and is
independent of $\mathbf{H}$, and therefore $\mathbf{Q}$ can be used as a pivotal statistic for
\eqref{eq:intro_null_strong}.

In addition to the novel test statistic construction inspired by classical results in
multivariate statistics \citep{Chikuse2003}, we also make algorithmic contributions to
conditional selective inference. Most existing selective inference frameworks reduce
the computation of the selective $p$-value to a one-dimensional truncation problem;
however, in our case, we need to sample $\mathbf{Q}$ truncated to a set. We propose a
reversible Markov-chain-based algorithm for sampling and use the serial construction of
\citet{BesagClifford1989}, which makes the Monte Carlo $p$-value exactly valid for any
number of draws. Finally, for $k$-means clustering, we show that each proposal can be
characterized exactly using analytical solutions. Conditioning on more information gives a
simpler alternative: fixing all of $\mathbf{Q}$ except the angle that determines
Hotelling's $T^2$ yields a one-dimensional truncated $p$-value $p_{\mathrm{polar,1d}}$, which is exact,
needs only a one-dimensional Monte Carlo sample for a general clustering algorithm, and needs none
for $k$-means when we condition on the Lloyd trajectory (Sections~\ref{sec:truncated_F_main}
and~\ref{sec:kmeans}). Tests based on our proposal
$p_{\mathrm{polar}}$ control the selective Type I error \eqref{eq:sel_typeI} in finite samples without any
knowledge of $\boldsymbol\Sigma$ (Figure~\ref{fig:motivation}(b)). In Figure~\ref{fig:motivation}(c) they even have at least as
much power as the test of \citet{ChenWitten2023} with the true $\boldsymbol\Sigma$, because that test conditions on the
whole sequence of $k$-means assignments and ours only on the returned clusters.

The methods are implemented in the \textsf{R} package \texttt{selectiveHotelling}, available at
\url{https://github.com/yiqunchen/selectiveHotelling}, and its tutorials are at
\url{https://yiqunchen.github.io/selectiveHotelling/}. The code that reproduces our results is at
\url{https://github.com/yiqunchen/selectiveHotelling-paper}.

\subsection{Related work}
\label{sec:related}

The methodological origin of this work, and the argument behind its validity, go
back to the general framework for conditional selective inference of
\citet{LeeSunSunTaylor2016} and \citet{FithianSunTaylor2017}. Beyond the
three works above, selective inference for clustering has been extended to testing
several pairs of clusters at once \citep{YunHe2024} and to dependent observations
\citep{GonzalezDelgadoEtAl2026}. It has also been extended to other procedures:
convex clustering \citep{BachocMaugisNeuvial2025}, deep clustering in a learned
latent space \citep{MizuiEtAl2026}, and multi-step clustering pipelines
\citep{MiyataEtAl2026}. All of these works
either assume a known covariance or plug in an estimate computed about the sample
mean. Such an estimate is conservative when the true means differ, because the
between-cluster spread inflates it \citep{GaoBienWitten2024,GonzalezDelgadoEtAl2026}.

One way to extend the conditional framework to an unknown covariance is to plug in
an estimate of it. \citet{GonzalezDelgadoEtAl2026} show that plugging the sample covariance into the tests
of \citet{GaoBienWitten2024} and \citet{ChenWitten2023} gives an asymptotically
conservative test that controls the selective Type I error, because the sample
covariance overestimates $\boldsymbol\Sigma$ asymptotically. The guarantee, however, needs the estimate to come
from data independent of the clustered sample, such as an independent copy. It also needs at least two well-separated true clusters. Moreover, when there is a true clustering structure, i.e., when the $\boldsymbol\mu_i$ are
not all equal, the sample covariance can lead to a substantial loss of power
(Figure~\ref{fig:motivation}(c)). An estimator that is universally consistent for
$\boldsymbol\Sigma$ without knowledge of $\boldsymbol\mu$ is a hard problem and remains an
active area of research, especially when $p$ is close to $n$
\citep{CaiRenZhou2016,FanLiaoLiu2016,LedoitWolf2020}.

Another line of work avoids conditioning by adding auxiliary randomness. It splits
one data set into two independent parts, one to choose the hypothesis and the other
to test it \citep{RasinesYoung2023}. Data fission
\citep{LeinerDuanWassermanRamdas2025} and data thinning
\citep{NeufeldDharamshiGaoWitten2024,DharamshiEtAl2024,NeufeldEtAl2024} construct such a
split systematically for a single realization from many parametric families,
including the multivariate Gaussian with unknown mean but \emph{known}
$\boldsymbol\Sigma$. Without $\boldsymbol\Sigma$, \citet{DharamshiEtAl2024Decomposing} prove that a
single multivariate Gaussian cannot be split into independent parts. With $n > 1$
observations they do construct independent parts, but these mix the rows of
$\mathbf{X}$, so an observation clustered in one part has no counterpart to test in the
other. We refer to \citet{NeufeldPerryWitten2026} for a comparison of the
conditioning and splitting approaches.

Clustering has also been studied from angles other than testing a selected pair.
One question is how many clusters there are, addressed by tests for the order of a
finite mixture \citep{ChenChenKalbfleisch2004,ChenLi2009,LiChen2010}, by permutation
methods for the number of factors \citep{Dobriban2020}, and by the mixture-model
literature reviewed in \citet{McLachlanLeeRathnayake2019}. Another is whether a
cluster, or the variables that drive it, are reproducible, assessed by resampling
\citep{KerrChurchill2001,SuzukiShimodaira2006}, by significance tests for
hierarchical clustering and for cluster membership \citep{KimesEtAl2017,Chung2020},
and by related tests of systematic variation and exchangeability
\citep{ChungStorey2015,JinWang2016,AwSpenceSong2021}. Bayesian work studies the
posterior of the number of components and of the mixture parameters
\citep{Nobile2004,GuhaHoNguyen2019,CaiCampbellBroderick2020}, and since $k$-means is a
limiting case of the expectation-maximization algorithm, statistical guarantees for
that algorithm are also relevant
\citep{WangEtAl2015,YiCaramanis2015,BalakrishnanWainwrightYu2017,CaiMaZhang2019}.
Many of these results rely on a consistent estimate of the cluster memberships, or at
least on the correct number of clusters. By contrast, we follow the conditional
selective inference framework and give a test that is valid whether or not the
estimated clusters correspond to true ones.

\subsection{Notation and organization}
\label{sec:organization}

The data are $\mathbf{X} \in \R^{n \times p}$ with rows $\mathbf{X}_1, \ldots, \mathbf{X}_n$, and $\mathbf{x}$
denotes a realization of $\mathbf{X}$; for any matrix, the subscript $i$ denotes its $i$th row, and
$\|\cdot\|_F$ is the Frobenius norm. We write $\MN_{n \times p}(\boldsymbol\mu, \mathbf{U}, \mathbf{V})$ for the matrix
normal distribution with mean $\boldsymbol\mu$, row covariance $\mathbf{U}$, and column covariance $\mathbf{V}$;
$\one_k$, $\mathbf{I}_k$, and $\mathbf 0$ for the $k$-vector of ones, the $k \times k$ identity, and a zero vector or
matrix; $\mathrm{dir}(\cdot)$ for a nonzero vector or matrix divided by its $\ell_2$ or Frobenius norm; and $\one\{\cdot\}$ for the indicator function. The clustering rule applied to $\mathbf{x}$ returns the partition
$\mathcal C(\mathbf{x}) = \{\hat{\mathcal C}_1, \ldots, \hat{\mathcal C}_K\}$ of $\{1, \ldots, n\}$. Finally, $\St(r-1, p)$ is the
set of $(r-1) \times p$ matrices with orthonormal columns.

The remainder of the paper is organized as follows. Section~\ref{sec:setup} reviews
the test of \citet{GaoBienWitten2024} for a known covariance, shows why estimating the
covariance trades Type I error control against power, and writes the test of
\citet{YunFoygelBarber2023} as a polar decomposition of the centered tested rows.
Section~\ref{sec:method} replaces that decomposition by the matrix polar decomposition,
writes Hotelling's $T^2$ as an angle, and gives two exact tests for any clustering
algorithm: the one-dimensional test $p_{\mathrm{polar,1d}}$ and the test
$p_{\mathrm{polar}}$. Section~\ref{sec:computation} computes both for any clustering
algorithm by combining a Markov chain with the construction of \citet{BesagClifford1989}, and
then discusses closed forms for $k$-means and hierarchical clustering. Section~\ref{sec:sim} reports the simulation study,
including the comparison with plug-in covariance estimates and data thinning, and
Section~\ref{sec:realdata} applies the test to leukemia, single-cell RNA sequencing, and drug review
data. Section~\ref{sec:discussion} concludes, and proofs and additional results are in
the appendices.

\section{Testing for a difference in means between two estimated clusters}
\label{sec:setup}

This section reviews selective inference for clustering when the covariance is
known (Section~\ref{sec:gao_review}) and explains why estimating the covariance does
not work (Section~\ref{sec:challenges}). We then end with the intuition behind our
approach (Section~\ref{sec:intuition}).

Throughout, we work under data model \eqref{eq:intro_model}.

\subsection{Selective inference for clustering with a known covariance}
\label{sec:gao_review}

We first review the test of \citet{GaoBienWitten2024} in the setting where
$\boldsymbol\Sigma$ in \eqref{eq:intro_model} is known. If the clusters $\hat{\mathcal C}_1$ and $\hat{\mathcal C}_2$
had been chosen before the data were seen, we would test \eqref{eq:intro_null} with the
difference in sample means $\mathbf{X}^\top\boldsymbol\nu = \bar{\mathbf{X}}_{\hat{\mathcal C}_1} - \bar{\mathbf{X}}_{\hat{\mathcal C}_2}$, where
$\boldsymbol\nu$ is the contrast vector defined in Section~\ref{sec:intro}. Under \eqref{eq:intro_null},
$\mathbf{X}^\top\boldsymbol\nu \sim \N_p(\mathbf 0, \|\boldsymbol\nu\|_2^2\,\boldsymbol\Sigma)$, so the whitened norm
$\|\boldsymbol\Sigma^{-1/2}\mathbf{X}^\top\boldsymbol\nu\|_2$ follows the scaled distribution
$\|\boldsymbol\nu\|_2 \cdot \chi_p$.

\begin{figure}[t]
\centering
\includegraphics[width=0.9\linewidth]{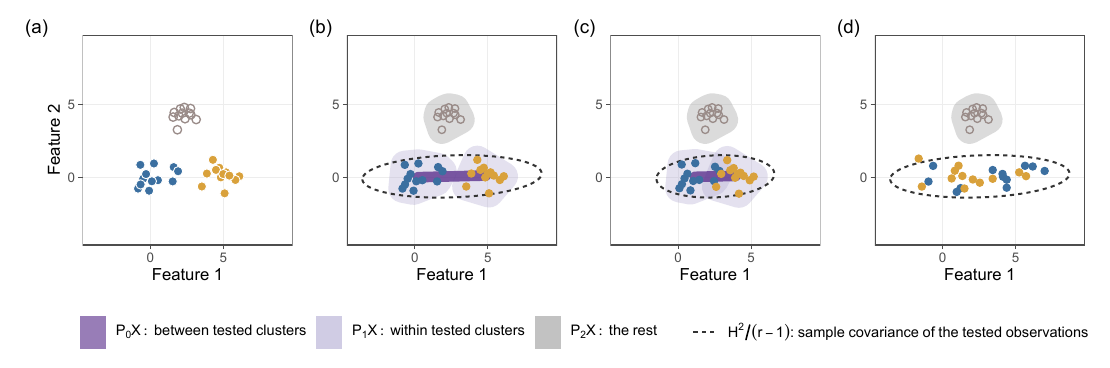}
\caption{The split $\mathbf{X} = \mathbf P_0\mathbf{X} + \mathbf P_1\mathbf{X} + \mathbf P_2\mathbf{X}$ for $n = 36$ observations
from three bivariate Gaussian groups, clustered into $K = 3$ clusters by $k$-means; open circles are the cluster that is
not tested. \textbf{(a)} The data. \textbf{(b)} The three parts: $\mathbf P_0\mathbf{X}$ joins the two tested cluster means,
$\mathbf P_1\mathbf{X}$ covers the deviations within each tested cluster, and $\mathbf P_2\mathbf{X}$ covers the untested
observations (it also holds the overall mean of the two tested clusters). \textbf{(c)} A data set with $\mathbf P_1\mathbf{X}$ and
$\mathbf P_2\mathbf{X}$ held fixed, as in \citet{GaoBienWitten2024}: the two tested clusters keep their shapes and move along the
line between their means. \textbf{(d)} A data set with $\mathbf P_2\mathbf{X}$ and $\mathbf{H}$ held fixed, as in our test: the
tested observations are redrawn around the same overall mean with the same sample covariance, and the untested observations do
not move. The dashed ellipse in (b)--(d) is the $95\%$ ellipse of the sample covariance matrix $\mathbf{H}^2/(r-1)$ of the tested
observations;
it is the same in (b) and (d) and smaller in (c).}
\label{fig:projections}
\end{figure}

Because the clusters were chosen from the data, this is not the right null
distribution, and the selective $p$-value $p_{\mathrm{sel}}$ in \eqref{eq:intro_psel} conditions on
the event that the clustering returns $\hat{\mathcal C}_1$ and $\hat{\mathcal C}_2$. That probability still depends on
the unknown means. It helps to split the data with three projections that we use throughout the paper,
$\mathbf{X} = \mathbf P_0\mathbf{X} + \mathbf P_1\mathbf{X} + \mathbf P_2\mathbf{X}$. Here
$\mathbf P_0 = \boldsymbol\nu\boldsymbol\nu^\top/\|\boldsymbol\nu\|_2^2$ projects onto the difference of the two cluster
means; $\mathbf P_1 = (\mathbf{I}_{\hat{\mathcal C}_1} - \one_{\hat{\mathcal C}_1}\one_{\hat{\mathcal C}_1}^\top/|\hat{\mathcal C}_1|)
+ (\mathbf{I}_{\hat{\mathcal C}_2} - \one_{\hat{\mathcal C}_2}\one_{\hat{\mathcal C}_2}^\top/|\hat{\mathcal C}_2|)$ projects onto the
deviations of each observation from the mean of its own cluster, with $\mathbf{I}_{\mathcal G}$
the diagonal matrix with ones in the positions of $\mathcal G$ and $\one_{\mathcal G}$ the
indicator vector of $\mathcal G$; and $\mathbf P_2 = \mathbf{I}_n - \mathbf P_0 - \mathbf P_1$ projects
onto the rest, namely the overall mean of $\hat{\mathcal C}_1 \cup \hat{\mathcal C}_2$ and the observations
outside the two clusters. In words, $\mathbf P_0\mathbf{X}$ is the variation between the two tested clusters,
$\mathbf P_1\mathbf{X}$ the variation within them, and $\mathbf P_2\mathbf{X}$ everything else;
Figure~\ref{fig:projections}(a) shows a clustering and Figure~\ref{fig:projections}(b) its three parts. We write every test in
this paper, theirs and ours, in terms of this split, so that the tests differ only in which parts they hold
fixed and which part they let vary. \citet{GaoBienWitten2024} decompose the data into the part along
$\boldsymbol\nu$ and the part orthogonal to it,
\[
\mathbf{X} = \mathbf P_1\mathbf{X} + \mathbf P_2\mathbf{X}
+ \frac{\|\boldsymbol\Sigma^{-1/2}\mathbf{X}^\top\boldsymbol\nu\|_2}{\|\boldsymbol\nu\|_2^2}\,
\boldsymbol\nu\,\mathrm{dir}\big(\boldsymbol\Sigma^{-1/2}\mathbf{X}^\top\boldsymbol\nu\big)^\top\boldsymbol\Sigma^{1/2},
\]
where the last term is $\mathbf P_0\mathbf{X}$ and $\mathbf P_1 + \mathbf P_2 = \mathbf{I}_n - \mathbf P_0$
is the projection orthogonal to $\boldsymbol\nu$. They compute instead
\begin{equation}
\label{eq:intro_gao}
\begin{aligned}
p_{\boldsymbol\Sigma}
= \Prob_{H_0}\Big( &\|\boldsymbol\Sigma^{-1/2}\mathbf{X}^\top\boldsymbol\nu\|_2
\ge \|\boldsymbol\Sigma^{-1/2}\mathbf{x}^\top\boldsymbol\nu\|_2
\;\Big|\; \hat{\mathcal C}_1, \hat{\mathcal C}_2 \in \mathcal C(\mathbf{X}),\\
&\mathbf P_1\mathbf{X} = \mathbf P_1\mathbf{x},\; \mathbf P_2\mathbf{X} = \mathbf P_2\mathbf{x},\;
\mathrm{dir}\big(\boldsymbol\Sigma^{-1/2}\mathbf{X}^\top\boldsymbol\nu\big)
= \mathrm{dir}\big(\boldsymbol\Sigma^{-1/2}\mathbf{x}^\top\boldsymbol\nu\big) \Big).
\end{aligned}
\end{equation}

The conditions on $\mathbf P_1\mathbf{X}$ and $\mathbf P_2\mathbf{X}$
remove the unknown means $(\mathbf P_1 + \mathbf P_2)\boldsymbol\mu$, which are nuisance parameters under $H_0$, and the
condition on $\mathrm{dir}(\cdot)$ is additional conditioning used to enable efficient
computation of the $p$-value. Only the scalar $\|\boldsymbol\Sigma^{-1/2}\mathbf{X}^\top\boldsymbol\nu\|_2$
then remains random. Because $p_{\boldsymbol\Sigma}$ is uniform under $H_0$ given more information
than the clustering, it also controls the selective Type I error in the sense of \eqref{eq:sel_typeI}
(Proposition~\ref{prop:gbw}).
To compute $p_{\boldsymbol\Sigma}$, replace the observed value of the statistic by $\phi \ge 0$ and define
the perturbed data and the values of $\phi$ that reproduce the clustering,
\begin{equation}
\label{eq:gbw_S}
\mathbf{x}'(\phi) = (\mathbf P_1 + \mathbf P_2)\mathbf{x}
+ \frac{\phi}{\|\boldsymbol\nu\|_2^2}\,\boldsymbol\nu\,
\mathrm{dir}\big(\boldsymbol\Sigma^{-1/2}\mathbf{x}^\top\boldsymbol\nu\big)^\top\boldsymbol\Sigma^{1/2},
\qquad
\mathcal S_{\boldsymbol\Sigma} = \big\{\phi \ge 0 : \hat{\mathcal C}_1, \hat{\mathcal C}_2 \in \mathcal C(\mathbf{x}'(\phi))\big\}.
\end{equation}

\begin{proposition}[\citealp{GaoBienWitten2024}, Theorem~1, applied to $\mathbf{X}\boldsymbol\Sigma^{-1/2}$]
\label{prop:gbw}
Let $\mathbf{X}$ follow \eqref{eq:intro_model} with $\boldsymbol\Sigma$ known. For any realization
$\mathbf{x}$ and any two non-overlapping groups $\hat{\mathcal C}_1, \hat{\mathcal C}_2$,
\[
p_{\boldsymbol\Sigma} = 1 - \mathbb F\big(\|\boldsymbol\Sigma^{-1/2}\mathbf{x}^\top\boldsymbol\nu\|_2;\,
\|\boldsymbol\nu\|_2,\, \mathcal S_{\boldsymbol\Sigma}\big),
\]
where $\mathbb F(\,\cdot\,; c, \mathcal S)$ is the distribution function of $c \cdot \chi_p$ truncated
to $\mathcal S$. If \eqref{eq:intro_null} holds, then
$\Prob_{H_0}\big(p_{\boldsymbol\Sigma} \le \alpha \mid \hat{\mathcal C}_1, \hat{\mathcal C}_2 \in \mathcal C(\mathbf{X})\big) = \alpha$
for every $\alpha \in [0,1]$.
\end{proposition}

The perturbed data set $\mathbf{x}'(\phi)$ keeps $\mathbf P_1\mathbf{x}$ and $\mathbf P_2\mathbf{x}$ and rescales only
$\mathbf P_0\mathbf{x}$, so it leaves every observation outside $\hat{\mathcal C}_1 \cup \hat{\mathcal C}_2$ in place and
keeps the deviations within each tested cluster. It moves the two clusters apart when $\phi$ exceeds the
observed statistic and together when $\phi$ is below it, along the direction of
$\bar{\mathbf{x}}_{\hat{\mathcal C}_1} - \bar{\mathbf{x}}_{\hat{\mathcal C}_2}$ (Figure~\ref{fig:projections}(c)). The set
$\mathcal S_{\boldsymbol\Sigma}$ therefore collects every value of the statistic that would have
produced the same clusters had it been observed instead. Proposition~\ref{prop:gbw}
says that the scaled $\chi_p$ distribution remains the right null distribution once
it is truncated to this set. For hierarchical clustering with common linkages,
$\mathcal S_{\boldsymbol\Sigma}$ is available in closed form \citep{GaoBienWitten2024}. For $k$-means,
\citet{ChenWitten2023} condition on the full $k$-means clustering trajectory of Lloyd's algorithm instead of on
$\hat{\mathcal C}_1, \hat{\mathcal C}_2 \in \mathcal C(\mathbf{X})$, and the corresponding set is a finite union of
intervals whose endpoints solve quadratic equations. With $\boldsymbol\Sigma$ known, the weaker
null \eqref{eq:intro_null} suffices, because the nuisance $(\mathbf P_1 + \mathbf P_2)\boldsymbol\mu$ is
removed by conditioning on $(\mathbf P_1 + \mathbf P_2)\mathbf{X}$.

\subsection{Challenges in estimating the covariance}
\label{sec:challenges}

In practice $\boldsymbol\Sigma$ is unknown and must be estimated from the same data. Two natural estimates
both fail, in ways that the clustering explains (Figure~\ref{fig:motivation}).
\begin{itemize}
\item \emph{The within-cluster covariance}
\[
\hat{\boldsymbol\Sigma}_{\mathrm{cluster}} = \frac{1}{n-K}\sum_{k=1}^K \sum_{i \in \hat{\mathcal C}_k}
(\mathbf{x}_i - \bar{\mathbf{x}}_{\hat{\mathcal C}_k})(\mathbf{x}_i - \bar{\mathbf{x}}_{\hat{\mathcal C}_k})^\top
\]
is consistent when the clusters are fixed in advance. But $k$-means minimizes the within-cluster sum of squares, so
when it splits a true group, $\hat{\boldsymbol\Sigma}_{\mathrm{cluster}}$ is too small along the tested direction
$\bar{\mathbf{x}}_{\hat{\mathcal C}_1} - \bar{\mathbf{x}}_{\hat{\mathcal C}_2}$. With a single Gaussian group, the test of
\citet{ChenWitten2023} with $\hat{\boldsymbol\Sigma}_{\mathrm{cluster}}$, which we write
$p_{\mathrm{CW}}(\hat{\boldsymbol\Sigma}_{\mathrm{cluster}})$, rejects $16.1\%$ of true null hypotheses at level $0.05$
(Figure~\ref{fig:motivation}(b)).
\item \emph{The sample covariance}
\[
\hat{\boldsymbol\Sigma}_{\mathrm{sample}} = \frac{1}{n-1}\sum_{i=1}^n (\mathbf{x}_i - \bar{\mathbf{x}})(\mathbf{x}_i - \bar{\mathbf{x}})^\top
\]
\citep{GonzalezDelgadoEtAl2026} overestimates $\boldsymbol\Sigma$ when there are separated groups. This protects the
Type I error but costs power, because the separation between the tested clusters inflates
$\hat{\boldsymbol\Sigma}_{\mathrm{sample}}$ along the tested direction. At the separation $\delta = 5$ of
Figure~\ref{fig:motivation}(a), $p_{\mathrm{CW}}(\hat{\boldsymbol\Sigma}_{\mathrm{sample}})$ rejects $18\%$ of the pairs from
\looseness=-1 different true groups, against $78\%$ with the true $\boldsymbol\Sigma$ (Figure~\ref{fig:motivation}(c)).
\end{itemize}
Our test does not estimate $\boldsymbol\Sigma$ (Section~\ref{sec:method}) and avoids this trade-off: it rejects $5.7\%$
of $900$ true null hypotheses at level $0.05$, and $99\%$ of the pairs from different groups at $\delta = 5$
(Figure~\ref{fig:motivation}(b) and (c)).

\subsection{Extension to an unknown variance}
\label{sec:intuition}

When $\boldsymbol\Sigma$ is unknown, the length $\|\boldsymbol\Sigma^{-1/2}\mathbf{X}^\top\boldsymbol\nu\|_2$ in
Section~\ref{sec:gao_review} has an unknown distribution, so we need a test statistic
whose null distribution does not depend on the covariance. Two-sample tests have this
property, because they depend on the data only through an angle. We explain the angle
with the decomposition of \citet{YunFoygelBarber2023} for $\boldsymbol\Sigma = \sigma^2\mathbf{I}_p$,
and then describe how we move to an arbitrary $\boldsymbol\Sigma$.

\citet{YunFoygelBarber2023} use the same decomposition
$\mathbf{X} = \mathbf P_0\mathbf{X} + \mathbf P_1\mathbf{X} + \mathbf P_2\mathbf{X}$ of Section~\ref{sec:gao_review}. Their
test statistic is
\begin{equation}
\label{eq:yb_R}
R = (r-2)\,\frac{\|\mathbf P_0\mathbf{X}\|_F^2}{\|\mathbf P_1\mathbf{X}\|_F^2},
\qquad r = |\hat{\mathcal C}_1| + |\hat{\mathcal C}_2|,
\end{equation}
which compares the spread between the two clusters with the spread within them.

To see why the distribution of $R$ does not involve $\sigma^2$, consider
$(\mathbf P_0 + \mathbf P_1)\mathbf{X}$, the deviations of the tested observations from their common
mean. The matrix $\mathbf P_0 + \mathbf P_1$ is the projection onto an $(r-1)$-dimensional
subspace of $\R^n$, so under null \eqref{eq:intro_null_strong},
$(\mathbf P_0 + \mathbf P_1)\mathbf{X} \sim \MN_{n \times p}(\mathbf 0, \mathbf P_0 + \mathbf P_1, \sigma^2\mathbf{I}_p)$; that is, $\mathrm{vec}\big((\mathbf P_0 + \mathbf P_1)\mathbf{X}\big)$ has independent $\N(0, \sigma^2)$
coordinates in an $(r-1)p$-dimensional subspace of $\R^{np}$. Such a Gaussian vector splits
into its length and its direction. The squared length,
$\|(\mathbf P_0 + \mathbf P_1)\mathbf{X}\|_F^2 \sim \sigma^2\chi^2_{(r-1)p}$, is a sufficient statistic
for $\sigma^2$. The direction
$(\mathbf P_0 + \mathbf P_1)\mathbf{X}/\|(\mathbf P_0 + \mathbf P_1)\mathbf{X}\|_F$ is uniformly distributed on the unit
sphere of that subspace, whatever $\sigma^2$ is. Moreover, the length and the direction are
independent \citep{Muirhead1982,Chikuse2003}.

The statistic $R$ is a function of the direction only. Write
$w = \|\mathbf P_0\mathbf{X}\|_F^2/\|(\mathbf P_0 + \mathbf P_1)\mathbf{X}\|_F^2$ for the fraction of the squared length of $(\mathbf P_0 + \mathbf P_1)\mathbf{X}$ that comes from the
difference between the two cluster means. The remaining fraction,
$1 - w = \|\mathbf P_1\mathbf{X}\|_F^2/\|(\mathbf P_0 + \mathbf P_1)\mathbf{X}\|_F^2$, comes from the deviations of
the observations from their own cluster means, and $R = (r-2)\,w/(1-w)$. A value of $w$ close to one
means that the observations sit close to their own cluster means and the two means are far
apart relative to that spread, so the clusters look well separated and $R$ is large. In
that case most of the variation of the tested observations is between the two clusters
rather than within them, which we do not expect when \eqref{eq:intro_null_strong} holds. A
value of $w$ close to zero means that the spread within the clusters dominates the difference
between their means. Because the direction is uniformly distributed on the sphere and the
range of $\mathbf P_0$ is a $p$-dimensional part of the $(r-1)p$-dimensional subspace,
$w \sim \mathrm{Beta}(p/2, (r-2)p/2)$ for every $\sigma^2$, and therefore
$R \sim F_{p,\,(r-2)p}$. It turns out that $R$ is just one such choice. Any other ratio of
$\|\mathbf P_0\mathbf{X}\|_F^2$, $\|\mathbf P_1\mathbf{X}\|_F^2$, and $\|(\mathbf P_0 + \mathbf P_1)\mathbf{X}\|_F^2$, such as $w$,
is a monotone function of $R$ and gives an equivalent test.

The construction rests on two ingredients, which Section~\ref{sec:method} reuses for an
arbitrary $\boldsymbol\Sigma$. First, we condition on statistics that are sufficient for the
nuisance parameters under the null, here the length for $\sigma^2$ and $\mathbf P_2\mathbf{X}$ for the
means, so that the conditional distribution of the data does not involve them
\citep[Chapter~4]{LehmannRomano2005}. Second, the test statistic is a function of the
remaining part, here the direction, which is independent of the sufficient statistics; its
conditional distribution is therefore its known unconditional one. The direction also
splits further: $w$ and the two directions $\mathrm{dir}(\mathbf P_0\mathbf{X})$ and
$\mathrm{dir}(\mathbf P_1\mathbf{X})$ are independent. Testing with a direction that is uniformly
distributed on a sphere has also been applied to testing in sparse linear models
\citep{SenguptaJanson2024,PaulsonSenguptaJanson2025}.

After clustering, $R$ no longer has an $F$ distribution, and conditioning on the
clustering alone is not enough to compute its distribution. The event
$\hat{\mathcal C}_1, \hat{\mathcal C}_2 \in \mathcal C(\mathbf{X})$ depends on the data through $\mathbf P_2\mathbf{X}$, the length,
and the whole direction, not only through $R$, so the conditional distribution of $R$
given the event would require an integral over all directions that reproduce the
clusters, which has no closed form. \citet{YunFoygelBarber2023} therefore condition on
every part of the data other than $R$:
\begin{equation}
\label{eq:yb_pvalue}
\begin{aligned}
p_{\mathrm{YB}} = \Prob_{H_0'}\Big( &R(\mathbf{X}) \ge R(\mathbf{x}) \;\Big|\; \hat{\mathcal C}_1, \hat{\mathcal C}_2 \in \mathcal C(\mathbf{X}),\
\mathbf P_2\mathbf{X} = \mathbf P_2\mathbf{x},\\
&\|(\mathbf P_0 + \mathbf P_1)\mathbf{X}\|_F = \|(\mathbf P_0 + \mathbf P_1)\mathbf{x}\|_F,\
\mathrm{dir}(\mathbf P_0\mathbf{X}) = \mathrm{dir}(\mathbf P_0\mathbf{x}),\ \mathrm{dir}(\mathbf P_1\mathbf{X}) = \mathrm{dir}(\mathbf P_1\mathbf{x}) \Big).
\end{aligned}
\end{equation}
The conditions on $\mathbf P_2\mathbf{X}$ and on the length remove the nuisance parameters, which
are the means and $\sigma^2$; this uses the first ingredient. The conditions on the two
directions are additional conditioning, as is the condition on $\mathrm{dir}(\cdot)$ in
\eqref{eq:intro_gao}. Given all four, the data are a function of $R$ alone: replacing the observed $R$ by
$\rho \ge 0$ gives the perturbed data set
\begin{equation}
\label{eq:yb_perturb}
\mathbf{x}'(\rho) = \mathbf P_2\mathbf{x} + \|(\mathbf P_0 + \mathbf P_1)\mathbf{x}\|_F\Big(\sqrt{\tfrac{\rho}{\rho + r - 2}}\,\mathrm{dir}(\mathbf P_0\mathbf{x})
+ \sqrt{\tfrac{r-2}{\rho + r - 2}}\,\mathrm{dir}(\mathbf P_1\mathbf{x})\Big),
\end{equation}
as $\mathbf{x}'(\phi)$ in \eqref{eq:gbw_S}, and by the second ingredient $R$ keeps its $F_{p,\,(r-2)p}$ distribution given the
length and the two directions. So $p_{\mathrm{YB}}$ is the upper tail of the
$F_{p,\,(r-2)p}$ distribution truncated to the set
$\mathcal S_{\mathrm{YB}} = \{\rho \ge 0 : \hat{\mathcal C}_1, \hat{\mathcal C}_2 \in \mathcal C(\mathbf{x}'(\rho))\}$. Section~\ref{sec:conditioning_less} takes the other
route in our setting: it drops the additional conditioning and computes the integral
over directions by Monte Carlo.

What does it take to extend this argument to a general $\boldsymbol\Sigma$? Both ingredients
rely on $\mathrm{vec}((\mathbf P_0 + \mathbf P_1)\mathbf{X})$ being an isotropic Gaussian vector. For a
general $\boldsymbol\Sigma$, the entries within each row are correlated, so the direction is not
uniformly distributed and the distribution of $R$ depends on $\boldsymbol\Sigma$. The
intuition still applies. In
Section~\ref{sec:method}, the polar decomposition of a matrix splits the distribution
of $(\mathbf P_0 + \mathbf P_1)\mathbf{X}$ into a part whose distribution involves $\boldsymbol\Sigma$ and
an independent part whose distribution does not depend on $\boldsymbol\Sigma$, and
an analogue of $R$ in \eqref{eq:yb_R} leads to the familiar Hotelling's $T^2$ statistic.

\begin{remark}[The null hypothesis we test]
\label{rem:which_null}
Both \citet{YunFoygelBarber2023} and our work test \eqref{eq:intro_null_strong} rather than
\eqref{eq:intro_null}. The reason is that both tests use the distribution of
$(\mathbf P_0 + \mathbf P_1)\mathbf{X}$, which has mean zero exactly when all $r$ tested observations
share one mean, because the only vectors on the tested observations that $\mathbf P_0 + \mathbf P_1$ maps to zero are the
constant ones. Under \eqref{eq:intro_null}, $\mathbf P_1\mathbf{X}$ has a
nonzero mean whenever the observations within a cluster have different means.
\end{remark}

\section{Selective $p$-value under an unknown covariance}
\label{sec:method}

This section builds on the decomposition in Section~\ref{sec:intuition} and proposes a
selective test for an arbitrary clustering algorithm under an unknown covariance
matrix. Section~\ref{subsec:polar} replaces the length $\|(\mathbf P_0 + \mathbf P_1)\mathbf{X}\|_F$ by
a matrix and shows that the remaining factor $\mathbf{Q}$ is uniformly distributed for every
$\boldsymbol\Sigma$. Section~\ref{sec:hotelling} shows that Hotelling's $T^2$ is a function of
one angle of $\mathbf{Q}$, which is the angle of Section~\ref{sec:intuition} when $p = 1$.
Section~\ref{subsec:selection} conditions on the clustering and obtains the selective
$p$-value $p_{\mathrm{polar}}$, and Section~\ref{sec:onedim} conditions on more to obtain the
one-dimensional test $p_{\mathrm{polar,1d}}$. Section~\ref{sec:computation} explains how to
compute both.

Throughout, the tested clusters have sizes $n_1 = |\hat{\mathcal C}_1|$ and $n_2 = |\hat{\mathcal C}_2|$, with
$r = n_1 + n_2 \ge p + 2$, which the pooled covariance in Hotelling's $T^2$ requires even
without clustering. Theorem~\ref{thm:polar_uniform} and Proposition~\ref{prop:T2_share} hold for any two fixed groups of
observations. From Section~\ref{subsec:selection} on, $\hat{\mathcal C}_1$ and $\hat{\mathcal C}_2$ are clusters returned by a clustering
algorithm that is a deterministic function of the data (for example, $k$-means from fixed initial centroids), and the
tested pair is fixed in advance or every pair is tested.

\subsection{The polar decomposition of $(\mathbf P_0 + \mathbf P_1)\mathbf{X}$}
\label{subsec:polar}
\label{subsec:uniform}

Section~\ref{sec:intuition} split $(\mathbf P_0 + \mathbf P_1)\mathbf{X}$, the deviations of the tested
observations from their common mean, into a length and a direction. The polar
decomposition makes the same split with a matrix in place of the scalar length. When
$(\mathbf P_0 + \mathbf P_1)\mathbf{X}$ has rank $p$, which holds with probability one when $r - 1 \ge p$,
\begin{equation}
\label{eq:polar_dec}
(\mathbf P_0 + \mathbf P_1)\mathbf{X} = \mathbf{Q}\,\mathbf{H},
\end{equation}
where $\mathbf{H}$ is symmetric positive definite and the polar factor $\mathbf{Q}$ is an
$n \times p$ matrix with orthonormal columns. The columns of $\mathbf{Q}$ lie in the
$(r-1)$-dimensional range of $\mathbf P_0 + \mathbf P_1$, so $\mathbf{Q}$ belongs to the set $\mathcal V$ of
$n \times p$ matrices with orthonormal columns in that range, which is a copy of the
Stiefel manifold $\St(r-1, p)$. The matrix $\mathbf{H}$ is the matrix square root of $r - 1$ times the sample covariance
matrix of the tested observations. The polar factor $\mathbf{Q}$ plays
the role of the direction, and $\mathbf{H}$ plays the role of the length; when $p = 1$ the
two decompositions are the same (Appendix~\ref{appendix:polar}).

\begin{theorem}[Distribution of the polar factor]
\label{thm:polar_uniform}
\label{cor:conditional}
Under data model \eqref{eq:intro_model} and null \eqref{eq:intro_null_strong}, $\mathbf{Q}$ is
uniformly distributed on $\mathcal V$ and is independent of $(\mathbf{H}, \mathbf P_2\mathbf{X})$.
\end{theorem}

This follows from a classical result on the polar decomposition of a Gaussian matrix
\citep{Muirhead1982,Chikuse2003}; Appendix~\ref{appendix:polar} gives a proof in our
notation. The two ingredients of Section~\ref{sec:intuition} therefore hold for every
$\boldsymbol\Sigma$: the means and $\boldsymbol\Sigma$ enter only through $(\mathbf P_2\mathbf{X}, \mathbf{H})$, and $\mathbf{Q}$
is independent of them with a distribution that involves no unknown parameter. Figure~\ref{fig:projections}(d) shows a data set
with $\mathbf P_2\mathbf{X}$ and $\mathbf{H}$ held fixed and $\mathbf{Q}$ drawn from its uniform distribution. This suggests that
we can construct a pivotal statistic based on $\mathbf{Q}$.

\subsection{Hotelling's $T^2$ as a function of $\mathbf{Q}$}
\label{sec:hotelling}

Which function of $\mathbf{Q}$ should we use? In Section~\ref{sec:intuition}, $R$ compares the
part of the direction of $(\mathbf P_0 + \mathbf P_1)\mathbf{X}$ that lies along $\mathbf P_0$, the between-cluster
part, with the part along $\mathbf P_1$, the within-cluster part. The matrix $\mathbf P_0$ projects onto
the one-dimensional column space of $\mathbf P_0\mathbf{X}$, the direction of between-cluster variation.
The analogue of $w$ for the polar factor is the squared length of the projection of $\mathbf{Q}$
onto that direction, $u(\mathbf{Q}) = \|\mathbf P_0\mathbf{Q}\|_F^2 \in [0, 1]$. As with $R$, a large value of $u(\mathbf{Q})/\{1 - u(\mathbf{Q})\}$ means that the polar factor is
dominated by between-cluster variation, which is unlikely under the null. This ratio is a
scaled version of the usual two-sample Hotelling statistic.

\begin{proposition}[Hotelling's $T^2$ as a function of $u(\mathbf{Q})$]
\label{prop:T2_share}
\label{lem:angle_beta}
The two-sample Hotelling statistic,
\begin{equation*}
T^2 = \frac{n_1 n_2}{r}\,
(\bar{\mathbf{X}}_{\hat{\mathcal C}_1} - \bar{\mathbf{X}}_{\hat{\mathcal C}_2})^\top \mathbf{S}_{\mathrm{pool}}^{-1}
(\bar{\mathbf{X}}_{\hat{\mathcal C}_1} - \bar{\mathbf{X}}_{\hat{\mathcal C}_2}),
\qquad
\mathbf{S}_{\mathrm{pool}} = \frac{\mathbf{X}^\top\mathbf P_1\mathbf{X}}{r - 2},
\end{equation*}
where $\mathbf{S}_{\mathrm{pool}}$ is the pooled within-cluster covariance matrix, satisfies
\begin{equation}
\label{eq:T2_share}
T^2 = (r-2)\,\frac{u(\mathbf{Q})}{1 - u(\mathbf{Q})} .
\end{equation}
When $p = 1$, $T^2 = R$ in \eqref{eq:yb_R}. Under data model \eqref{eq:intro_model} and null
\eqref{eq:intro_null_strong}, $u(\mathbf{Q}) \sim \mathrm{Beta}\big(\tfrac p2, \tfrac{r-1-p}2\big)$;
equivalently, $\frac{r-p-1}{(r-2)p}\,T^2 \sim F_{p,\,r-p-1}$.
\end{proposition}

The proof is in Appendix~\ref{appendix:polar}. The connection is classical: $T^2$ is the
largest squared two-sample $t$-statistic over all one-dimensional projections of the data,
attained at the Fisher discriminant direction \citep{Hotelling1931,Anderson2003}.

\subsection{The selective $p$-value}
\label{subsec:selection}

Proposition~\ref{prop:T2_share} describes $T^2$ before clustering. After clustering, as in
\eqref{eq:intro_gao} and \eqref{eq:yb_pvalue}, we condition on the clustering and on the
sufficient statistics for the nuisance parameters, and propose the selective $p$-value
\begin{equation}
\label{eq:pstar}
p_{\mathrm{polar}} = \Prob_{H_0'}\Big(T^2(\mathbf{X}) \ge T^2(\mathbf{x}) \;\Big|\;
\hat{\mathcal C}_1, \hat{\mathcal C}_2 \in \mathcal C(\mathbf{X}),\
\mathbf P_2\mathbf{X} = \mathbf P_2\mathbf{x},\ \mathbf{H}(\mathbf{X}) = \mathbf{H}(\mathbf{x})\Big).
\end{equation}
The conditioning fixes $\hat{\mathcal C}_1$ and $\hat{\mathcal C}_2$, and with them $\mathbf P_0$, $\mathbf P_1$, and $\mathbf P_2$.
As in \eqref{eq:gbw_S} and \eqref{eq:yb_perturb}, replacing the observed polar factor by
$\mathbf{Q} \in \mathcal V$ gives the perturbed data set
\begin{equation*}
\mathbf{x}'(\mathbf{Q}) = \mathbf P_2\mathbf{x} + \mathbf{Q}\,\mathbf{H}(\mathbf{x}),
\qquad
\mathcal S_{\mathbf{Q}} = \big\{\mathbf{Q} \in \mathcal V : \hat{\mathcal C}_1, \hat{\mathcal C}_2 \in \mathcal C(\mathbf{x}'(\mathbf{Q}))\big\},
\end{equation*}
where $\mathcal S_{\mathbf{Q}}$ is the set of polar factors that reproduce the tested clusters.

\begin{proposition}[The selective $p$-value]
\label{prop:conditional_selection}
Under data model \eqref{eq:intro_model} and null \eqref{eq:intro_null_strong}, conditional on
$\mathbf P_2\mathbf{X}$, $\mathbf{H}$, and $\hat{\mathcal C}_1, \hat{\mathcal C}_2 \in \mathcal C(\mathbf{X})$, the polar factor $\mathbf{Q}$ is uniformly
distributed on $\mathcal S_{\mathbf{Q}}$. Consequently, with $\mathbf{Q}_{\text{obs}}$ the observed polar factor and
$d\mathbf{Q}$ integration against the uniform distribution on $\mathcal V$,
\begin{equation}
\label{eq:ppolar_integral}
p_{\mathrm{polar}} = \frac{\int_{\mathcal S_{\mathbf{Q}} \cap \{u(\mathbf{Q}) \ge u(\mathbf{Q}_{\text{obs}})\}} d\mathbf{Q}}{\int_{\mathcal S_{\mathbf{Q}}} d\mathbf{Q}},
\qquad
\Prob_{H_0'}\big(p_{\mathrm{polar}} \le \alpha \mid \hat{\mathcal C}_1, \hat{\mathcal C}_2 \in \mathcal C(\mathbf{X})\big) = \alpha
\ \text{ for all } \alpha \in [0, 1].
\end{equation}
\end{proposition}

The proof is in Appendix~\ref{appendix:polar}. The second part of \eqref{eq:ppolar_integral} says that $p_{\mathrm{polar}}$ controls the
selective Type I error in the sense of \eqref{eq:sel_typeI}, with equality and with $H_0'$ in place of $H_0$. In
words, among all data sets with the same
$\mathbf P_2\mathbf{X}$ and $\mathbf{H}$ on which the clustering returns $\hat{\mathcal C}_1$ and $\hat{\mathcal C}_2$,
$p_{\mathrm{polar}}$ is the fraction on which the two clusters look at least as separated, as
measured by $T^2$, as in the observed data. 

Given \eqref{eq:intro_gao} and \eqref{eq:yb_pvalue}, it is tempting to expect
$p_{\mathrm{polar}}$ to be the tail of an $F$ distribution truncated to the values of $T^2$ that
reproduce the clusters. It is not, because whether the clustering returns $\hat{\mathcal C}_1$ and
$\hat{\mathcal C}_2$ depends on all of $\mathbf{Q}$, not only on $T^2$: polar factors with the same $T^2$ arrange
the tested observations differently in the other directions, so some of them reproduce the
clusters and some do not. The clustering therefore reweights the null distribution of $T^2$ by
the fraction of polar factors with each value of $T^2$ that reproduce the clusters, and this fraction has no closed
form. In contrast, \eqref{eq:intro_gao} and \eqref{eq:yb_pvalue} condition on everything except one scalar. Each value of
that scalar corresponds to a single data set, which either reproduces the clusters or does not. Computing
$p_{\mathrm{polar}}$ in \eqref{eq:pstar} therefore needs Monte Carlo sampling, and Section~\ref{sec:computation} gives sampling
methods whose $p$-values are exact for any number of draws.

\subsection{Conditioning on more to obtain a one-dimensional pivot}
\label{sec:onedim}
\label{sec:truncated_F_main}

We now work toward a $p$-value that, as in \eqref{eq:intro_gao} and \eqref{eq:yb_pvalue}, is the
survival function of a known one-dimensional distribution truncated to a set. The high-level
strategy is to identify a one-dimensional function of $\mathbf{Q}$ and condition on all the rest. By
Proposition~\ref{prop:T2_share}, $T^2$ depends on $\mathbf{Q}$ only through $u(\mathbf{Q})$, so $u(\mathbf{Q})$ is
the function we keep random. To say what the rest of $\mathbf{Q}$ is, we use the Fisher discriminant
direction of Section~\ref{sec:hotelling}.

Since $\mathbf P_0 = \bar{\boldsymbol\nu}\bar{\boldsymbol\nu}^\top$ with $\bar{\boldsymbol\nu} = \boldsymbol\nu/\|\boldsymbol\nu\|_2$, the
between-cluster part of $\mathbf{Q}$ is the rank-one matrix
\begin{equation}
\label{eq:P0Q}
\mathbf P_0\mathbf{Q} = \sqrt{u(\mathbf{Q})}\;\bar{\boldsymbol\nu}\,\boldsymbol\gamma^\top,
\qquad
\boldsymbol\gamma = \mathrm{dir}(\mathbf{Q}^\top\boldsymbol\nu) = \mathrm{dir}(\mathbf{H}^{-1}\mathbf{X}^\top\boldsymbol\nu).
\end{equation}
The vector $\mathbf{q} = \mathbf{Q}\boldsymbol\gamma = (\mathbf P_0 + \mathbf P_1)\mathbf{X}\,\mathbf{H}^{-1}\boldsymbol\gamma$ is the tested data projected
onto the direction $\mathbf{H}^{-1}\boldsymbol\gamma$, which is proportional to
$\{\mathbf{X}^\top(\mathbf P_0 + \mathbf P_1)\mathbf{X}\}^{-1}\mathbf{X}^\top\boldsymbol\nu$, the Fisher discriminant direction, and
scaled so that $\|\mathbf{q}\|_2 = 1$. So $\mathbf{q}$ is a data set with one feature. By
\eqref{eq:P0Q}, its between-cluster part is $\mathbf P_0\mathbf{q} = \sqrt{u(\mathbf{Q})}\,\bar{\boldsymbol\nu}$, so $u(\mathbf{Q})$
is the fraction $w$ of Section~\ref{sec:intuition} computed on $\mathbf{q}$, and
$T^2 = (r-2)\|\mathbf P_0\mathbf{q}\|_2^2/\|\mathbf P_1\mathbf{q}\|_2^2$ is the statistic $R$ in \eqref{eq:yb_R} computed
on $\mathbf{q}$.

This suggests applying the conditioning of \citet{YunFoygelBarber2023} to $\mathbf{q}$ and holding
the rest of $\mathbf{Q}$ fixed. \citet{YunFoygelBarber2023} fix the directions of the between-cluster
and within-cluster parts and keep only their relative size random. The between-cluster part of
$\mathbf{q}$ always points along $\bar{\boldsymbol\nu}$, so we fix the following three quantities:
\begin{itemize}
  \item the direction $\boldsymbol\gamma$, which fixes the direction onto which the data are projected;
  \item the direction $\mathrm{dir}(\mathbf P_1\mathbf{q})$ of the within-cluster part of $\mathbf{q}$;
  \item the part of $\mathbf{Q}$ orthogonal to $\boldsymbol\gamma$, $\mathbf{Q}(\mathbf{I}_p - \boldsymbol\gamma\boldsymbol\gamma^\top)$, which
  describes the tested data in the directions orthogonal to the Fisher direction.
\end{itemize}
We write $\boldsymbol\Gamma = \big(\boldsymbol\gamma, \mathrm{dir}(\mathbf P_1\mathbf{q}), \mathbf{Q}(\mathbf{I}_p - \boldsymbol\gamma\boldsymbol\gamma^\top)\big)$. Since $\mathbf{q}$ has unit
length, $\|\mathbf P_1\mathbf{q}\|_2 = \sqrt{1 - u(\mathbf{Q})}$, and writing $\mathbf{Q} = \mathbf{q}\boldsymbol\gamma^\top + \mathbf{Q}(\mathbf{I}_p - \boldsymbol\gamma\boldsymbol\gamma^\top)$
gives
\begin{equation}
\label{eq:angle_coords}
\mathbf{Q} = \Big(\sqrt{u(\mathbf{Q})}\,\bar{\boldsymbol\nu} + \sqrt{1 - u(\mathbf{Q})}\,\mathrm{dir}(\mathbf P_1\mathbf{q})\Big)\boldsymbol\gamma^\top
+ \mathbf{Q}(\mathbf{I}_p - \boldsymbol\gamma\boldsymbol\gamma^\top).
\end{equation}
So $\mathbf{Q}$ is determined by $u(\mathbf{Q})$ and $\boldsymbol\Gamma$, and conditioning on $\boldsymbol\Gamma$ leaves $u(\mathbf{Q})$ as the
only random quantity. It is convenient to write this quantity as an angle. For $\phi \in (0, \pi/2)$, replacing
$\sqrt{u(\mathbf{Q})}$ by $\cos\phi$ and $\sqrt{1 - u(\mathbf{Q})}$ by $\sin\phi$ in \eqref{eq:angle_coords} gives
\begin{equation}
\label{eq:QM_path}
\mathbf{Q}_{\boldsymbol\Gamma}(\phi) = \big(\cos\phi\,\bar{\boldsymbol\nu} + \sin\phi\,\mathrm{dir}(\mathbf P_1\mathbf{q})\big)\boldsymbol\gamma^\top
+ \mathbf{Q}(\mathbf{I}_p - \boldsymbol\gamma\boldsymbol\gamma^\top),
\end{equation}
the polar factor that has the same $\boldsymbol\Gamma$ as $\mathbf{Q}$ and whose projected data have between-cluster
fraction $\cos^2\phi$. It lies in $\mathcal V$ for every $\phi$ (Proposition~\ref{lem:angle_fiber}). We write $\theta$ for the
observed angle, $u(\mathbf{Q}) = \cos^2\theta$; along the path, $T^2 = (r-2)\cot^2\phi$. As $\phi$ varies, the unit
vector $\cos\phi\,\bar{\boldsymbol\nu} + \sin\phi\,\mathrm{dir}(\mathbf P_1\mathbf{q})$ in \eqref{eq:QM_path} turns at constant speed in the
plane spanned by $\bar{\boldsymbol\nu}$ and $\mathrm{dir}(\mathbf P_1\mathbf{q})$. The path is therefore a rotation of $\mathbf{Q}$, the form used in
Section~\ref{sec:computation}. We write
\begin{equation}
\label{eq:S_theta}
\mathcal S_\theta = \big\{\phi \in (0, \pi/2) : \hat{\mathcal C}_1, \hat{\mathcal C}_2 \in
\mathcal C\big(\mathbf P_2\mathbf{x} + \mathbf{Q}_{\boldsymbol\Gamma}(\phi)\,\mathbf{H}(\mathbf{x})\big)\big\}
\end{equation}
for the set of angles at which the clustering returns the tested clusters. The one-dimensional selective $p$-value conditions on
$\boldsymbol\Gamma$ in addition to the clustering and the nuisance statistics,
\begin{equation}
\label{eq:onedim_def}
\begin{aligned}
p_{\mathrm{polar,1d}} = \Prob_{H_0'}\Big(&T^2(\mathbf{X}) \ge T^2(\mathbf{x}) \;\Big|\;
\hat{\mathcal C}_1, \hat{\mathcal C}_2 \in \mathcal C(\mathbf{X}),\\
&\mathbf P_2\mathbf{X} = \mathbf P_2\mathbf{x},\ \mathbf{H}(\mathbf{X}) = \mathbf{H}(\mathbf{x}),\
\boldsymbol\Gamma(\mathbf{X}) = \boldsymbol\Gamma(\mathbf{x})\Big).
\end{aligned}
\end{equation}

\begin{proposition}[The one-dimensional selective $p$-value]
\label{thm:angle_independent}
\label{cor:truncated_F}
Under data model \eqref{eq:intro_model} and null \eqref{eq:intro_null_strong}, $u(\mathbf{Q})$ is
independent of $(\boldsymbol\Gamma, \mathbf{H}, \mathbf P_2\mathbf{X})$. Consequently, given $\boldsymbol\Gamma$, $\mathbf P_2\mathbf{X}$, $\mathbf{H}$,
and the clustering, the scaled statistic
$F = \frac{r-p-1}{(r-2)p}T^2 = \frac{r-p-1}{p}\,\frac{u(\mathbf{Q})}{1 - u(\mathbf{Q})}$ has the
$F_{p,\,r-p-1}$ distribution truncated to the set
$\mathcal S_{\boldsymbol\Gamma} = \{\tfrac{r-p-1}{p}\cot^2\phi : \phi \in \mathcal S_\theta\}$ of values that reproduce the clusters, and $p_{\mathrm{polar,1d}}$ is the survival function of
this truncated distribution evaluated at the observed value of $F$. Under the null,
$p_{\mathrm{polar,1d}}$ is uniformly distributed conditional on the clustering.
\end{proposition}

The proof is in Appendix~\ref{appendix:hotelling_angle}. Fixing $\boldsymbol\Gamma$ is additional conditioning,
like the direction in \eqref{eq:intro_gao} and the two directions in \eqref{eq:yb_pvalue}: its
distribution involves no unknown parameter, and we fix it only so that the resulting distribution is easy to characterize
computationally. In
terms of the data, moving along the path \eqref{eq:QM_path} keeps $\mathbf P_2\mathbf{X}$, $\mathbf{H}$, and the
data in the directions orthogonal to the Fisher direction, and moves each tested observation
along a line in the direction of the difference in cluster means. So $p_{\mathrm{polar,1d}}$ is the
test of \citet{YunFoygelBarber2023} applied to the projected data $\mathbf{q}$, with the rest of the
data held fixed. Figure~\ref{fig:perturbation}(a) and Table~\ref{tab:nuisance} in Appendix~\ref{appendix:onedim_compare}
compare the three one-dimensional tests. Because it fixes $\boldsymbol\Gamma$, $p_{\mathrm{polar,1d}}$ uses less of the data than
\looseness=-1 $p_{\mathrm{polar}}$ and has less power (Section~\ref{sec:sim}).

\begin{figure}[t]
\centering
\includegraphics[width=\linewidth]{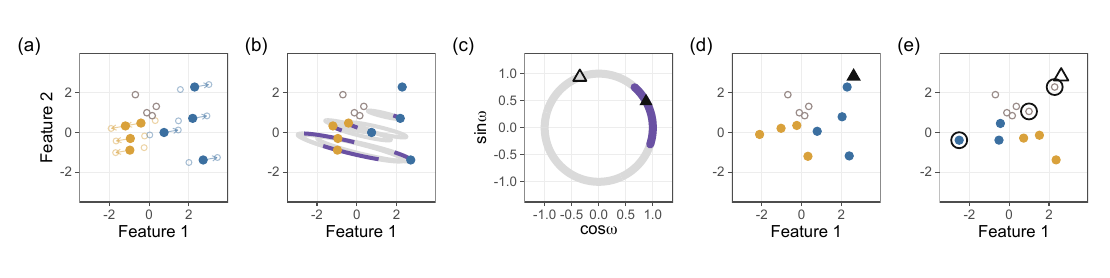}
\caption{Two ways to perturb the data, for $n = 12$ observations from one bivariate Gaussian split into $K = 3$
clusters by $k$-means; open circles are the cluster that is not tested. \textbf{(a)} The perturbation of
\citet{GaoBienWitten2024} and \citet{ChenWitten2023} slides the two tested clusters apart along $\boldsymbol\nu$, so
every perturbed data set lies on a line. \textbf{(b)} Our perturbation $\mathbf{Q} \mapsto \mathbf{G}_{ij}(\omega)\mathbf{Q}$
for one pair of rows $(i, j)$, where $\mathbf{G}_{ij}(\omega)$ rotates rows $i$ and $j$ of $\mathbf{Q}$ by the angle $\omega$ (Section~\ref{sec:mc}): each tested observation moves on an ellipse, purple where $k$-means still returns the tested
clusters. \textbf{(c)} The angles $\omega$ on the unit circle, purple where $k$-means returns the tested clusters,
with an accepted proposal (filled triangle) and a rejected proposal (open triangle); the observed data are at
$\omega = 0$. \textbf{(d)} The data at the accepted angle, colored by the clusters that $k$-means returns on them. \textbf{(e)} The
same at the rejected angle; the ringed observations have changed cluster.}
\label{fig:perturbation}
\end{figure}

\section{Computing the selective $p$-values}
\label{sec:computation}
\label{sec:conditioning_less}

Both selective $p$-values, $p_{\mathrm{polar}}$ in \eqref{eq:pstar} and $p_{\mathrm{polar,1d}}$ in
\eqref{eq:onedim_def}, are ratios of integrals over sets that the clustering algorithm defines. By Proposition~\ref{prop:conditional_selection}, $p_{\mathrm{polar}}$
is the fraction of the volume of $\mathcal S_{\mathbf{Q}}$ on which $u(\mathbf{Q}) \ge u(\mathbf{Q}_{\text{obs}})$, and by
Proposition~\ref{cor:truncated_F}, $p_{\mathrm{polar,1d}}$ is a ratio of two masses of a truncated
$F$ distribution over $\mathcal S_{\boldsymbol\Gamma}$. For a general clustering algorithm, neither set has a
closed form, but we can check whether a given point belongs to it by running the algorithm
once: $\mathbf{Q} \in \mathcal S_{\mathbf{Q}}$ exactly when the algorithm, applied to $\mathbf{x}'(\mathbf{Q})$, returns
$\hat{\mathcal C}_1$ and $\hat{\mathcal C}_2$. Sections~\ref{sec:mc} to~\ref{sec:onedim_mc} use only this check,
and so apply to any algorithm. Section~\ref{sec:extensions_km_hc}
\looseness=-1 shows what more can be done for $k$-means and hierarchical clustering.

\subsection{Monte Carlo sampling for any clustering algorithm}
\label{sec:mc}
\label{sec:comp_general}

A direct approach, following \citet{GaoBienWitten2024}, is to estimate the two integrals in
\eqref{eq:ppolar_integral} by importance sampling. We draw
$\mathbf{Q}_1, \ldots, \mathbf{Q}_B$ independently from a proposal distribution on $\mathcal V$ with density $\eta$
relative to the uniform distribution, run the clustering algorithm on each perturbed data set
$\mathbf{x}'(\mathbf{Q}_b)$, and estimate
\begin{equation}
\label{eq:is_pvalue}
\hat p_{\mathrm{IS}} = \frac{\sum_{b=1}^{B} w_b\,\one\{\mathbf{Q}_b \in \mathcal S_{\mathbf{Q}}\}\,\one\{u(\mathbf{Q}_b) \ge u(\mathbf{Q}_{\text{obs}})\}}
{\sum_{b=1}^{B} w_b\,\one\{\mathbf{Q}_b \in \mathcal S_{\mathbf{Q}}\}},
\qquad w_b = \frac{1}{\eta(\mathbf{Q}_b)}.
\end{equation}
For $p_{\mathrm{polar,1d}}$ the random quantity is one angle and this approach is workable
(Section~\ref{sec:onedim_mc}). For $p_{\mathrm{polar}}$ the random quantity is the whole polar factor
$\mathbf{Q}$, which has $(r-1)p - p(p+1)/2$ free coordinates, and $\mathcal S_{\mathbf{Q}}$ is typically a small region
around $\mathbf{Q}_{\text{obs}}$. Points drawn uniformly from $\mathcal V$ ($\eta = 1$) rarely land in $\mathcal S_{\mathbf{Q}}$, so
both sums in \eqref{eq:is_pvalue} are usually zero. A proposal concentrated near $\mathbf{Q}_{\text{obs}}$ has
the opposite problem. Its density is small away from $\mathbf{Q}_{\text{obs}}$, so the few draws that land in
$\mathcal S_{\mathbf{Q}}$ far from $\mathbf{Q}_{\text{obs}}$ get much larger weights than the others, and $\hat p_{\mathrm{IS}}$ is
close to the indicator $\one\{u(\mathbf{Q}_b) \ge u(\mathbf{Q}_{\text{obs}})\}$ of one or two draws. It is then
close to $0$ or to $1$ depending on which side of $u(\mathbf{Q}_{\text{obs}})$ those draws happen to fall,
and the values close to $0$ are false rejections.

An alternative is to draw $\mathbf{Q}_1, \ldots, \mathbf{Q}_B$ from a Markov chain whose stationary distribution
is the uniform distribution on $\mathcal S_{\mathbf{Q}}$, such as a random-walk Metropolis chain or
Hamiltonian Monte Carlo \citep{Neal2011}. Every draw then lies in $\mathcal S_{\mathbf{Q}}$, and the proposal
is the target itself, so all weights in \eqref{eq:is_pvalue} are equal and the estimate becomes
the fraction of draws with $u(\mathbf{Q}_b) \ge u(\mathbf{Q}_{\text{obs}})$. Used naively, however, this estimate is
valid only as $B \to \infty$, and that limit is far away here. The chain moves slowly through
$\mathcal S_{\mathbf{Q}}$, and its draws are strongly correlated with $\mathbf{Q}_{\text{obs}}$ and with each other, so the
resulting $p$-values are anticonservative in our experiments. For example, the selective Type~I
error at level $0.05$ reaches $0.19$ at $n = 200$ and $p = 10$ (Appendix~\ref{appendix:mc-budget}).

We therefore combine a Markov chain with the construction of \citet{BesagClifford1989}, which
turns a chain into an exact $p$-value however slowly it mixes (Section~\ref{sec:bc}). The
construction needs a Markov kernel that is reversible with respect to the uniform distribution
on $\mathcal S_{\mathbf{Q}}$. Since each new point must again have orthonormal columns, we propose moves by
Givens rotations \citep[Section~5.1.8]{GolubVanLoan2013}. Let $\mathbf{a}_1, \ldots, \mathbf{a}_{r-1}$ be a fixed
orthonormal basis of the range of $\mathbf P_0 + \mathbf P_1$. For $1 \le i < j \le r-1$ and an angle $\omega$,
the Givens rotation
\begin{equation}
\label{eq:givens_main}
\mathbf{G}_{ij}(\omega) = \mathbf{I}_n + (\cos\omega - 1)(\mathbf{a}_i\mathbf{a}_i^\top + \mathbf{a}_j\mathbf{a}_j^\top)
+ \sin\omega\,(\mathbf{a}_j\mathbf{a}_i^\top - \mathbf{a}_i\mathbf{a}_j^\top)
\end{equation}
rotates the plane spanned by $\mathbf{a}_i$ and $\mathbf{a}_j$ by the angle $\omega$ and leaves every
vector orthogonal to that plane unchanged. Since $\mathbf{G}_{ij}(\omega)$ is orthogonal and maps the
range of $\mathbf P_0 + \mathbf P_1$ to itself, $\mathbf{G}_{ij}(\omega)\mathbf{Q} \in \mathcal V$ whenever $\mathbf{Q} \in \mathcal V$.
In the coordinates $\mathbf{A}^\top\mathbf{Q}$, with $\mathbf{A} = (\mathbf{a}_1, \ldots, \mathbf{a}_{r-1})$, the rotation changes
only rows $i$ and $j$. As $\omega$ varies, each tested observation moves along an ellipse and the
untested observations do not move (Figure~\ref{fig:perturbation}(b) and (c)). We call the following kernel $\mathcal K$ the \emph{random Givens kernel}.
Given the current point $\mathbf{Q}$ and a number of steps $L$:
\begin{enumerate}
  \item draw a step size $\epsilon$ with $\log\epsilon$ uniform on $[\log 10^{-3}, 0]$,
  independently of $\mathbf{Q}$;
  \item repeat $L$ times:
  \begin{enumerate}
    \item draw a pair $i < j$ uniformly at random and an angle $\omega \sim \N\big(0, (\pi\epsilon)^2\big)$;
    \item set $\mathbf{Q}' = \mathbf{G}_{ij}(\omega)\mathbf{Q}$ and run the clustering algorithm on
    $\mathbf{x}'(\mathbf{Q}') = \mathbf P_2\mathbf{x} + \mathbf{Q}'\mathbf{H}(\mathbf{x})$;
    \item if the algorithm returns $\hat{\mathcal C}_1$ and $\hat{\mathcal C}_2$, set $\mathbf{Q} = \mathbf{Q}'$; otherwise keep $\mathbf{Q}$.
  \end{enumerate}
\end{enumerate}
Drawing $\epsilon$ from a range of scales lets the chain take short steps where $\mathcal S_{\mathbf{Q}}$ is
narrow and long steps where it is wide, without tuning $\epsilon$ on the observed data, which the
construction of Section~\ref{sec:bc} does not allow. One application of $\mathcal K$ runs the clustering
algorithm $L$ times. Figure~\ref{fig:perturbation}(d) and (e) show an accepted and a rejected proposal on a small example.

\begin{proposition}[The random Givens kernel is reversible]
\label{prop:givens_reversible}
For every $L \ge 1$, the random Givens kernel is reversible with respect to the uniform
distribution on $\mathcal S_{\mathbf{Q}}$.
\end{proposition}

The proof is in Appendix~\ref{appendix:cayley}. The reason is that the proposal
$\mathbf{Q} \mapsto \mathbf{G}_{ij}(\omega)\mathbf{Q}$ is a random rotation that is as likely as its inverse, because
$\mathbf{G}_{ij}(\omega)^{-1} = \mathbf{G}_{ij}(-\omega)$ and $\omega$ is symmetric about zero. Such a proposal does not change the
uniform distribution on $\mathcal V$, and keeping only the proposals that reproduce the clusters restricts it to
$\mathcal S_{\mathbf{Q}}$. The same argument applies to any random rotation
with this symmetry. One example is the Cayley random walk, which rotates all $r - 1$
coordinates at once by the Cayley transform of a random skew-symmetric matrix
\citep{WenYin2013}; it is described in Appendix~\ref{appendix:cayley}. Both kernels give exact
$p$-values, and in a pilot comparison (Appendix~\ref{appendix:kernels}) the random Givens kernel
had at least the power of the Cayley random walk at a lower cost per $p$-value.

\subsection{An exact $p$-value from a Markov chain}
\label{sec:bc}
\label{sec:bc_construction}
\label{sec:validity_principle}

\citet{BesagClifford1989} showed how to obtain an exact $p$-value from a reversible Markov
chain without waiting for it to mix: place the observed point at a uniformly random position
along a path of the chain and use its rank on that path. Let $\mathcal K$ be the random Givens
kernel of Section~\ref{sec:mc}. Build a path of length $M + 1$ as follows.
\begin{enumerate}
    \item Draw $J$ uniformly from $\{0, 1, \ldots, M\}$ and set $\mathbf{Q}_J = \mathbf{Q}_{\text{obs}}$.
    \item Run the chain forward from $\mathbf{Q}_J$: $\mathbf{Q}_{j+1} = \mathcal K(\mathbf{Q}_j)$ for $j = J, \ldots, M-1$.
    \item Run the chain backward from $\mathbf{Q}_J$: $\mathbf{Q}_{j-1} = \mathcal K(\mathbf{Q}_j)$ for $j = J, \ldots, 1$.
\end{enumerate}
The backward run uses the same kernel as the forward run because $\mathcal K$ is reversible. With
$u_j = u(\mathbf{Q}_j)$, the Monte Carlo version of $p_{\mathrm{polar}}$ is the randomized rank of $u_J$ among
$u_0, \ldots, u_M$,
\begin{equation}
\label{eq:bc_pvalue}
\hat p_{\mathrm{polar}} = \frac{1}{M+1}\Big(\sum_{j=0}^{M}\one\{u_j > u_J\}
+ \xi\sum_{j=0}^{M}\one\{u_j = u_J\}\Big),
\qquad \xi \sim \text{Uniform}(0,1),
\end{equation}
with $\xi$ drawn independently of the path to break ties.

\begin{proposition}[The Monte Carlo $p$-value is exact]
\label{prop:bc_validity}
Suppose that the kernel $\mathcal K$ is reversible with respect to the uniform distribution on
$\mathcal S_{\mathbf{Q}}$, and that $\mathcal K$ and $M$ depend on the data only through $\mathbf P_2\mathbf{X}$, $\mathbf{H}$, and
the clustering. Under data model \eqref{eq:intro_model} and null \eqref{eq:intro_null_strong},
\[
\Prob_{H_0'}\big(\hat p_{\mathrm{polar}} \le \alpha \,\big|\, \hat{\mathcal C}_1, \hat{\mathcal C}_2 \in \mathcal C(\mathbf{X}),\
\mathbf P_2\mathbf{X},\ \mathbf{H}\big) = \alpha \qquad \text{for all } \alpha \in [0,1],
\]
for every $M \ge 1$, however slowly the chain mixes.
\end{proposition}

The proof is in Appendix~\ref{appendix:bc_proof}. Proposition~\ref{prop:bc_validity} says that
\eqref{eq:bc_pvalue} gives a valid test for any $M$, any $L$, and any reversible kernel, in the
sense that it controls the selective Type~I error \eqref{eq:sel_typeI}. Different choices, however, lead to different
power when $H_0'$ does not hold. Power grows with the path length $M$ and levels off
(Appendix~\ref{appendix:bc_tuning}), and it grows with how far the chain moves between
consecutive points of the path, which increases with $L$ and with the size of the accepted
steps. The path then spreads $u_0, \ldots, u_M$ over a wider range, so that under the alternative
the observed $u_J$ can stand out at the top of that range. At the other extreme, a kernel that never moves
gives $\hat p_{\mathrm{polar}} = \xi$, which is exact and has no power. The one restriction is that these
choices may not depend on $\mathbf{Q}_{\text{obs}}$: tuning $\epsilon$ on trial moves from $\mathbf{Q}_{\text{obs}}$ can
break exactness (Appendix~\ref{appendix:bc_tuning}), whereas drawing $\epsilon$ from a fixed
distribution, as in Section~\ref{sec:mc}, does not. We summarize the test based on
$\hat p_{\mathrm{polar}}$ in Algorithm~\ref{alg:bc}, which runs the clustering algorithm $L$ times at each of the
$M$ steps of the path.

\begin{algorithm}[t]
\caption{The selective test $\hat p_{\mathrm{polar}}$ for any clustering algorithm}
\label{alg:bc}
\begin{algorithmic}[1]
\Require data $\mathbf{x} \in \R^{n \times p}$, clustering algorithm $\mathcal C$, tested clusters
  $\hat{\mathcal C}_1, \hat{\mathcal C}_2 \in \mathcal C(\mathbf{x})$ (fixed in advance, or each pair tested in turn), path
  length $M$, number of steps $L$
\State Form $\mathbf P_0$, $\mathbf P_1$, $\mathbf P_2$ from $\hat{\mathcal C}_1$ and $\hat{\mathcal C}_2$, and an orthonormal basis
  $\mathbf{a}_1, \ldots, \mathbf{a}_{r-1}$ of the range of $\mathbf P_0 + \mathbf P_1$
\State $\mathbf{H} \gets \big(\mathbf{x}^\top(\mathbf P_0 + \mathbf P_1)\mathbf{x}\big)^{1/2}$ and
  $\mathbf{Q}_{\text{obs}} \gets (\mathbf P_0 + \mathbf P_1)\mathbf{x}\,\mathbf{H}^{-1}$
  \Comment{polar decomposition, Section~\ref{subsec:polar}}
\State Draw $J$ uniformly from $\{0, 1, \ldots, M\}$ and set $\mathbf{Q}_J \gets \mathbf{Q}_{\text{obs}}$
\For{$j = J, J+1, \ldots, M-1$}
  \State $\mathbf{Q}_{j+1} \gets \Call{RandomGivens}{\mathbf{Q}_j}$ \Comment{forward}
\EndFor
\For{$j = J, J-1, \ldots, 1$}
  \State $\mathbf{Q}_{j-1} \gets \Call{RandomGivens}{\mathbf{Q}_j}$ \Comment{backward}
\EndFor
\State $u_j \gets \|\mathbf P_0\mathbf{Q}_j\|_F^2$ for $j = 0, \ldots, M$, and draw $\xi \sim \text{Uniform}(0,1)$
\Ensure $\hat p_{\mathrm{polar}} = \frac{1}{M+1}\big(\sum_{j=0}^{M}\one\{u_j > u_J\} + \xi\sum_{j=0}^{M}\one\{u_j = u_J\}\big)$
\Statex
\Function{RandomGivens}{$\mathbf{Q}$} \Comment{the random Givens kernel $\mathcal K$}
  \State Draw $\log\epsilon$ uniformly on $[\log 10^{-3}, 0]$
  \For{$\ell = 1, \ldots, L$}
    \State Draw $i < j$ uniformly from $\{1, \ldots, r-1\}$ and $\omega \sim \N\big(0, (\pi\epsilon)^2\big)$
    \State $\mathbf{Q}' \gets \mathbf{G}_{ij}(\omega)\,\mathbf{Q}$ \Comment{Givens rotation \eqref{eq:givens_main}}
    \If{$\hat{\mathcal C}_1, \hat{\mathcal C}_2 \in \mathcal C\big(\mathbf P_2\mathbf{x} + \mathbf{Q}'\mathbf{H}\big)$}
      \State $\mathbf{Q} \gets \mathbf{Q}'$
    \EndIf
  \EndFor
  \State \Return $\mathbf{Q}$
\EndFunction
\end{algorithmic}
\end{algorithm}

\subsection{The one-dimensional test for any clustering algorithm}
\label{sec:onedim_mc}

Recall that for $p_{\mathrm{polar,1d}}$ only the angle is random, and recall the set $\mathcal S_\theta$
in \eqref{eq:S_theta} of angles at which the clustering returns the tested clusters. Under the
null, the angle has density $f(\phi) \propto \cos^{p-1}\phi\,\sin^{r-p-2}\phi$ on $(0, \pi/2)$, which
is the $\mathrm{Beta}(p/2, (r-1-p)/2)$ distribution of $u(\mathbf{Q}) = \cos^2\phi$
(Proposition~\ref{prop:T2_share}). Given $\boldsymbol\Gamma$, $\mathbf P_2\mathbf{X}$, $\mathbf{H}$, and the clustering, the
angle therefore has density
\begin{equation}
\label{eq:truncated_law}
\frac{f(\phi)\,\one\{\phi \in \mathcal S_\theta\}}{\int_{\mathcal S_\theta} f(\psi)\,d\psi},
\end{equation}
and since a smaller angle means a larger $T^2$, Proposition~\ref{cor:truncated_F} says
\begin{equation}
\label{eq:pF}
p_{\mathrm{polar,1d}} = \frac{\int_{\mathcal S_\theta \cap (0, \theta]} f(\phi)\,d\phi}{\int_{\mathcal S_\theta} f(\phi)\,d\phi}.
\end{equation}
We compute \eqref{eq:pF} with the construction of Section~\ref{sec:bc}. Algorithm~\ref{alg:bc1d} in
Appendix~\ref{appendix:alg_onedim} is Algorithm~\ref{alg:bc} with two changes. First, the path moves the angle $\phi$ in
$(0, \pi/2)$ instead of the polar factor $\mathbf{Q}$. Second, each step is a Metropolis step for the density
\eqref{eq:truncated_law}: it proposes a nearby angle, accepts it with the usual density ratio, and keeps it only if
the clustering returns the tested clusters. This step is reversible with respect to \eqref{eq:truncated_law}, so the
resulting $p$-value is exactly uniform under the null by the argument of Proposition~\ref{prop:bc_validity}. Each
\looseness=-1 step runs the clustering only if the proposed angle passes the density check.

Alternatives exist, such as a slice sampler with shrinkage \citep{Neal2003}, which is also
reversible with respect to \eqref{eq:truncated_law}, and importance sampling as in
\citet{GaoBienWitten2024}, with the angle drawn from a mixture of $f$ and a normal distribution
centered at $\theta$. In our preliminary experiments, the slice sampler gave similar $p$-values at
about ten times the number of clustering runs. Importance sampling agreed with the exact test in
Figure~\ref{fig:motivation}, but it is consistent rather than exact, and it was unreliable when
$\theta$ was far in the tail of $f$: with well-separated clusters, $2{,}000$ draws had an effective
sample size of $2$ to $5$.

\subsection{Extensions to $k$-means and hierarchical clustering}
\label{sec:extensions_km_hc}

\paragraph{$k$-means: closed-form constraints from the trajectory.}
\label{sec:kmeans}
\label{sec:kmeans_review}
\label{sec:the_test}

Algorithms~\ref{alg:bc} and~\ref{alg:bc1d} apply to selective inference after any clustering
algorithm. For hierarchical and $k$-means clustering, \citet{GaoBienWitten2024} and
\citet{ChenWitten2023} also gave closed-form descriptions of their truncation sets. For
$k$-means, $\mathcal S_{\mathbf{Q}}$ has no closed form, because many sequences of intermediate assignments
end in the same final clusters (Appendix~\ref{appendix:feasible}). Following
\citet{ChenWitten2023}, however, we can condition on the whole sequence of assignments made by
Lloyd's algorithm. The set of polar factors that reproduce this sequence then has a closed
form, and so does its intersection with any rotation of $\mathbf{Q}$, which gives a kernel that needs
no rejections and no step size.

Lloyd's algorithm starts from the centroids $\mathbf{m}_\ell^{(0)} = \mathbf{x}_{\iota_\ell}$,
$\ell = 1, \ldots, K$, for fixed indices $\iota = (\iota_1, \ldots, \iota_K)$, and alternates
two steps until the assignments stop changing: assign each observation to its nearest
centroid, giving labels $c_i^{(t)}$, and recompute each centroid as the mean of the
observations assigned to it. Write $\tau = (c^{(0)}, \ldots, c^{(t_{\max})})$ for the observed
sequence of assignments. For a polar factor $\mathbf{Q}$, let $\mathbf{m}_\ell^{(t)}(\mathbf{Q})$ be the centroids of
$\mathbf{x}'(\mathbf{Q})$ under the \emph{observed} labels,
\[
\mathbf{m}_\ell^{(0)}(\mathbf{Q}) = \mathbf{x}'_{\iota_\ell}(\mathbf{Q}), \qquad
\mathbf{m}_\ell^{(t)}(\mathbf{Q}) = \frac{1}{|\{a : c_a^{(t-1)} = \ell\}|}\sum_{a : c_a^{(t-1)} = \ell}\mathbf{x}'_a(\mathbf{Q}),
\quad t \ge 1.
\]
Lloyd's algorithm applied to $\mathbf{x}'(\mathbf{Q})$ produces the same sequence $\tau$ exactly when, at every
iteration, every observation is closer to its observed centroid than to the others:
\begin{equation}
\label{eq:selection_event}
\|\mathbf{x}'_i(\mathbf{Q}) - \mathbf{m}_{c_i^{(t)}}^{(t)}(\mathbf{Q})\|_2^2 \le \|\mathbf{x}'_i(\mathbf{Q}) - \mathbf{m}_\ell^{(t)}(\mathbf{Q})\|_2^2
\quad \text{for all } t \le t_{\max},\ i \le n,\ \ell \ne c_i^{(t)}.
\end{equation}
Let $\mathcal S_{\mathbf{Q},\tau}$ be the set of $\mathbf{Q} \in \mathcal V$ that satisfy \eqref{eq:selection_event}. It is contained in $\mathcal S_{\mathbf{Q}}$, because the same
sequence of assignments ends in the same clusters, and Sections~\ref{subsec:selection}
to~\ref{sec:bc} hold with $\mathcal S_{\mathbf{Q},\tau}$ in place of $\mathcal S_{\mathbf{Q}}$.

\begin{proposition}[The trajectory event is quadratic in $\mathbf{Q}$]
\label{prop:kmeans_quadratic}
For each $t$, $i$, and $\ell \ne c_i^{(t)}$, the difference
\begin{equation}
\label{eq:feasible_g_main}
g_{t,i,\ell}(\mathbf{Q}) = \|\mathbf{x}'_i(\mathbf{Q}) - \mathbf{m}_{c_i^{(t)}}^{(t)}(\mathbf{Q})\|_2^2
- \|\mathbf{x}'_i(\mathbf{Q}) - \mathbf{m}_\ell^{(t)}(\mathbf{Q})\|_2^2
\end{equation}
is a quadratic polynomial in the entries of $\mathbf{Q}$, with coefficients determined by
$\mathbf P_2\mathbf{x}$, $\mathbf{H}$, and $\tau$. Hence
$\mathcal S_{\mathbf{Q},\tau} = \{\mathbf{Q} \in \mathcal V : g_{t,i,\ell}(\mathbf{Q}) \le 0 \text{ for all } t, i, \ell \ne c_i^{(t)}\}$
is defined by the $(t_{\max}+1)n(K-1)$ quadratic inequalities $g_{t,i,\ell}(\mathbf{Q}) \le 0$.
\end{proposition}

The reason is that $\mathbf{x}'(\mathbf{Q}) = \mathbf P_2\mathbf{x} + \mathbf{Q}\mathbf{H}$ is linear in $\mathbf{Q}$ apart from a constant,
and with the labels fixed by $\tau$ each centroid is a fixed average of rows of $\mathbf{x}'(\mathbf{Q})$, so
each squared distance is quadratic (Appendix~\ref{appendix:feasible}). The quadratic structure
also describes $\mathcal S_{\mathbf{Q},\tau}$ along any curve on which $\mathbf{Q}$ rotates.

\begin{proposition}[The trajectory event along a rotation]
\label{prop:kmeans_arcs}
Let $\mathbf{Q}(\omega) = \mathbf{D}_0 + \cos\omega\,\mathbf{D}_1 + \sin\omega\,\mathbf{D}_2$ be a curve in $\mathcal V$ with fixed
$n \times p$ matrices $\mathbf{D}_0$, $\mathbf{D}_1$, $\mathbf{D}_2$. Then along the curve each $g_{t,i,\ell}$ in
\eqref{eq:feasible_g_main} has the form
\begin{equation}
\label{eq:wave_main}
g_{t,i,\ell}(\mathbf{Q}(\omega)) = \beta_1\cos 2\omega + \beta_2\sin 2\omega + \beta_3\cos\omega + \beta_4\sin\omega + \beta_5,
\end{equation}
with constants $\beta_1, \ldots, \beta_5$ that have closed forms. Unless it is constant, such a
function has at most four zeros in $[0, 2\pi)$, which are the roots of a quartic polynomial.
Hence the set of $\omega$ with $\mathbf{Q}(\omega) \in \mathcal S_{\mathbf{Q},\tau}$ is a finite union of intervals
whose endpoints have closed forms.
\end{proposition}

Two curves of this form are the ones we need. The path of Section~\ref{sec:onedim} is one, with
$\mathbf{D}_0 = \mathbf{Q}(\mathbf{I}_p - \boldsymbol\gamma\boldsymbol\gamma^\top)$, $\mathbf{D}_1 = \bar{\boldsymbol\nu}\boldsymbol\gamma^\top$,
$\mathbf{D}_2 = \mathrm{dir}(\mathbf P_1\mathbf{q})\,\boldsymbol\gamma^\top$, and $\omega = \phi$.
The Givens circle $\omega \mapsto \mathbf{G}_{ij}(\omega)\mathbf{Q}$ of \eqref{eq:givens_main} is another. The
proofs and the formulas for $\beta_1, \ldots, \beta_5$ are in Appendix~\ref{appendix:arcs}.
Proposition~\ref{prop:kmeans_arcs} has two consequences for computation.
\begin{itemize}
  \item \textbf{The one-dimensional test needs no Monte Carlo.} Conditioning on $\tau$, the set
  $\mathcal S_\theta$ in \eqref{eq:pF} is a finite union of intervals with known endpoints, so both
  integrals in \eqref{eq:pF} are differences of the $\mathrm{Beta}(p/2, (r-1-p)/2)$ distribution
  function at $\cos^2$ of those endpoints.
  \item \textbf{A kernel without rejections or step sizes.} On the Givens circle through the
  current $\mathbf{Q}$, we can compute the whole set
  $\mathcal S_\omega = \{\omega : \mathbf{G}_{ij}(\omega)\mathbf{Q} \in \mathcal S_{\mathbf{Q},\tau}\}$ and draw $\omega$
  uniformly from it, instead of proposing one angle and checking it. This is a hit-and-run step
  \citep{Smith1984,BelisleRomeijnSmith1993}. It is reversible with respect to the uniform
  distribution on $\mathcal S_{\mathbf{Q},\tau}$ (Proposition~\ref{prop:K_reversible} in
  Appendix~\ref{appendix:sampler_valid}), it moves at every step, and it never runs Lloyd's
  algorithm.
\end{itemize}
Conditioning on $\tau$ is conditioning on more than the event that the clusters are returned,
and it costs power (Figure~\ref{fig:motivation}(c) and Section~\ref{sec:sim}).

\paragraph{Extension to hierarchical clustering.}
\label{sec:hclust}
\label{sec:extension}
\label{sec:ext_geometry}
\label{sec:sim_hclust}
\label{sec:ext_study}
We sought to follow \citet{GaoBienWitten2024} to obtain a more efficient computation for agglomerative hierarchical
clustering cut at $K$ clusters. Recall that \citet{GaoBienWitten2024} characterized the set $\mathcal S_{\boldsymbol\Sigma}$ in
\eqref{eq:gbw_S}, defined through the perturbation $\mathbf{x}'(\phi)$, as the solution set of $O(n^2)$ inequalities,
each quadratic in $\phi$, obtained by conditioning on the order of all merges down to $K$ clusters. A remarkable
consequence is that, because conditioning on $\mathbf P_1\mathbf{X}$ and $\mathbf P_2\mathbf{X}$ fixes the distances between
observations \emph{within the same cluster} \citep[Lemma~1]{GaoBienWitten2024}, conditioning on the order of all
merges is equivalent to conditioning on $\hat{\mathcal C}_1, \hat{\mathcal C}_2 \in \mathcal C(\mathbf{X})$ alone. The extra
conditioning therefore costs no power, unlike in other problems where conditioning on more than the selection
event does \citep{JewellFearnheadWitten2022, LeDuyTakeuchi2022, ChenJewellWitten2023}.

This is \emph{no longer the case} for our perturbation $\mathbf{x}'(\mathbf{Q})$, even for the one-dimensional path
$\mathbf{x}'(\mathbf{Q}_{\boldsymbol\Gamma}(\phi))$ of Section~\ref{sec:onedim}. To preserve the sample covariance, $\mathbf{x}'(\mathbf{Q})$ and
$\mathbf{x}'(\mathbf{Q}_{\boldsymbol\Gamma}(\phi))$ rotate the observations within each cluster, so the distances between observations in the same
cluster change, and many different merge orders end in the same tested clusters. Each merge of hierarchical clustering
with a commonly used linkage, such as average linkage on squared Euclidean distances, can still be characterized by
quadratic inequalities in $\mathbf{Q}$ (or in $\phi$), because it compares averages of squared distances between rows of
$\mathbf{x}'(\mathbf{Q})$. But conditioning on the merge order is now conditioning on more than the event that the clustering returns
the tested clusters, and so it costs power. In fact, in a pilot study, the resulting test had almost no power. We therefore do not present that alternative, and we compute $p_{\mathrm{polar}}$ and
$p_{\mathrm{polar,1d}}$ for hierarchical clustering with Algorithms~\ref{alg:bc} and~\ref{alg:bc1d}.

\section{Simulation study}
\label{sec:sim}

We simulate data from \eqref{eq:intro_model}, cluster them by $k$-means or by average-linkage hierarchical
clustering with $K = 3$, and test the null hypothesis $H_0'$ in \eqref{eq:intro_null_strong} for two of the estimated
clusters at level $\alpha = 0.05$; $H_0'$ implies the null $H_0$ in \eqref{eq:intro_null}, which the comparators
test. We compare these $p$-values:
\begin{itemize}
  \item the naive $p$-value $p_{\mathrm{naive}}$;
  \item our proposals $p_{\mathrm{polar}}$ (Algorithm~\ref{alg:bc}) and $p_{\mathrm{polar,1d}}$
  (Algorithm~\ref{alg:bc1d}), with $M = 5{,}000$ and $L = 2$, and, for $k$-means, $p_{\mathrm{polar,1d}}^{\mathrm{traj}}$,
  which conditions on the sequence of assignments of Lloyd's algorithm (Section~\ref{sec:kmeans});
  \item for $k$-means, the test of \citet{ChenWitten2023} with $\hat\sigma^2_{\mathrm{MED}}\mathbf{I}_p$ (their median-based
  estimate), with the sample covariance $\hat{\boldsymbol\Sigma}_{\mathrm{sample}}$ of $\mathbf{X}$ about its overall mean, or
  with the true $\boldsymbol\Sigma$: $p_{\mathrm{CW}}(\hat\sigma^2_{\mathrm{MED}}\mathbf{I})$,
  $p_{\mathrm{CW}}(\hat{\boldsymbol\Sigma}_{\mathrm{sample}})$, and $p_{\mathrm{CW}}(\boldsymbol\Sigma)$;
  \item for hierarchical clustering, the test of \citet{GaoBienWitten2024} with $\hat\sigma^2_{\mathrm{sample}}\mathbf{I}_p$
  (with $\hat\sigma^2_{\mathrm{sample}}$ estimated from the residuals of $\mathbf{X}$ about its overall mean),
  $\hat{\boldsymbol\Sigma}_{\mathrm{sample}}$, or $\boldsymbol\Sigma$:
  $p_{\mathrm{GBW}}(\hat\sigma^2_{\mathrm{sample}}\mathbf{I})$, $p_{\mathrm{GBW}}(\hat{\boldsymbol\Sigma}_{\mathrm{sample}})$, and
  $p_{\mathrm{GBW}}(\boldsymbol\Sigma)$.
\end{itemize}
Appendix~\ref{appendix:grid-extra} reports every cell, two more comparators, and Leiden community detection.

\subsection{Selective Type I error under the global null}
\label{sec:typeI}

Under the global null, every row of $\mathbf{X}$ has mean zero in \eqref{eq:intro_model}, so $H_0$ is
true whichever pair the clustering returns. The covariance $\boldsymbol\Sigma$ is isotropic
($\boldsymbol\Sigma = \mathbf{I}_p$), first-order autoregressive ($\boldsymbol\Sigma_{ij} = 0.6^{|i-j|}$), compound symmetric
($\boldsymbol\Sigma = 0.5\,\mathbf{I}_p + 0.5\,\one_p\one_p^\top$), or spiked
($\boldsymbol\Sigma = \mathbf{I}_p + 5\,\mathbf{v}\mathbf{v}^\top$ with $\mathbf{v} = p^{-1/2}\one_p$). For $k$-means, we simulate $600$ data sets for each
$n \in \{30, 50, 100, 200\}$ and $p \in \{2, 5, 10, 20, 50, 100\}$ with $p \le n/2$, apply $k$-means
with $K = 3$ from one fixed random initialization (Section~\ref{sec:kmeans}), and compute the
$p$-values for a randomly chosen pair of clusters with $|\hat{\mathcal C}_1| + |\hat{\mathcal C}_2| \ge p + 2$, so that the
pooled sample covariance in Hotelling's $T^2$ is invertible. For hierarchical clustering,
we simulate $600$ data sets for $n = 200$ and $p \in \{2, 10, 20, 50\}$ under the first three
covariances, cut the average-linkage tree on squared Euclidean distances at $K = 3$, and test every
pair of clusters with at least three members and $|\hat{\mathcal C}_1| + |\hat{\mathcal C}_2| \ge p + 2$.

\begin{figure}[t]
\centering
\includegraphics[width=\textwidth]{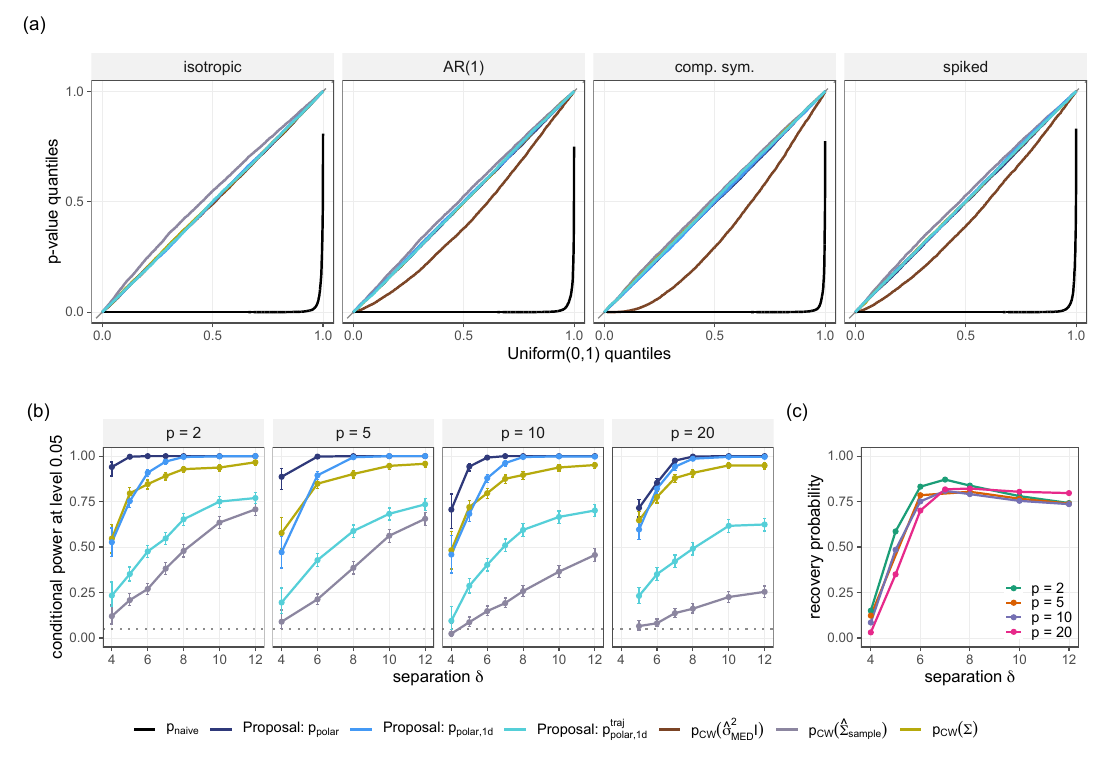}
\caption{$k$-means with $K = 3$. \textbf{(a)} Quantile-quantile plots of the $p$-values under
\eqref{eq:intro_model} with $\boldsymbol\mu = \mathbf 0_{n \times p}$, stratified by $\boldsymbol\Sigma$ and
pooled over $n$ and $p$.
\textbf{(b)} The conditional power \eqref{eq:cond_power} at $\alpha = 0.05$ under
\eqref{eq:intro_model} with $\boldsymbol\mu$ defined in \eqref{eq:sim_means}, $n = 50$, and
$\boldsymbol\Sigma = \mathbf{I}_p$, stratified by $p$, for the tests that control the selective Type~I
error. \textbf{(c)} The recovery probability \eqref{eq:recovery} in the same settings.}
\label{fig:kmeans}
\end{figure}

Figures~\ref{fig:kmeans}(a) and \ref{fig:hier}(a) plot the quantiles of the null $p$-values, pooled
over $n$ and $p$, against those of the Uniform$(0,1)$ distribution. Three patterns hold for both
clustering algorithms and every $\boldsymbol\Sigma$: (i) the curve of $p_{\mathrm{naive}}$ lies far below
the diagonal, so the naive test rejects true null hypotheses almost always; (ii)
$p_{\mathrm{CW}}(\hat\sigma^2_{\mathrm{MED}}\mathbf{I})$ and
$p_{\mathrm{GBW}}(\hat\sigma^2_{\mathrm{sample}}\mathbf{I})$ are uniform under isotropic noise and
stochastically smaller otherwise, because they assume $\boldsymbol\Sigma = \sigma^2\mathbf{I}_p$; (iii)
$p_{\mathrm{polar}}$, $p_{\mathrm{polar,1d}}$, and $p_{\mathrm{polar,1d}}^{\mathrm{traj}}$ control the selective Type~I error \eqref{eq:sel_typeI}, as do
$p_{\mathrm{CW}}(\boldsymbol\Sigma)$, $p_{\mathrm{GBW}}(\boldsymbol\Sigma)$, $p_{\mathrm{CW}}(\hat{\boldsymbol\Sigma}_{\mathrm{sample}})$, and
$p_{\mathrm{GBW}}(\hat{\boldsymbol\Sigma}_{\mathrm{sample}})$; the last two lead to conservative tests. In Appendix~\ref{appendix:grid-extra}, the unknown-scale test of
\citet{YunFoygelBarber2023} behaves like $p_{\mathrm{CW}}(\hat\sigma^2_{\mathrm{MED}}\mathbf{I})$, and data
thinning \citep{NeufeldDharamshiGaoWitten2024} with the sample covariance controls the selective
\looseness=-1 Type~I error but is conservative, rejecting about $2\%$ of true null hypotheses at level $0.05$.

\subsection{Conditional power and recovery probability}
\label{sec:powerstudy}

We generate data from \eqref{eq:intro_model} with three true clusters of equal size,
\begin{equation}
\label{eq:sim_means}
\boldsymbol\mu_1 = \cdots = \boldsymbol\mu_{n/3} = -\delta\,\mathbf{e}_1, \quad
\boldsymbol\mu_{n/3+1} = \cdots = \boldsymbol\mu_{2n/3} = \mathbf 0_p, \quad
\boldsymbol\mu_{2n/3+1} = \cdots = \boldsymbol\mu_n = \delta\,\mathbf{e}_1,
\end{equation}
where $\mathbf{e}_1$ is the first standard basis vector. The clusters that the algorithm returns need
not be true clusters, and a rejection is informative only when they are. Following
\citet{GaoBienWitten2024} and \citet{ChenWitten2023}, we therefore report two quantities over
$B$ simulated data sets with true clusters $\mathcal C_1, \mathcal C_2, \mathcal C_3$: the conditional
power, the rejection rate among the data sets in which both tested clusters are true clusters, and
the recovery probability, the fraction of data sets in which the two tested clusters are true clusters:
\begin{equation}
\label{eq:cond_power}
\text{Conditional power} =
\frac{\sum_{b=1}^B \one\big\{\{\hat{\mathcal C}_1^{(b)}, \hat{\mathcal C}_2^{(b)}\} \subseteq \{\mathcal C_1, \mathcal C_2, \mathcal C_3\},\ p^{(b)} \le \alpha\big\}}
{\sum_{b=1}^B \one\big\{\{\hat{\mathcal C}_1^{(b)}, \hat{\mathcal C}_2^{(b)}\} \subseteq \{\mathcal C_1, \mathcal C_2, \mathcal C_3\}\big\}},
\end{equation}
\begin{equation}
\label{eq:recovery}
\text{Recovery probability} =
\frac{1}{B}\sum_{b=1}^B \one\big\{\{\hat{\mathcal C}_1^{(b)}, \hat{\mathcal C}_2^{(b)}\} \subseteq \{\mathcal C_1, \mathcal C_2, \mathcal C_3\}\big\},
\end{equation}
where $p^{(b)}$, $\hat{\mathcal C}_1^{(b)}$, and $\hat{\mathcal C}_2^{(b)}$ are the $p$-value and the
clusters tested in the $b$th data set. We show power only for the tests and covariances under which
the test controls the selective Type~I error in Section~\ref{sec:typeI}. For $k$-means, we simulate
$B = 1{,}000$ data sets for each cell of the null design and each
$\delta \in \{3, 4, 5, 6, 7, 8, 10, 12\}$; for hierarchical clustering, $B = 300$ for each
$\delta \in \{4, 6, 8, 10, 12\}$.

Figure~\ref{fig:kmeans}(b) displays the conditional power of $k$-means at $n = 50$ and
Figure~\ref{fig:kmeans}(c) its recovery probability. The conditional power of every test increases
with $\delta$. In every panel, $p_{\mathrm{polar}}$ has the highest conditional power among the tests
that do not use $\boldsymbol\Sigma$ (tied with $p_{\mathrm{polar,1d}}$ at $p = 20$ and $\delta = 4$), and it is also at least as
powerful as $p_{\mathrm{CW}}(\boldsymbol\Sigma)$, which uses the true $\boldsymbol\Sigma$. This is possible because
$p_{\mathrm{CW}}(\boldsymbol\Sigma)$ conditions on more: on the whole sequence of assignments of Lloyd's algorithm rather than
only on the event that the clustering returns the tested pair, and on all of the data except one scalar. Each extra
condition costs power, as the gaps from $p_{\mathrm{polar}}$ to $p_{\mathrm{polar,1d}}$ and from $p_{\mathrm{polar,1d}}$ to
$p_{\mathrm{polar,1d}}^{\mathrm{traj}}$ show. Using
$\hat{\boldsymbol\Sigma}_{\mathrm{sample}}$ in $p_{\mathrm{CW}}$ leads to the least powerful test, because
the sample covariance about the overall mean includes the spread between the true cluster means, which
grows with $\delta$, and so overstates $\boldsymbol\Sigma$ along the direction that separates them. The recovery probability increases up to $\delta \approx 7$
and then decreases slightly, because with collinear means some initializations converge to a
partition that merges the middle cluster with an end cluster. At $n = 100$ and $200$, the conditional
power of $p_{\mathrm{polar}}$ exceeds $0.98$ in every cell (Appendix~\ref{appendix:grid-extra}).

\begin{figure}[t]
\centering
\includegraphics[width=\textwidth]{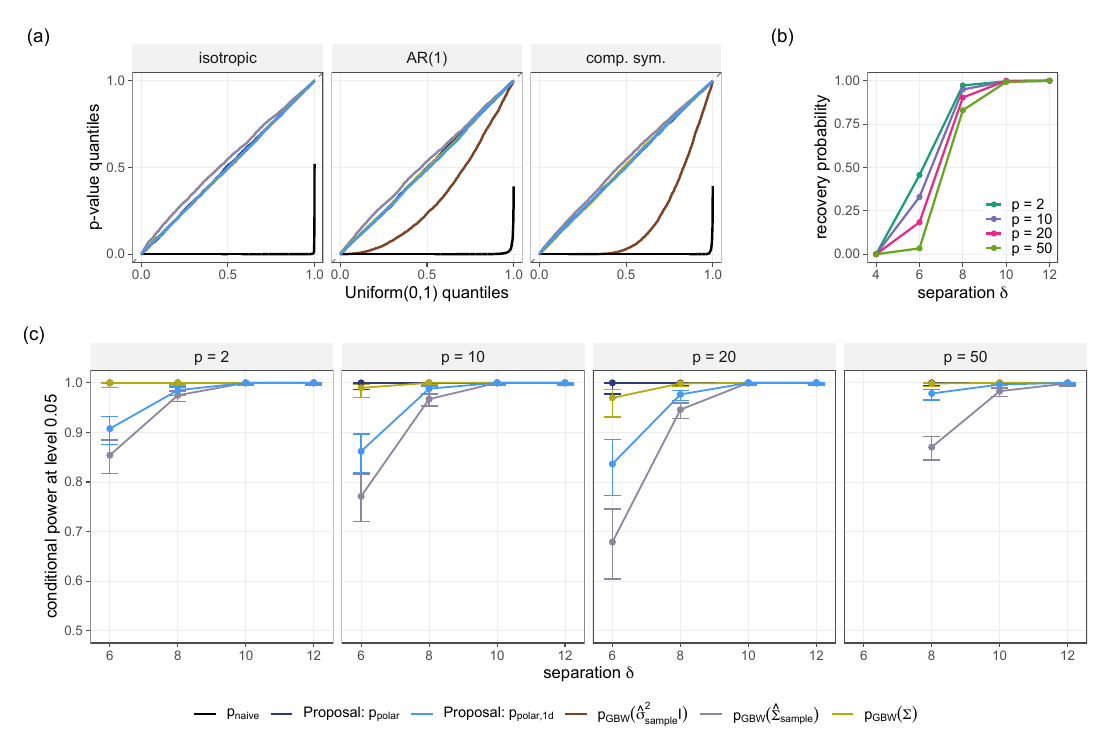}
\caption{Average-linkage hierarchical clustering with $K = 3$ and $n = 200$. \textbf{(a)}
Quantile-quantile plots of the $p$-values under \eqref{eq:intro_model} with
$\boldsymbol\mu = \mathbf 0_{n \times p}$, stratified by $\boldsymbol\Sigma$ and pooled over $p$. \textbf{(b)} The
recovery probability \eqref{eq:recovery} under \eqref{eq:intro_model} with $\boldsymbol\mu$ defined in
\eqref{eq:sim_means} and $\boldsymbol\Sigma = \mathbf{I}_p$. \textbf{(c)} The conditional power \eqref{eq:cond_power} at
$\alpha = 0.05$ in the same settings, stratified by $p$, for the tests that control the selective Type~I error
under every $\boldsymbol\Sigma$; at each $\delta$ the tests are drawn side by side, with $95\%$ Wilson intervals.}
\label{fig:hier}
\end{figure}

Figure~\ref{fig:hier}(b) and (c) display the recovery probability and the conditional power for
hierarchical clustering under isotropic noise. Average linkage rarely returns two true clusters at
$\delta = 4$. From $\delta = 6$ on, $p_{\mathrm{polar}}$ rejects $H_0$ for every pair of true clusters, and
$p_{\mathrm{GBW}}(\boldsymbol\Sigma)$, which knows $\boldsymbol\Sigma$, for almost every pair, while $p_{\mathrm{polar,1d}}$ and
$p_{\mathrm{GBW}}(\hat{\boldsymbol\Sigma}_{\mathrm{sample}})$ are less powerful at $\delta = 6$, increasingly so as $p$
grows.

\paragraph{Computing time.} At $n = 200$, one value of $p_{\mathrm{polar}}$ takes a median of $2.1$ seconds
on one core of an Apple M4 Max processor at $p = 2$ and $5.4$ seconds at $p = 100$, and one value of
$p_{\mathrm{polar,1d}}$ takes $1.0$ to $2.3$ seconds. The closed-form $p_{\mathrm{polar,1d}}^{\mathrm{traj}}$
and the tests of \citet{ChenWitten2023} take about $0.1$ seconds (Figure~\ref{fig:kmeans_timing} in
Appendix~\ref{appendix:grid-extra}). Each of the $M = 5{,}000$ steps of the path reruns the clustering
$L = 2$ times.

\subsection{Sensitivity to the Gaussian assumption}
\label{sec:robust}

The validity of our proposals in Section~\ref{sec:method} depends on the rows of $\mathbf{X}$ being Gaussian. We study
the sensitivity to moderate deviations from this assumption in Appendix~\ref{appendix:grid-extra}. The null data have
$n = 200$ and compound-symmetric $\boldsymbol\Sigma$, and their rows are drawn from a multivariate $t$ distribution, drawn
as Gaussian and censored at zero, or drawn as Gaussian with $p \ge r$ and reduced to their leading principal components
before clustering, a common step in the analysis of high-dimensional data. The tests are less sensitive after
$k$-means than after average linkage. After $k$-means, $p_{\mathrm{polar}}$ and $p_{\mathrm{polar,1d}}$ control the
selective Type~I error with $t$ rows ($5$ or $10$ degrees of freedom) and after the reduction to principal components,
and reject at most $7.1\%$ of true null hypotheses with $5\%$ of the entries censored. After average linkage, they
reject at most $7.4\%$ with $5\%$ censoring but are anticonservative with $t$ rows.

\section{Data application}
\label{sec:realdata}

We apply the proposed tests to three data sets spanning gene expression microarrays, text embeddings, and single-cell RNA
sequencing: leukemia samples (Section~\ref{sec:realdata_golub}) and drug reviews (Section~\ref{sec:realdata_reviews}), both
clustered by $k$-means, and cells of the mouse brain, clustered by Leiden community detection
(Section~\ref{sec:realdata_brain}). We use the known labels in each data set to construct ``alternative'' data sets, whose
observations come from several labeled groups, and, where possible, ``null'' data sets, whose observations come from a
single group, and we evaluate the proposed tests against existing ones on them. We compute $p_{\mathrm{polar}}$ and $p_{\mathrm{polar,1d}}$ with $M = 5{,}000$ and $L = 2$ for every pair of
estimated clusters, together with the naive test and, after $k$-means, the test of \citet{ChenWitten2023} with
$\hat{\boldsymbol\Sigma}_{\mathrm{sample}}$. Because the tests require $|\hat{\mathcal C}_1| + |\hat{\mathcal C}_2| \ge p + 2$, we reduce each data set to its
leading principal components before clustering: ten for the leukemia data and twenty for the drug reviews and the
mouse brain. The figures display each data set with UMAP \citep{McInnesHealyMelville2018}.

\subsection{Leukemia microarray data}
\label{sec:realdata_golub}

\citet{GolubEtAl1999} measured gene expression in $72$ bone-marrow and blood samples from patients with acute lymphoblastic leukemia
(ALL, $47$ samples: $38$ B-cell acute lymphoblastic leukemia and $9$ T-cell acute lymphoblastic leukemia) or acute
myeloid leukemia (AML, $25$ samples). We preprocess the data as in \citet{DudoitFridlyandSpeed2002} and keep the first ten
principal components.

\begin{figure}[t]
\centering
\begin{minipage}[c]{0.4\textwidth}\centering
\includegraphics[width=\textwidth]{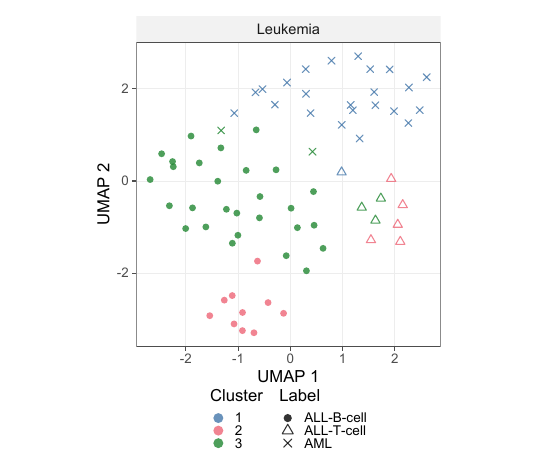}
\end{minipage}\hfill
\begin{minipage}[c]{0.58\textwidth}\centering
{\small\setlength{\tabcolsep}{4pt}\resizebox{\textwidth}{!}{\begin{tabular}{lrrr}
\toprule
Pair & 1-2 & 1-3 & 2-3 \\
\midrule
Majority labels & AML 96\% / B-ALL 67\% & AML 96\% / B-ALL 85\% & B-ALL 67\% / B-ALL 85\% \\
Expected & ``alternative'' & ``alternative'' & ``null'' \\
$(n_1, n_2)$ & $(24, 15)$ & $(24, 33)$ & $(15, 33)$ \\
\midrule
$p_{\mathrm{naive}}$ & $5.8\times 10^{-12}$ & $1.4\times 10^{-17}$ & $1.7\times 10^{-9}$ \\
$p_{\mathrm{polar}}$ & 0.005 & 0.015 & 0.316 \\
$p_{\mathrm{polar,1d}}$ & 0.969 & 0.208 & 0.990 \\
$p_{\mathrm{CW}}(\hat{\boldsymbol\Sigma}_{\mathrm{sample}})$ & 0.941 & 0.201 & 0.966 \\
\bottomrule
\end{tabular}
}}
\end{minipage}
\caption{The leukemia data of Section~\ref{sec:realdata_golub}. Left: a UMAP embedding, colored by cluster, with the known
label as the plotting symbol. Right: naive and selective $p$-values for each pair of clusters. The row Expected is
``alternative'' when the two clusters have different majority labels and ``null'' when they share one.}
\label{fig:realdata_golub}
\end{figure}

We apply $k$-means with $K = 3$ and obtain clusters of $24$, $15$, and $33$ samples: cluster~1 is $96\%$ AML, cluster~2 is
$67\%$ B-cell ALL, and cluster~3 is $85\%$ B-cell ALL (Figure~\ref{fig:realdata_golub}, left). The naive test rejects all three
null hypotheses, with $p$-values below $10^{-8}$ (Figure~\ref{fig:realdata_golub}, right). The test $p_{\mathrm{polar}}$ rejects the
two contrasts between the AML cluster and a B-cell ALL cluster ($p = 0.005$ and $0.015$), but not the contrast between the
two B-cell ALL clusters ($p = 0.32$), which we expect not to differ much biologically. Neither $p_{\mathrm{polar,1d}}$ nor the
test of \citet{ChenWitten2023} with $\hat{\boldsymbol\Sigma}_{\mathrm{sample}}$ rejects any of the three, including the two contrasts
between the AML cluster and a B-cell ALL cluster. This reflects the power lost to extra conditioning and to a
conservative covariance estimate, respectively.

\subsection{Drug reviews}
\label{sec:realdata_reviews}

We apply clustering to group patients by their reviews of the drugs they take, in data collected by
\citet{GraesserEtAl2018} on Drugs.com. Each review is labeled with the condition that the drug was taken for and with a
rating from $1$ to $10$. We use a random sample of $3{,}000$ reviews of
at least $20$ words and represent each review by its $384$-dimensional text embedding from the pretrained model
\texttt{bge-small-en-v1.5} \citep{XiaoEtAl2023}. For each data set below, the features are the first twenty principal components
of the embeddings, computed from the reviews that are not in that data set.

\begin{figure}[t]
\centering
\includegraphics[width=0.8\textwidth]{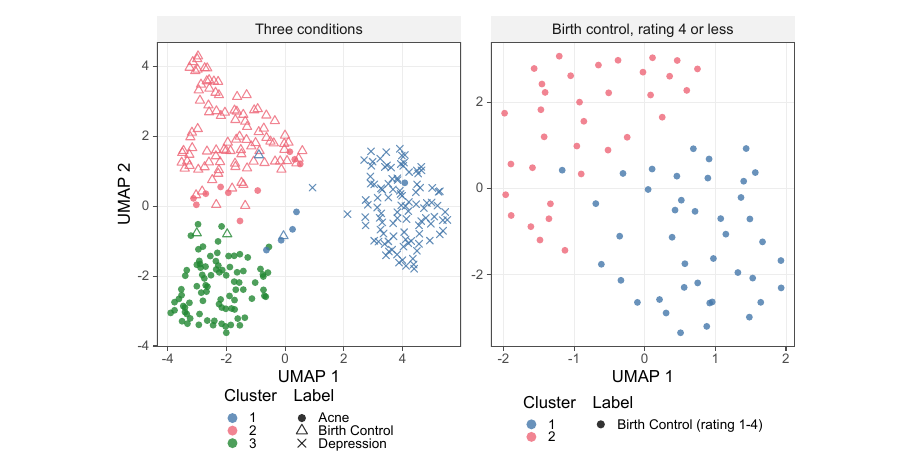}

\vspace{4pt}
{\small\setlength{\tabcolsep}{4pt}\resizebox{1\textwidth}{!}{\begin{tabular}{lllllrrrr}
\toprule
Data set & Pair & Majority labels & Expected & $(n_1,n_2)$ & $p_{\mathrm{naive}}$ & $p_{\mathrm{polar}}$ & $p_{\mathrm{polar,1d}}$ & $p_{\mathrm{CW}}(\hat{\boldsymbol\Sigma}_{\mathrm{sample}})$ \\
\midrule
``Alternative'' & 1-2 & Dep 93\% / BC 91\% & ``alternative'' & $(107, 106)$ & $2.6\times 10^{-86}$ & 0.005 & 0.021 & 0.145 \\
 & 1-3 & Dep 93\% / Acne 98\% & ``alternative'' & $(107, 87)$ & $2.1\times 10^{-82}$ & 0.005 & 0.366 & 0.617 \\
 & 2-3 & BC 91\% / Acne 98\% & ``alternative'' & $(106, 87)$ & $1.3\times 10^{-67}$ & $4.8\times 10^{-4}$ & 0.031 & 0.965 \\
\addlinespace
``Null'' & 1-2 & BC- 100\% / BC- 100\% & ``null'' & $(41, 34)$ & $8.7\times 10^{-14}$ & 0.136 & 0.218 & 0.299 \\
\bottomrule
\end{tabular}
}}
\caption{\looseness=-1 The drug reviews of Section~\ref{sec:realdata_reviews}. Top: a UMAP embedding of their principal components, colored by cluster, with the condition as the plotting symbol; left, the ``alternative'' data set; right, the ``null'' data set. Bottom: naive and selective $p$-values for each pair of clusters. The column Expected is ``alternative'' when the two clusters have different majority labels and ``null'' when they share one. Majority labels: Dep, depression; BC, birth control; BC-, birth control with a rating of $4$ or less.}
\label{fig:realdata_reviews}
\end{figure}

We construct an ``alternative'' data set by drawing $100$ reviews each of drugs taken for birth control, depression, and
acne, so $n = 300$. We apply $k$-means with $K = 3$ and obtain clusters of $107$, $106$, and $87$ reviews: cluster~1 is $93\%$
depression, cluster~2 is $91\%$ birth control, and cluster~3 is $98\%$ acne (Figure~\ref{fig:realdata_reviews}, top left). The
naive test gives $p$-values below $10^{-66}$ for all three pairs (Figure~\ref{fig:realdata_reviews}, bottom). The test
$p_{\mathrm{polar}}$ also rejects all three null hypotheses ($p$ between $4.8 \times 10^{-4}$ and $0.005$), as it should for drugs
taken for three different conditions, and $p_{\mathrm{polar,1d}}$ rejects two of them. The test of \citet{ChenWitten2023} with
$\hat{\boldsymbol\Sigma}_{\mathrm{sample}}$ rejects none ($p$ between $0.15$ and $0.97$).

We construct a ``null'' data set by drawing $75$ reviews of drugs taken for birth control with a rating of $4$ or less. We
apply $k$-means with $K = 2$ and obtain clusters of $41$ and $34$ reviews. The naive test rejects
($p = 8.7 \times 10^{-14}$), but $p_{\mathrm{polar}}$ and $p_{\mathrm{polar,1d}}$ do not ($p = 0.14$ and $0.22$).

\subsection{Single-cell RNA sequencing of the mouse brain}
\label{sec:realdata_brain}

In single-cell RNA sequencing, biologists routinely cluster cells to find putative cell types and then test the
clusters for differences in gene expression. \citet{ZeiselEtAl2015} sequenced cells from the mouse cortex and hippocampus
and labeled each cell with a cell type and a finer subclass. For each data set below, we choose the $2{,}000$ most
variable genes and the first twenty principal components from the cells that are not in that data set and project the
analyzed cells onto them. We cluster with Leiden community detection \citep{TraagWaltmanVanEck2019} on the graph that joins
each cell to its $15$ nearest neighbors, with the resolution fixed at $0.5$. No other selective test applies to this
clustering, so we compare our tests with the naive test only.

\begin{figure}[t]
\centering
\includegraphics[width=0.8\textwidth]{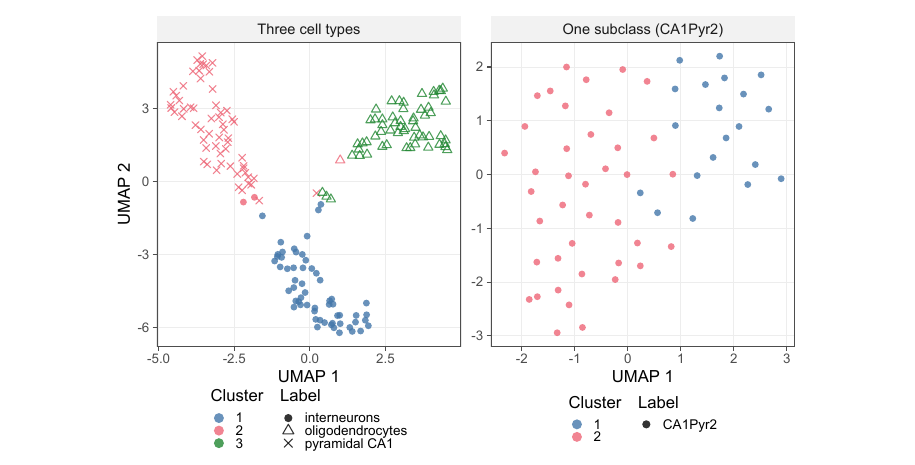}

\vspace{4pt}
{\small\setlength{\tabcolsep}{4pt}\resizebox{0.85\textwidth}{!}{\begin{tabular}{lllllrrr}
\toprule
Data set & Pair & Majority labels & Expected & $(n_1,n_2)$ & $p_{\mathrm{naive}}$ & $p_{\mathrm{polar}}$ & $p_{\mathrm{polar,1d}}$ \\
\midrule
``Alternative'' & 1-2 & IN 100\% / CA1 95\% & ``alternative'' & $(58, 63)$ & $1.5\times 10^{-55}$ & 0.002 & 0.969 \\
 & 1-3 & IN 100\% / OL 100\% & ``alternative'' & $(58, 59)$ & $6.4\times 10^{-59}$ & 0.003 & $7.0\times 10^{-4}$ \\
 & 2-3 & CA1 95\% / OL 100\% & ``alternative'' & $(63, 59)$ & $2.2\times 10^{-65}$ & $2.9\times 10^{-4}$ & 0.961 \\
\addlinespace
``Null'' & 1-2 & CA1Pyr2 100\% / CA1Pyr2 100\% & ``null'' & $(20, 40)$ & $3.4\times 10^{-8}$ & 0.757 & 0.461 \\
\bottomrule
\end{tabular}
}}
\caption{The mouse brain cells of Section~\ref{sec:realdata_brain}. Top: a UMAP embedding, colored by cluster, with the known label as the plotting symbol; left, the ``alternative'' data set; right, the ``null'' data set. Bottom: naive and selective $p$-values for each pair of clusters. The column Expected is ``alternative'' when the two clusters have different majority labels and ``null'' when they share one. Majority labels: IN, interneurons; CA1, CA1 pyramidal cells; OL, oligodendrocytes. No existing selective inference test applies to Leiden clustering.}
\label{fig:realdata_brain}
\end{figure}

We construct an ``alternative'' data set by drawing $60$ cells each of interneurons, CA1 pyramidal cells, and
oligodendrocytes, so $n = 180$. Leiden community detection returns clusters of $58$, $63$, and $59$ cells: cluster~1 is
$100\%$ interneurons, cluster~2 is $95\%$ CA1 pyramidal cells, and cluster~3 is $100\%$ oligodendrocytes
(Figure~\ref{fig:realdata_brain}, top left). The naive test gives $p$-values below $10^{-54}$, and $p_{\mathrm{polar}}$ rejects all
three null hypotheses ($p$ between $2.9 \times 10^{-4}$ and $0.003$), as it should for three distinct cell types. The test
$p_{\mathrm{polar,1d}}$ rejects only the contrast between interneurons and oligodendrocytes.

We construct a ``null'' data set by drawing $60$ cells of the single subclass CA1Pyr2. Leiden community detection
splits them into clusters of $20$ and $40$ cells. The naive test rejects the null hypothesis ($p = 3.4 \times 10^{-8}$), while
$p_{\mathrm{polar}}$ and $p_{\mathrm{polar,1d}}$ do not ($p = 0.76$ and $0.46$). With $180$ CA1Pyr2 cells instead of $60$, Leiden
community detection returns three clusters, the naive test again rejects all three null hypotheses
($p < 10^{-20}$), and $p_{\mathrm{polar}}$ rejects none ($p$ between $0.29$ and $0.79$).

\section{Discussion}
\label{sec:discussion}

In this paper, we proposed a selective test for a difference in means between two
estimated clusters when the covariance matrix $\boldsymbol\Sigma$ in \eqref{eq:intro_model} is
\emph{unknown and unstructured}. The test does not estimate $\boldsymbol\Sigma$; instead, it uses the
polar decomposition of the tested observations, which splits their distribution into
a part whose distribution depends on $\boldsymbol\Sigma$ and an independent part whose
distribution is known.
As a result, the test avoids the trade-off between Type I error control and power that
comes with a plug-in estimate of $\boldsymbol\Sigma$ (Section~\ref{sec:challenges}), and it is
exact in finite samples for any clustering algorithm. We gave two versions of the test.
The test $p_{\mathrm{polar}}$ conditions on less, has higher power, and is computed by an exact Monte Carlo
construction that takes longer to run. The one-dimensional test $p_{\mathrm{polar,1d}}$ conditions on more
information, is faster to compute, and has a closed form for $k$-means when we also condition on the sequence of
assignments of Lloyd's algorithm.

These results suggest a number of extensions and open questions. First, we test the
null hypothesis \eqref{eq:intro_null_strong} that all observations in the two clusters
share a mean, which is stronger than the null hypothesis \eqref{eq:intro_null} that the
two cluster means are equal (Remark~\ref{rem:which_null}). Testing
\eqref{eq:intro_null} without knowledge of $\boldsymbol\Sigma$ remains open. Second, we have taken the tested pair
as given; when the pair is chosen from the data, for example as the two clusters that look most separated, the
conditional distribution of $\mathbf{Q}$ must be reweighted by the probability of that choice in the Monte Carlo
simulation. Third, any reversible kernel gives a $p$-value that controls the selective Type~I error, but different
kernels and different values of their parameters lead to different power, and the best choice for a given $n$,
$p$, and $\boldsymbol\Sigma$ remains an open question. Fourth, we assumed that the rows of $\mathbf{X}$ are independent;
testing data with correlated observations could build on the techniques of \citet{GonzalezDelgadoEtAl2026}.
Fifth, we would like to extend the test to non-Gaussian data, such as the counts that arise in single-cell RNA
sequencing, using techniques from data fission \citep{LeinerDuanWassermanRamdas2025} and data thinning
\citep{NeufeldDharamshiGaoWitten2024, DharamshiEtAl2024}.

\section*{AI assistance statement}

During manuscript development, the authors used AI assistants, including GPT-5.5
and coding agents, as interactive tools for mathematical brainstorming, software
development, and editorial review. This assistance included proposing and
stress-testing proof strategies, checking notation and edge cases, drafting code
scaffolds for simulations and manuscript packaging, and suggesting revisions to
the exposition. Several proof directions that shaped the final presentation were
developed through this workflow: the covariance-free polar-factor formulation, the
conditioning argument on the Stiefel feasible set, the Besag--Clifford random-index
validity proof, the algorithm-agnostic extension to hierarchical clustering, and
the oracle-geometry statement separating validity from power. AI outputs were
treated as suggestions rather than evidence: all mathematical derivations, code,
simulation outputs, citations, and prose were checked, revised, and verified by the
authors, who take full responsibility for the final content.

\bibliographystyle{plainnat}
\bibliography{references}

\appendix
\counterwithin{figure}{section}
\counterwithin{table}{section}
\counterwithin{algorithm}{section}

\section{Proofs for Sections~\ref{subsec:polar} to~\ref{subsec:selection}}
\label{appendix:polar}

This appendix proves Theorem~\ref{thm:polar_uniform} and Propositions~\ref{prop:T2_share}
and~\ref{prop:conditional_selection}, one per subsection, and Appendix~\ref{appendix:onedim_compare} compares the three
one-dimensional tests. Appendix~\ref{appendix:feasible} uses the same notation.

\paragraph{Preliminaries and notation.} Fix a matrix $\mathbf{A} \in \R^{n \times (r-1)}$ whose columns are an
orthonormal basis of the range of $\mathbf P_0 + \mathbf P_1$, so that
\[
\mathbf{A}\mathbf{A}^\top = \mathbf P_0 + \mathbf P_1, \qquad \mathbf{A}^\top\mathbf{A} = \mathbf{I}_{r-1}.
\]
The matrix $\mathbf{Z} = \mathbf{A}^\top\mathbf{X} \in \R^{(r-1) \times p}$ holds the coordinates of $(\mathbf P_0 + \mathbf P_1)\mathbf{X}$ in this basis:
$(\mathbf P_0 + \mathbf P_1)\mathbf{X} = \mathbf{A}\mathbf{Z}$ and $\mathbf{H}^2 = \mathbf{X}^\top(\mathbf P_0 + \mathbf P_1)\mathbf{X} = \mathbf{Z}^\top\mathbf{Z}$. Every matrix in $\mathcal V$
is $\mathbf{A}$ times a matrix in $\St(r-1, p)$, and multiplying by $\mathbf{A}$ preserves inner products, so we identify the
polar factor with its coordinates $\mathbf{A}^\top\mathbf{Q} \in \St(r-1, p)$ and write $\mathbf{Q}$ for both. In coordinates,
\[
\mathbf{Z} = \mathbf{Q}\mathbf{H}, \qquad \mathbf{x}'(\mathbf{Q}) = \mathbf P_2\mathbf{x} + \mathbf{A}\mathbf{Q}\mathbf{H}, \qquad u(\mathbf{Q}) = \|\mathbf{Q}^\top\mathbf{a}\|_2^2
\ \text{ with } \ \mathbf{a} = \mathbf{A}^\top\mathrm{dir}(\boldsymbol\nu),
\]
where $\mathbf{a}$ is a unit vector, the uniform distribution on $\mathcal V$ becomes the uniform distribution on $\St(r-1, p)$,
written $d\mathbf{Q}$, and the Givens rotation \eqref{eq:givens_main} becomes the $(r-1) \times (r-1)$ rotation of rows $i$ and
$j$, still written $\mathbf{G}_{ij}(\omega)$.

\subsection{Proof of Theorem~\ref{thm:polar_uniform}}
\label{appendix:proof_polar}

The proof uses the distribution of the coordinates $\mathbf{Z}$.

\begin{proposition}[Distribution of the coordinates]
\label{lem:M_dist}
Under data model \eqref{eq:intro_model} and null \eqref{eq:intro_null_strong}, the matrix $\mathbf{Z}$ follows
$\MN(\mathbf 0, \mathbf{I}_{r-1}, \boldsymbol\Sigma)$ and is independent of $\mathbf P_2\mathbf{X}$.
\end{proposition}

\begin{proof}
Multiplying \eqref{eq:intro_model} on the left by $\mathbf{A}^\top$ gives $\mathbf{Z} \sim \MN(\mathbf{A}^\top\boldsymbol\mu, \mathbf{A}^\top\mathbf{A}, \boldsymbol\Sigma)
= \MN(\mathbf{A}^\top\boldsymbol\mu, \mathbf{I}_{r-1}, \boldsymbol\Sigma)$. Under \eqref{eq:intro_null_strong}, the rows of $\boldsymbol\mu$ in
$\hat{\mathcal C}_1 \cup \hat{\mathcal C}_2$ are equal, so $(\mathbf P_0 + \mathbf P_1)\boldsymbol\mu = \mathbf 0$ and
$\mathbf{A}^\top\boldsymbol\mu = \mathbf{A}^\top(\mathbf P_0 + \mathbf P_1)\boldsymbol\mu = \mathbf 0$. The matrices $\mathbf{Z}$ and $\mathbf P_2\mathbf{X}$ are jointly
Gaussian, and their cross-covariance is proportional to $\mathbf{A}^\top\mathbf P_2 = \mathbf 0$, so they are independent.
\end{proof}

\begin{proof}[Proof of Theorem~\ref{thm:polar_uniform}]
We must show that $\mathbf{Q}$ is uniformly distributed on $\St(r-1, p)$ and independent of $(\mathbf{H}, \mathbf P_2\mathbf{X})$. Because
$(\mathbf{Q}, \mathbf{H})$ is a function of $\mathbf{Z}$, which is independent of $\mathbf P_2\mathbf{X}$ by Proposition~\ref{lem:M_dist}, it is enough
to show that for all bounded functions $h_1$ of $\mathbf{Q}$ and $h_2$ of $\mathbf{H}$,
\begin{equation}
\label{eq:thm31_goal}
\E\big[h_1(\mathbf{Q})\,h_2(\mathbf{H})\big] = \E\big[h_2(\mathbf{H})\big]\int h_1(\mathbf{Q}')\,d\mathbf{Q}'.
\end{equation}
The matrix $\mathbf{Z}$ has rank $p$ with probability one when $r - 1 \ge p$, so $\mathbf{H} = (\mathbf{Z}^\top\mathbf{Z})^{1/2}$ is invertible and
$\mathbf{Q} = \mathbf{Z}\mathbf{H}^{-1}$.

Next, let $\mathbf{O}$ be a uniformly distributed $(r-1) \times (r-1)$ orthogonal matrix, drawn independently of $\mathbf{X}$. By
Proposition~\ref{lem:M_dist},
\[
\mathbf{O}\mathbf{Z} \sim \MN(\mathbf 0, \mathbf{O}\mathbf{O}^\top, \boldsymbol\Sigma) = \MN(\mathbf 0, \mathbf{I}_{r-1}, \boldsymbol\Sigma),
\]
so $\mathbf{O}\mathbf{Z}$ has the same distribution as $\mathbf{Z}$. The polar decomposition of $\mathbf{O}\mathbf{Z}$ is $(\mathbf{O}\mathbf{Q})\mathbf{H}$, because
$\mathbf{O}\mathbf{Q}$ has orthonormal columns and $(\mathbf{O}\mathbf{Z})^\top(\mathbf{O}\mathbf{Z}) = \mathbf{Z}^\top\mathbf{Z} = \mathbf{H}^2$. So $(\mathbf{O}\mathbf{Q}, \mathbf{H})$ has the
same distribution as $(\mathbf{Q}, \mathbf{H})$, and
\[
\E\big[h_1(\mathbf{Q})\,h_2(\mathbf{H})\big] = \E\big[h_1(\mathbf{O}\mathbf{Q})\,h_2(\mathbf{H})\big]
= \E\Big[h_2(\mathbf{H})\;\E\big\{h_1(\mathbf{O}\mathbf{Q}) \mid \mathbf{Q}, \mathbf{H}\big\}\Big].
\]
Finally, for a fixed $\mathbf{Q}_0 \in \St(r-1, p)$, the matrix $\mathbf{O}\mathbf{Q}_0$ is uniformly distributed on $\St(r-1, p)$, so the inner
expectation equals $\int h_1(\mathbf{Q}')\,d\mathbf{Q}'$ whatever the values of $\mathbf{Q}$ and $\mathbf{H}$. This gives
\eqref{eq:thm31_goal}.
\end{proof}

\subsection{Proof of Proposition~\ref{prop:T2_share}}
\label{appendix:proof_T2}

Proposition~\ref{prop:T2_share} makes two claims: the identity \eqref{eq:T2_share}, which holds for any data with an
invertible pooled covariance, and the Beta distribution of $u(\mathbf{Q})$ under the null.

\begin{proof}[Proof of the identity \eqref{eq:T2_share}]
Write $\bar{\boldsymbol\nu} = \mathrm{dir}(\boldsymbol\nu)$ and $\mathbf{b} = \mathbf{Q}^\top\bar{\boldsymbol\nu} \in \R^p$, with $\mathbf{Q}$ the $n \times p$ polar factor of
Section~\ref{subsec:polar}. Since $\mathbf P_0 = \bar{\boldsymbol\nu}\bar{\boldsymbol\nu}^\top$, we have $u(\mathbf{Q}) = \|\mathbf P_0\mathbf{Q}\|_F^2 = \|\mathbf{b}\|_2^2$. We
write both ingredients of $T^2$, the difference in cluster means and the pooled covariance, in terms of $\mathbf{H}$ and $\mathbf{b}$.

The difference in cluster means is $\bar{\mathbf{X}}_{\hat{\mathcal C}_1} - \bar{\mathbf{X}}_{\hat{\mathcal C}_2} = \mathbf{X}^\top\boldsymbol\nu = \|\boldsymbol\nu\|_2\,\mathbf{X}^\top\bar{\boldsymbol\nu}$,
and $n_1 n_2/r = \|\boldsymbol\nu\|_2^{-2}$, so
\begin{equation}
\label{eq:T2_step1}
T^2 = (r-2)\,\bar{\boldsymbol\nu}^\top\mathbf{X}\,(\mathbf{X}^\top\mathbf P_1\mathbf{X})^{-1}\,\mathbf{X}^\top\bar{\boldsymbol\nu}.
\end{equation}
The vector $\bar{\boldsymbol\nu}$ lies in the range of $\mathbf P_0$, so $\bar{\boldsymbol\nu} = (\mathbf P_0 + \mathbf P_1)\bar{\boldsymbol\nu}$, and by \eqref{eq:polar_dec},
\begin{equation}
\label{eq:T2_step2}
\mathbf{X}^\top\bar{\boldsymbol\nu} = \mathbf{X}^\top(\mathbf P_0 + \mathbf P_1)\bar{\boldsymbol\nu} = \mathbf{H}\mathbf{Q}^\top\bar{\boldsymbol\nu} = \mathbf{H}\mathbf{b}.
\end{equation}
Since $\mathbf P_1 = (\mathbf P_0 + \mathbf P_1) - \mathbf P_0$, the same identity gives
\begin{equation}
\label{eq:T2_step3}
\mathbf{X}^\top\mathbf P_1\mathbf{X} = \mathbf{H}^2 - \mathbf{X}^\top\bar{\boldsymbol\nu}\bar{\boldsymbol\nu}^\top\mathbf{X} = \mathbf{H}^2 - \mathbf{H}\mathbf{b}\mathbf{b}^\top\mathbf{H}
= \mathbf{H}\big(\mathbf{I}_p - \mathbf{b}\mathbf{b}^\top\big)\mathbf{H}.
\end{equation}
The matrix $\mathbf{I}_p - \mathbf{b}\mathbf{b}^\top$ has determinant $1 - \|\mathbf{b}\|_2^2 = 1 - u(\mathbf{Q})$, so the pooled covariance is invertible
exactly when $u(\mathbf{Q}) < 1$. Substituting \eqref{eq:T2_step2} and \eqref{eq:T2_step3} into \eqref{eq:T2_step1}, the factors
$\mathbf{H}$ cancel:
\[
T^2 = (r-2)\,\mathbf{b}^\top\mathbf{H}\,\mathbf{H}^{-1}\big(\mathbf{I}_p - \mathbf{b}\mathbf{b}^\top\big)^{-1}\mathbf{H}^{-1}\,\mathbf{H}\mathbf{b}
= (r-2)\,\mathbf{b}^\top\big(\mathbf{I}_p - \mathbf{b}\mathbf{b}^\top\big)^{-1}\mathbf{b}.
\]
By the Sherman--Morrison formula,
\[
\big(\mathbf{I}_p - \mathbf{b}\mathbf{b}^\top\big)^{-1} = \mathbf{I}_p + \frac{\mathbf{b}\mathbf{b}^\top}{1 - \|\mathbf{b}\|_2^2},
\qquad\text{so}\qquad
\mathbf{b}^\top\big(\mathbf{I}_p - \mathbf{b}\mathbf{b}^\top\big)^{-1}\mathbf{b} = \|\mathbf{b}\|_2^2 + \frac{\|\mathbf{b}\|_2^4}{1 - \|\mathbf{b}\|_2^2}
= \frac{\|\mathbf{b}\|_2^2}{1 - \|\mathbf{b}\|_2^2}.
\]
With $\|\mathbf{b}\|_2^2 = u(\mathbf{Q})$, this is \eqref{eq:T2_share}. When $p = 1$, $u(\mathbf{Q}) = \|\mathbf P_0\mathbf{X}\|_F^2/\|(\mathbf P_0 + \mathbf P_1)\mathbf{X}\|_F^2 = w$
and $1 - u(\mathbf{Q}) = \|\mathbf P_1\mathbf{X}\|_F^2/\|(\mathbf P_0 + \mathbf P_1)\mathbf{X}\|_F^2$, which gives $T^2 = R$.
\end{proof}

\begin{proof}[Proof of the distribution in Proposition~\ref{prop:T2_share}]
By Theorem~\ref{thm:polar_uniform}, under the null $\mathbf{Q}$ is uniformly distributed on $\St(r-1, p)$, and in coordinates
$u(\mathbf{Q}) = \|\mathbf{Q}^\top\mathbf{a}\|_2^2$ for the fixed unit vector $\mathbf{a}$. We find the distribution in three steps.

First, a uniformly distributed $\mathbf{Q}$ has the same distribution as the first $p$ columns of a uniformly distributed
$(r-1) \times (r-1)$ orthogonal matrix $\mathbf{O}$. Then $\mathbf{Q}^\top\mathbf{a}$ consists of the first $p$ entries of the vector
$\mathbf{O}^\top\mathbf{a}$.

Second, $\mathbf{O}^\top\mathbf{a}$ is a fixed unit vector turned by a uniformly distributed orthogonal matrix, so it is uniformly
distributed on the unit sphere of $\R^{r-1}$. A standard Gaussian vector $\mathbf{z} \sim \N(\mathbf 0, \mathbf{I}_{r-1})$ has a
distribution that is unchanged by orthogonal matrices, so its direction $\mathbf{z}/\|\mathbf{z}\|_2$ is also uniformly distributed on
the sphere. Hence $\mathbf{O}^\top\mathbf{a}$ has the same distribution as $\mathbf{z}/\|\mathbf{z}\|_2$.

Third, therefore,
\[
u(\mathbf{Q}) = \sum_{j=1}^{p}\frac{z_j^2}{\|\mathbf{z}\|_2^2} = \frac{U}{U + V},
\qquad U = \sum_{j=1}^{p} z_j^2 \sim \chi^2_p, \quad V = \sum_{j=p+1}^{r-1} z_j^2 \sim \chi^2_{r-1-p},
\]
with $U$ and $V$ independent, and such a ratio has the $\mathrm{Beta}\big(\tfrac p2, \tfrac{r-1-p}2\big)$ distribution.
The statement for $T^2$ follows from \eqref{eq:T2_share}, because
$\frac{r-p-1}{(r-2)p}\,T^2 = \frac{r-1-p}{p}\cdot\frac{U}{V}$ is a ratio of independent $\chi^2$ variables divided by their degrees of
freedom. The substitution $u(\mathbf{Q}) = \cos^2\phi$ gives the density $f$ of Section~\ref{sec:onedim_mc}.
\end{proof}

\subsection{Proof of Proposition~\ref{prop:conditional_selection}}
\label{appendix:proof_selection}

Proposition~\ref{prop:conditional_selection} makes three claims. (a) Given $\mathbf P_2\mathbf{X}$, $\mathbf{H}$, and the event that the
clustering returns $\hat{\mathcal C}_1$ and $\hat{\mathcal C}_2$, the polar factor $\mathbf{Q}$ is uniformly distributed on $\mathcal S_{\mathbf{Q}}$. (b) The
$p$-value $p_{\mathrm{polar}}$ is the ratio of integrals in \eqref{eq:ppolar_integral}. (c) Under the null, $p_{\mathrm{polar}}$ is
uniformly distributed given that the clustering returns $\hat{\mathcal C}_1$ and $\hat{\mathcal C}_2$. Throughout, $\mathcal S_{\mathbf{Q}}$ is the set of
Section~\ref{subsec:selection} with the random $(\mathbf P_2\mathbf{X}, \mathbf{H})$ in place of their observed values,
\[
\mathcal S_{\mathbf{Q}} = \big\{\mathbf{Q}' \in \St(r-1, p) : \hat{\mathcal C}_1, \hat{\mathcal C}_2 \in \mathcal C(\mathbf P_2\mathbf{X} + \mathbf{A}\mathbf{Q}'\mathbf{H})\big\}.
\]

\begin{proof}[Proof of (a)]
The data are $\mathbf{X} = \mathbf P_2\mathbf{X} + \mathbf{A}\mathbf{Q}\mathbf{H}$, so the clustering returns $\hat{\mathcal C}_1$ and $\hat{\mathcal C}_2$ exactly when
$\mathbf{Q} \in \mathcal S_{\mathbf{Q}}$. By Theorem~\ref{thm:polar_uniform}, $\mathbf{Q}$ is uniformly distributed and independent of
$(\mathbf P_2\mathbf{X}, \mathbf{H})$, and $\mathcal S_{\mathbf{Q}}$ is fixed once $(\mathbf P_2\mathbf{X}, \mathbf{H})$ is. So for every set $B \subseteq \St(r-1, p)$,
\[
\Prob\big(\mathbf{Q} \in B \text{ and } \hat{\mathcal C}_1, \hat{\mathcal C}_2 \in \mathcal C(\mathbf{X}) \;\big|\; \mathbf P_2\mathbf{X}, \mathbf{H}\big)
= \Prob\big(\mathbf{Q} \in B \cap \mathcal S_{\mathbf{Q}} \;\big|\; \mathbf P_2\mathbf{X}, \mathbf{H}\big)
= \int_{B \cap \mathcal S_{\mathbf{Q}}} d\mathbf{Q}.
\]
Taking $B = \St(r-1, p)$ gives $\Prob(\hat{\mathcal C}_1, \hat{\mathcal C}_2 \in \mathcal C(\mathbf{X}) \mid \mathbf P_2\mathbf{X}, \mathbf{H}) = \int_{\mathcal S_{\mathbf{Q}}} d\mathbf{Q}$,
and dividing the two displays,
\[
\Prob\big(\mathbf{Q} \in B \;\big|\; \hat{\mathcal C}_1, \hat{\mathcal C}_2 \in \mathcal C(\mathbf{X}),\ \mathbf P_2\mathbf{X}, \mathbf{H}\big)
= \frac{\int_{B \cap \mathcal S_{\mathbf{Q}}} d\mathbf{Q}}{\int_{\mathcal S_{\mathbf{Q}}} d\mathbf{Q}},
\]
which is the uniform distribution on $\mathcal S_{\mathbf{Q}}$. The division is allowed: the clustering returns the tested clusters
while $\int_{\mathcal S_{\mathbf{Q}}} d\mathbf{Q} = 0$ with probability
$\E\big[\one\{\int_{\mathcal S_{\mathbf{Q}}} d\mathbf{Q} = 0\}\int_{\mathcal S_{\mathbf{Q}}} d\mathbf{Q}\big] = 0$.
\end{proof}

\begin{proof}[Proof of (b)]
By \eqref{eq:T2_share}, $T^2$ increases with $u(\mathbf{Q})$, so $T^2(\mathbf{X}) \ge T^2(\mathbf{x})$ exactly when
$u(\mathbf{Q}) \ge u(\mathbf{Q}_{\text{obs}})$. Taking $B = \{\mathbf{Q} : u(\mathbf{Q}) \ge u(\mathbf{Q}_{\text{obs}})\}$ in (a) gives
\eqref{eq:ppolar_integral}.
\end{proof}

\begin{proof}[Proof of (c)]
Fix $\mathbf P_2\mathbf{X}$ and $\mathbf{H}$. For a number $c$, the probability that $u(\mathbf{Q}) \ge c$ when $\mathbf{Q}$ is uniformly distributed on
$\mathcal S_{\mathbf{Q}}$ is
\[
\bar F(c) = \frac{\int_{\mathcal S_{\mathbf{Q}}} \one\{u(\mathbf{Q}) \ge c\}\,d\mathbf{Q}}{\int_{\mathcal S_{\mathbf{Q}}} d\mathbf{Q}} .
\]
By (b), $p_{\mathrm{polar}} = \bar F\big(u(\mathbf{Q}_{\text{obs}})\big)$, and by (a), under the null $\mathbf{Q}_{\text{obs}}$ is a draw from the uniform
distribution on $\mathcal S_{\mathbf{Q}}$.

The function $\bar F$ is continuous. Under the uniform distribution on $\St(r-1, p)$, $u(\mathbf{Q})$ has a Beta distribution
(Proposition~\ref{prop:T2_share}), so each value $c$ has probability zero. The uniform distribution on $\mathcal S_{\mathbf{Q}}$ is the
uniform distribution on $\St(r-1, p)$ restricted to a set of positive probability, so each value still has probability zero.

For a random variable with a continuous distribution, its upper tail probability evaluated at the variable itself is
uniformly distributed (the probability integral transform). Hence
\[
\Prob\big(p_{\mathrm{polar}} \le \alpha \;\big|\; \hat{\mathcal C}_1, \hat{\mathcal C}_2 \in \mathcal C(\mathbf{X}),\ \mathbf P_2\mathbf{X}, \mathbf{H}\big) = \alpha
\qquad \text{for all } \alpha \in [0, 1].
\]
The right side does not depend on $(\mathbf P_2\mathbf{X}, \mathbf{H})$, so averaging over them given the selection event gives the
statement conditional on the selection event alone.
\end{proof}

\subsection{Three one-dimensional tests side by side}
\label{appendix:onedim_compare}

Table~\ref{tab:nuisance} lists, for the test of \citet{GaoBienWitten2024}, the test of
\citet{YunFoygelBarber2023}, and $p_{\mathrm{polar,1d}}$, the null hypothesis, the nuisance parameters, the parts of
$\mathbf{X} = \mathbf P_0\mathbf{X} + \mathbf P_1\mathbf{X} + \mathbf P_2\mathbf{X}$ that each test holds fixed, and the scalar
that remains random.

\begin{table}[H]
\centering
\footnotesize
\caption{Nuisance parameters and conditioning in three one-dimensional selective tests.
Each test conditions on the selection event, on statistics that remove the nuisance
parameters under its null, and on further statistics that leave one scalar random.}
\label{tab:nuisance}
\begin{tabular}{@{}llll@{}}
\toprule
 & \citet{GaoBienWitten2024} & \citet{YunFoygelBarber2023} & $p_{\mathrm{polar,1d}}$ \\
 & ($\boldsymbol\Sigma$ known) & ($\boldsymbol\Sigma = \sigma^2\mathbf{I}_p$) & (any $\boldsymbol\Sigma$) \\
\midrule
null & \eqref{eq:intro_null} & \eqref{eq:intro_null_strong} & \eqref{eq:intro_null_strong} \\
nuisance parameters & $(\mathbf P_1 + \mathbf P_2)\boldsymbol\mu$ & means, $\sigma^2$ & means, $\boldsymbol\Sigma$ \\
removed by fixing & $\mathbf P_1\mathbf{X}$, $\mathbf P_2\mathbf{X}$ & $\mathbf P_2\mathbf{X}$, $\|(\mathbf P_0 + \mathbf P_1)\mathbf{X}\|_F$ & $\mathbf P_2\mathbf{X}$, $\mathbf{H}$ \\
fixed to reach one dimension & $\mathrm{dir}(\boldsymbol\Sigma^{-1/2}\mathbf{X}^\top\boldsymbol\nu)$ & $\mathrm{dir}(\mathbf P_0\mathbf{X})$, $\mathrm{dir}(\mathbf P_1\mathbf{X})$ & $\boldsymbol\gamma$, $\mathrm{dir}(\mathbf P_1\mathbf{q})$, $\mathbf{Q}(\mathbf{I}_p - \boldsymbol\gamma\boldsymbol\gamma^\top)$ \\
random scalar & $\|\boldsymbol\Sigma^{-1/2}\mathbf{X}^\top\boldsymbol\nu\|_2$ & $R$ & $T^2$ \\
null distribution & $\|\boldsymbol\nu\|_2\cdot\chi_p$ & $F_{p,\,(r-2)p}$ & $\frac{(r-2)p}{r-p-1}F_{p,\,r-p-1}$ \\
\bottomrule
\end{tabular}
\end{table}

\section{The $k$-means trajectory event and the Monte Carlo kernels}
\label{appendix:feasible}

This appendix proves the results of Sections~\ref{sec:onedim}, \ref{sec:mc}, \ref{sec:bc}, and~\ref{sec:kmeans}.
Appendices~\ref{appendix:quadratic} and~\ref{appendix:arcs} show that the event that Lloyd's algorithm repeats its
observed sequence of assignments is a set of quadratic inequalities in $\mathbf{Q}$, with closed-form solutions on a
Givens circle; Appendices~\ref{appendix:sampler_valid} to~\ref{appendix:bc_proof} prove the reversibility of the kernels, Proposition~\ref{cor:truncated_F} on the one-dimensional test, and the exactness of the Besag--Clifford
$p$-value.

We use the notation of Appendix~\ref{appendix:polar}: the basis $\mathbf{A}$, the polar factor $\mathbf{Q}$ written in coordinates
in $\St(r-1, p)$, and $d\mathbf{Q}$ for the uniform distribution on $\St(r-1, p)$. We write $\mathcal S$ for the set of polar
factors that the test conditions on ($\mathcal S_{\mathbf{Q}}$ or $\mathcal S_{\mathbf{Q},\tau}$).

\subsection{The trajectory event is quadratic in $\mathbf{Q}$}
\label{appendix:quadratic}

This subsection proves Proposition~\ref{prop:kmeans_quadratic}. For a polar factor $\mathbf{Q}$, the data set
\begin{equation}
\label{eq:recon}
\mathbf{x}'(\mathbf{Q}) = \mathbf P_2\mathbf{x} + \mathbf{A}\mathbf{Q}\mathbf{H}
\end{equation}
keeps $\mathbf P_2\mathbf{x}$ and $\mathbf{H}$ and replaces the polar factor. Writing $\boldsymbol\alpha_i^\top$ for the $i$th row of $\mathbf{A}$,
which is zero for observations outside $\hat{\mathcal C}_1 \cup \hat{\mathcal C}_2$, and $\mathbf{b}_i^\top$ for the $i$th row of
$\mathbf P_2\mathbf{x}$, each observation is an affine function of $\mathbf{Q}$,
\begin{equation}
\label{eq:xaffine}
\mathbf{x}'_i(\mathbf{Q})^\top = \mathbf{b}_i^\top + \boldsymbol\alpha_i^\top\mathbf{Q}\mathbf{H}.
\end{equation}

Let $\mathcal S_\ell^{(t)} = \{i : c_i^{(t)} = \ell\}$ be the observations that the observed sequence $\tau$ assigns
to cluster $\ell$ at iteration $t$; no cluster is empty at any iteration, since otherwise Lloyd's algorithm is not defined. With these assignments held
\looseness=-1 fixed, each centroid is a fixed average of the rows \eqref{eq:xaffine}, so it is also affine in $\mathbf{Q}$:
\begin{equation}
\label{eq:muaffine}
\mathbf{m}_\ell^{(t)}(\mathbf{Q})^\top = \bar{\mathbf{b}}_\ell^{(t)\top} + \bar{\boldsymbol\alpha}_\ell^{(t)\top}\mathbf{Q}\mathbf{H},
\qquad
\bar{\boldsymbol\alpha}_\ell^{(t)} = \frac{1}{|\mathcal S_\ell^{(t-1)}|}\sum_{i \in \mathcal S_\ell^{(t-1)}}\boldsymbol\alpha_i,
\quad
\bar{\mathbf{b}}_\ell^{(t)} = \frac{1}{|\mathcal S_\ell^{(t-1)}|}\sum_{i \in \mathcal S_\ell^{(t-1)}}\mathbf{b}_i,
\end{equation}
for $t \ge 1$, and the initial centroids $\mathbf{m}_\ell^{(0)}(\mathbf{Q}) = \mathbf{x}'_{\iota_\ell}(\mathbf{Q})$ are affine by
\eqref{eq:xaffine}.

\begin{proof}[Proof of Proposition~\ref{prop:kmeans_quadratic}]
Fix $t$, $i$, and a cluster $\ell \ne c = c_i^{(t)}$. Expanding the two squared distances in
\eqref{eq:feasible_g_main}, the terms $\|\mathbf{x}'_i(\mathbf{Q})\|_2^2$ cancel and
\[
g_{t,i,\ell}(\mathbf{Q}) = 2\,\big\langle \mathbf{x}'_i(\mathbf{Q}),\, \mathbf{m}_\ell^{(t)}(\mathbf{Q}) - \mathbf{m}_c^{(t)}(\mathbf{Q})\big\rangle
+ \|\mathbf{m}_c^{(t)}(\mathbf{Q})\|_2^2 - \|\mathbf{m}_\ell^{(t)}(\mathbf{Q})\|_2^2 .
\]
Substituting \eqref{eq:xaffine} and \eqref{eq:muaffine}, the terms with two factors of $\mathbf{Q}$ are
\[
2\,\boldsymbol\alpha_i^\top\mathbf{Q}\mathbf{H}^2\mathbf{Q}^\top(\bar{\boldsymbol\alpha}_\ell^{(t)} - \bar{\boldsymbol\alpha}_c^{(t)})
+ \bar{\boldsymbol\alpha}_c^{(t)\top}\mathbf{Q}\mathbf{H}^2\mathbf{Q}^\top\bar{\boldsymbol\alpha}_c^{(t)}
- \bar{\boldsymbol\alpha}_\ell^{(t)\top}\mathbf{Q}\mathbf{H}^2\mathbf{Q}^\top\bar{\boldsymbol\alpha}_\ell^{(t)},
\]
the terms with one factor of $\mathbf{Q}$ are linear in its entries, and the rest does not depend on $\mathbf{Q}$. All
coefficients are determined by $\mathbf P_2\mathbf{x}$, $\mathbf{H}$, and $\tau$. So $g_{t,i,\ell}$ is a quadratic polynomial in the
entries of $\mathbf{Q}$, and $\mathcal S_{\mathbf{Q},\tau}$ is the intersection of the sets $\{g_{t,i,\ell} \le 0\}$ over all
$t \le t_{\max}$, $i \le n$, and $\ell \ne c_i^{(t)}$, which are $(t_{\max}+1)\,n\,(K-1)$ quadratic inequalities.
\end{proof}

The proof uses the observed assignments at every iteration. If we condition only on the final clusters, the
intermediate assignments vary with $\mathbf{Q}$, and formally
\begin{equation}
\label{eq:final_union}
\mathcal S_{\mathbf{Q}} = \bigcup_{\tau'} \mathcal S_{\mathbf{Q},\tau'},
\end{equation}
where the union is over the sequences of assignments $\tau'$ that end in clusters containing $\hat{\mathcal C}_1$ and
$\hat{\mathcal C}_2$. Each $\mathcal S_{\mathbf{Q},\tau'}$ is an intersection of quadratic sets, so $\mathcal S_{\mathbf{Q}}$ is a union of
intersections of quadratic sets. The union is finite, but the number of sequences can be very large and they cannot be listed in advance, which is
why we condition on $\tau$ when we want closed forms. Techniques that compute such unions of truncation sets efficiently
have been developed for other selective tests \citep{LeDuyTakeuchi2022, ChenJewellWitten2023} and might apply here.

\subsection{Closed-form feasible arcs on a Givens circle}
\label{appendix:arcs}

This subsection proves Proposition~\ref{prop:kmeans_arcs} and gives the hit-and-run step. Fix the current
$\mathbf{Q} \in \mathcal S_{\mathbf{Q},\tau}$ and a pair of rows $i < j$. Then
\[
\mathbf{G}_{ij}(\omega)\mathbf{Q} = \mathbf{Q} + (\cos\omega - 1)\mathbf{D}_1 + \sin\omega\,\mathbf{D}_2,
\]
where $\mathbf{D}_1$ has rows $i$ and $j$ equal to those of $\mathbf{Q}$ and $\mathbf{D}_2$ has rows $i$ and $j$ equal to
$-\mathbf{Q}_{j\cdot}$ and $\mathbf{Q}_{i\cdot}$, all other rows being zero. By \eqref{eq:recon},
\begin{equation}
\label{eq:XPUW_main}
\begin{aligned}
&\mathbf{x}'(\mathbf{G}_{ij}(\omega)\mathbf{Q}) = \mathbf{Y}_0 + \cos\omega\,\mathbf{Y}_1 + \sin\omega\,\mathbf{Y}_2, \\
&\mathbf{Y}_0 = \mathbf P_2\mathbf{x} + \mathbf{A}(\mathbf{Q} - \mathbf{D}_1)\mathbf{H},\quad \mathbf{Y}_1 = \mathbf{A}\mathbf{D}_1\mathbf{H},\quad
\mathbf{Y}_2 = \mathbf{A}\mathbf{D}_2\mathbf{H}.
\end{aligned}
\end{equation}
The proof below uses only this form, $\mathbf{Y}_0 + \cos\omega\,\mathbf{Y}_1 + \sin\omega\,\mathbf{Y}_2$, so it covers any curve
$\mathbf{D}_0 + \cos\omega\,\mathbf{D}_1 + \sin\omega\,\mathbf{D}_2$ in Proposition~\ref{prop:kmeans_arcs}, with $\mathbf{Y}_0 = \mathbf P_2\mathbf{x} + \mathbf{A}\mathbf{D}_0\mathbf{H}$
and $\mathbf{Y}_s = \mathbf{A}\mathbf{D}_s\mathbf{H}$. The one-dimensional path of Section~\ref{sec:onedim} has the same form, with $\mathbf{Q}(\mathbf{I}_p - \boldsymbol\gamma\boldsymbol\gamma^\top)$,
$\mathbf{A}^\top\bar{\boldsymbol\nu}\boldsymbol\gamma^\top$, and $\mathbf{A}^\top\mathrm{dir}(\mathbf P_1\mathbf{q})\,\boldsymbol\gamma^\top$ in place of $\mathbf{Q} - \mathbf{D}_1$, $\mathbf{D}_1$, and $\mathbf{D}_2$. With the
assignments fixed by $\tau$, each centroid is the same fixed average of the rows of $\mathbf{Y}_0$, $\mathbf{Y}_1$, and $\mathbf{Y}_2$,
\[
\mathbf{m}_\ell^{(t)}(\omega) = \bar{\mathbf{y}}_{0,\ell}^{(t)} + \cos\omega\,\bar{\mathbf{y}}_{1,\ell}^{(t)} + \sin\omega\,\bar{\mathbf{y}}_{2,\ell}^{(t)}.
\]

\begin{proof}[Proof of Proposition~\ref{prop:kmeans_arcs}]
Fix $t$, an observation $k$ with observed label $c = c_k^{(t)}$, and a cluster $\ell \ne c$. The differences between
observation $k$ and the two centroids are
\[
\mathbf{x}'_k(\omega) - \mathbf{m}_c^{(t)}(\omega) = \mathbf{e}_0 + \cos\omega\,\mathbf{e}_1 + \sin\omega\,\mathbf{e}_2,
\qquad
\mathbf{x}'_k(\omega) - \mathbf{m}_\ell^{(t)}(\omega) = \mathbf{f}_0 + \cos\omega\,\mathbf{f}_1 + \sin\omega\,\mathbf{f}_2,
\]
with $\mathbf{e}_s = [\mathbf{Y}_s]_k - \bar{\mathbf{y}}_{s,c}^{(t)}$ and $\mathbf{f}_s = [\mathbf{Y}_s]_k - \bar{\mathbf{y}}_{s,\ell}^{(t)}$ for
$s = 0, 1, 2$. The difference of the two squared norms is
\[
g(\omega) = \kappa_1\cos^2\omega + \kappa_2\sin^2\omega + \kappa_3\cos\omega\sin\omega + \kappa_4\cos\omega
+ \kappa_5\sin\omega + \kappa_6,
\]
where
\[
\begin{aligned}
\kappa_1 &= \|\mathbf{e}_1\|_2^2 - \|\mathbf{f}_1\|_2^2, &
\kappa_2 &= \|\mathbf{e}_2\|_2^2 - \|\mathbf{f}_2\|_2^2, &
\kappa_3 &= 2(\mathbf{e}_1^\top\mathbf{e}_2 - \mathbf{f}_1^\top\mathbf{f}_2), \\
\kappa_4 &= 2(\mathbf{e}_0^\top\mathbf{e}_1 - \mathbf{f}_0^\top\mathbf{f}_1), &
\kappa_5 &= 2(\mathbf{e}_0^\top\mathbf{e}_2 - \mathbf{f}_0^\top\mathbf{f}_2), &
\kappa_6 &= \|\mathbf{e}_0\|_2^2 - \|\mathbf{f}_0\|_2^2 .
\end{aligned}
\]
Using $\cos^2\omega = (1 + \cos 2\omega)/2$, $\sin^2\omega = (1 - \cos 2\omega)/2$, and
$\cos\omega\sin\omega = \sin 2\omega/2$ gives \eqref{eq:wave_main},
\begin{equation}
\label{eq:harmonic_app}
\begin{aligned}
&g(\omega) = \beta_1\cos 2\omega + \beta_2\sin 2\omega + \beta_3\cos\omega + \beta_4\sin\omega + \beta_5, \\
&\beta_1 = \tfrac{\kappa_1 - \kappa_2}{2},\quad \beta_2 = \tfrac{\kappa_3}{2},\quad \beta_3 = \kappa_4,\quad
\beta_4 = \kappa_5,\quad \beta_5 = \tfrac{\kappa_1 + \kappa_2}{2} + \kappa_6 .
\end{aligned}
\end{equation}
To find the zeros, substitute $s = \tan(\omega/2)$, so that
\[
\cos\omega = \frac{1 - s^2}{1 + s^2}, \qquad \sin\omega = \frac{2s}{1 + s^2}, \qquad
\cos 2\omega = \frac{1 - 6s^2 + s^4}{(1 + s^2)^2}, \qquad \sin 2\omega = \frac{4s - 4s^3}{(1 + s^2)^2}.
\]
Multiplying $g$ by $(1 + s^2)^2$ gives the polynomial
\begin{equation}
\label{eq:weier}
\varphi(s) = (\beta_1 + \beta_3 + \beta_5) + (4\beta_2 + 2\beta_4)s + (2\beta_5 - 6\beta_1)s^2
+ (2\beta_4 - 4\beta_2)s^3 + (\beta_1 - \beta_3 + \beta_5)s^4 .
\end{equation}
Each real root $s_*$ of $\varphi$ gives a zero $\omega_* = 2\arctan s_*$ of $g$, and every zero of $g$ in $(-\pi, \pi)$
arises this way. The substitution misses only $\omega = \pi$, where $s = \tan(\omega/2)$ is infinite. At $\omega = \pi$,
\[
g(\pi) = \beta_1 - \beta_3 + \beta_5,
\]
which is the leading coefficient of $\varphi$. So $\omega = \pi$ is a zero exactly when $\varphi$ has degree at most three.

We can now count the zeros. If $g$ is identically zero, the constraint $g \le 0$ holds on the whole circle and plays no
role. Otherwise, either $g(\pi) \ne 0$ and the zeros of $g$ are the at most four real roots of the quartic $\varphi$, or
$g(\pi) = 0$ and the zeros are $\omega = \pi$ and the at most three real roots of the cubic $\varphi$. In both cases $g$ has at
most four zeros on the circle.

Finally, each $g_{t,k,\ell}$ is continuous in $\omega$, so it keeps its sign between two consecutive zeros. Take all the
zeros of all the constraints together and sort them by increasing angle in $[0, 2\pi)$. On each interval between two
consecutive zeros, every constraint keeps its sign, so either every point of the interval satisfies all the constraints
or none does. The feasible set on the circle is therefore a union of intervals whose endpoints are among these zeros.
There are at most four zeros per constraint, so the number of intervals is at most four times the number of
constraints; a single constraint alone gives at most two intervals.
\end{proof}

We also note two shortcuts that further reduce the computational cost. First, some constraints hold on the whole
circle and can be dropped. The two pairs of terms in
\eqref{eq:harmonic_app} are bounded, $|\beta_1\cos 2\omega + \beta_2\sin 2\omega| \le \sqrt{\beta_1^2 + \beta_2^2}$ and
$|\beta_3\cos\omega + \beta_4\sin\omega| \le \sqrt{\beta_3^2 + \beta_4^2}$, so if
\[
\beta_5 + \sqrt{\beta_1^2 + \beta_2^2} + \sqrt{\beta_3^2 + \beta_4^2} < 0,
\]
then $g(\omega) < 0$ for every $\omega$, and the constraint can be skipped. Second, the feasible set can be read off
from the zeros of the remaining constraints. Sort these zeros by increasing angle from $0$ to $2\pi$, and add $0$ and $2\pi$
as the first and last points. On each interval between two consecutive points, the proof above shows that either every point is feasible or none is, so it
is enough to check one point, the midpoint, against all the constraints. Keep the intervals whose midpoint is feasible,
and merge neighboring kept intervals into one. Algorithm~\ref{alg:har} gives one step of the hit-and-run kernel. We checked the closed-form
\looseness=-1 intervals against a dense grid of angles: they agreed at all $400{,}000$ grid points we tried.

\begin{algorithm}[t]
\caption{One step of the hit-and-run kernel on the trajectory event}
\label{alg:har}
\begin{algorithmic}[1]
\Require current $\mathbf{Q} \in \mathcal S_{\mathbf{Q},\tau}$; $\mathbf{H}$, $\mathbf P_2\mathbf{X}$, and the observed sequence $\tau$
\State Draw a pair of rows $i < j$ uniformly from $\{1, \ldots, r-1\}$
\State Form $\mathbf{Y}_0, \mathbf{Y}_1, \mathbf{Y}_2$ from \eqref{eq:XPUW_main} and the centroid coefficients
$\bar{\mathbf{y}}_{0,\ell}^{(t)}, \bar{\mathbf{y}}_{1,\ell}^{(t)}, \bar{\mathbf{y}}_{2,\ell}^{(t)}$
\For{each constraint $(t, k, \ell)$}
  \State Compute $\beta_1, \ldots, \beta_5$ from \eqref{eq:harmonic_app}; skip the constraint if it holds on the whole circle
  \State Add the real roots of \eqref{eq:weier}, mapped to angles, to the list of endpoints (and $\pi$ if $g(\pi) = 0$)
\EndFor
\State Keep the intervals between consecutive endpoints whose midpoints satisfy every constraint, giving
$\mathcal S_\omega = \bigcup_a [\omega_a^{\mathrm{lo}}, \omega_a^{\mathrm{hi}}]$
\State Draw $\omega$ uniformly on $\mathcal S_\omega$
\Ensure $\mathbf{G}_{ij}(\omega)\mathbf{Q}$
\end{algorithmic}
\end{algorithm}

\subsection{Reversibility of the hit-and-run kernel}
\label{appendix:sampler_valid}

This subsection proves that the hit-and-run kernel of Section~\ref{sec:kmeans} is reversible with respect to the
uniform distribution on $\mathcal S$, which is all that Proposition~\ref{prop:bc_validity} requires. The proof does not
use the quadratic form of $\mathcal S$; it holds for any set $\mathcal S$ with $\int \one_{\mathcal S}(\mathbf{Q})\,d\mathbf{Q} > 0$.

Fix a pair of rows $i < j$. Rotating rows $i$ and $j$ of $\mathbf{Q}$ by every angle $\omega \in [0, 2\pi)$ gives the points
$\mathbf{G}_{ij}(\omega)\mathbf{Q}$, which form a closed curve through $\mathbf{Q}$ in $\St(r-1, p)$; we call this curve the circle through $\mathbf{Q}$.
(Figure~\ref{fig:perturbation}(b) shows the data along such a circle, and Figure~\ref{fig:perturbation}(c) shows the
angles $\omega$.) The fraction of this circle that lies in $\mathcal S$ is
\[
z(\mathbf{Q}) = \frac{1}{2\pi}\int_0^{2\pi} \one_{\mathcal S}\big(\mathbf{G}_{ij}(\omega)\mathbf{Q}\big)\,d\omega .
\]
The kernel $\mathcal K_{ij}$ moves $\mathbf{Q}$ to $\mathbf{G}_{ij}(\omega)\mathbf{Q}$ with $\omega$ drawn uniformly from the feasible angles
$\{\omega : \mathbf{G}_{ij}(\omega)\mathbf{Q} \in \mathcal S\}$. So, when $z(\mathbf{Q}) > 0$, the expected value of a bounded function $h$ after one step is
\begin{equation}
\label{eq:heatbath}
(\mathcal K_{ij}h)(\mathbf{Q}) = \frac{\int_0^{2\pi} \one_{\mathcal S}\big(\mathbf{G}_{ij}(\omega)\mathbf{Q}\big)\,h\big(\mathbf{G}_{ij}(\omega)\mathbf{Q}\big)\,d\omega}
{\int_0^{2\pi} \one_{\mathcal S}\big(\mathbf{G}_{ij}(\omega)\mathbf{Q}\big)\,d\omega},
\end{equation}
and $\mathcal K_{ij}$ leaves $\mathbf{Q}$ in place when $z(\mathbf{Q}) = 0$. The hit-and-run kernel $\mathcal K$ draws the pair $i < j$ uniformly
and then applies $\mathcal K_{ij}$.

Two facts about circles drive the proof. First, a function that depends on $\mathbf{Q}$ only through its circle takes the same
value at $\mathbf{G}_{ij}(\psi)\mathbf{Q}$ as at $\mathbf{Q}$ for every $\psi$; examples are $z$ and $\mathcal K_{ij}h$, because replacing $\mathbf{Q}$ by
$\mathbf{G}_{ij}(\psi)\mathbf{Q}$ only shifts the angle of integration. Second, the uniform distribution on $\St(r-1, p)$ is unchanged by
multiplying on the left by an orthogonal matrix, so $\int h(\mathbf{G}_{ij}(\omega)\mathbf{Q})\,d\mathbf{Q} = \int h(\mathbf{Q})\,d\mathbf{Q}$ for every $\omega$.
Averaging this identity over $\omega$, and then applying it to a product $g\,h$ in which $g$ depends on $\mathbf{Q}$ only through its
circle, gives
\begin{equation}
\label{eq:circle_average}
\int g(\mathbf{Q})\,h(\mathbf{Q})\,d\mathbf{Q} = \int g(\mathbf{Q})\left\{\frac{1}{2\pi}\int_0^{2\pi} h\big(\mathbf{G}_{ij}(\omega)\mathbf{Q}\big)\,d\omega\right\}d\mathbf{Q}.
\end{equation}
In words, inside an integral against $d\mathbf{Q}$, $h$ can be replaced by its average over the circle.

\begin{proposition}[Reversibility of the hit-and-run kernel]
\label{prop:K_reversible}
\label{lem:orbit_conditional}
Each $\mathcal K_{ij}$, the kernel $\mathcal K$, and every power $\mathcal K^L$ are reversible with respect to the uniform
distribution on $\mathcal S$.
\end{proposition}

\begin{proof}
By definition, $\mathcal K_{ij}$ is reversible with respect to the uniform distribution on $\mathcal S$ if moving from $\mathbf{Q}$ to a new
point is as likely as moving back. Equivalently, for all bounded functions $h_1$ and $h_2$, the integral
$\int \one_{\mathcal S}(\mathbf{Q})\,h_1(\mathbf{Q})\,(\mathcal K_{ij}h_2)(\mathbf{Q})\,d\mathbf{Q}$ is unchanged by swapping $h_1$ and $h_2$.

\emph{Step 1: $z(\mathbf{Q}) > 0$ for almost every $\mathbf{Q} \in \mathcal S$.} Formula \eqref{eq:heatbath} needs $z(\mathbf{Q}) > 0$, that is, a circle
with a feasible arc of positive length. Applying \eqref{eq:circle_average} with $g = \one\{z = 0\}$ and $h = \one_{\mathcal S}$,
\[
\int \one\{z(\mathbf{Q}) = 0\}\,\one_{\mathcal S}(\mathbf{Q})\,d\mathbf{Q} = \int \one\{z(\mathbf{Q}) = 0\}\,z(\mathbf{Q})\,d\mathbf{Q} = 0 .
\]
So the other case has probability zero.

\emph{Step 2: symmetry.} The function $\mathcal K_{ij}h_2$ depends on $\mathbf{Q}$ only through its circle, so \eqref{eq:circle_average} with
$g = \mathcal K_{ij}h_2$ and $h = \one_{\mathcal S}h_1$, followed by \eqref{eq:heatbath}, gives
\[
\begin{aligned}
&\int \one_{\mathcal S}(\mathbf{Q})\,h_1(\mathbf{Q})\,(\mathcal K_{ij}h_2)(\mathbf{Q})\,d\mathbf{Q} \\
&\qquad = \int_{\{z > 0\}} \frac{1}{z(\mathbf{Q})}
\Big\{\frac{1}{2\pi}\int_0^{2\pi} (\one_{\mathcal S}h_1)\big(\mathbf{G}_{ij}(\omega)\mathbf{Q}\big)\,d\omega\Big\}
\Big\{\frac{1}{2\pi}\int_0^{2\pi} (\one_{\mathcal S}h_2)\big(\mathbf{G}_{ij}(\omega)\mathbf{Q}\big)\,d\omega\Big\}\,d\mathbf{Q}.
\end{aligned}
\]
The right side is symmetric in $h_1$ and $h_2$, so $\mathcal K_{ij}$ is reversible.

Finally, $\mathcal K$ averages the kernels $\mathcal K_{ij}$ with weights that do not depend on $\mathbf{Q}$, and an average of reversible
kernels is reversible. A power of a reversible kernel is reversible too, so $\mathcal K^L$ is.
\end{proof}

Two remarks complete the picture. First, reversibility is all that the exactness of the test needs: the kernel need
not reach every part of $\mathcal S$, which can be disconnected. How well it moves through $\mathcal S$ affects only the power,
which is why the path length $M$ and the number of steps $L$ matter (Section~\ref{sec:bc}). Second, the pair of rows is
drawn afresh at every step, independently of $\mathbf{Q}$, so it is part of the kernel and not an extra condition; keeping one
pair fixed for the whole path would keep the path on one circle.

\subsection{Proof of Proposition~\ref{cor:truncated_F}}
\label{appendix:hotelling_angle}

This subsection proves Proposition~\ref{cor:truncated_F}, the one-dimensional selective $p$-value, in three steps. We
first describe all polar factors with the same coordinates $\boldsymbol\Gamma$ (Proposition~\ref{lem:angle_fiber}), then show that
$u(\mathbf{Q})$ is independent of $\boldsymbol\Gamma$, and finally condition on the clustering.

In coordinates, write $u = u(\mathbf{Q}) = \|\mathbf{Q}^\top\mathbf{a}\|_2^2$ with $0 < u < 1$. Since $\boldsymbol\gamma = \mathrm{dir}(\mathbf{Q}^\top\mathbf{a})$, we have
$\mathbf{Q}^\top\mathbf{a} = \sqrt u\,\boldsymbol\gamma$. Let $\mathbf{q} = \mathbf{Q}\boldsymbol\gamma$, a unit vector with $\mathbf{a}^\top\mathbf{q} = \sqrt u$, and split it into its
parts along $\mathbf{a}$ and orthogonal to $\mathbf{a}$:
\begin{equation}
\label{eq:q_split}
\mathbf{q} = \sqrt u\,\mathbf{a} + \sqrt{1-u}\,\mathbf{w}, \qquad \mathbf{w} = \frac{\mathbf{q} - \sqrt u\,\mathbf{a}}{\sqrt{1-u}},
\end{equation}
where $\mathbf{w}$ is the within-cluster direction $\mathrm{dir}(\mathbf P_1\mathbf{q})$ in coordinates. With
$\mathbf{B} = \mathbf{Q}(\mathbf{I}_p - \boldsymbol\gamma\boldsymbol\gamma^\top)$, the coordinates of Section~\ref{sec:onedim} are $\boldsymbol\Gamma = (\boldsymbol\gamma, \mathbf{w}, \mathbf{B})$, and
\begin{equation}
\label{eq:Q_split}
\mathbf{Q} = \mathbf{Q}\boldsymbol\gamma\boldsymbol\gamma^\top + \mathbf{Q}(\mathbf{I}_p - \boldsymbol\gamma\boldsymbol\gamma^\top) = \mathbf{B} + \big(\sqrt u\,\mathbf{a} + \sqrt{1-u}\,\mathbf{w}\big)\boldsymbol\gamma^\top .
\end{equation}

\begin{proposition}[Polar factors with the same $\boldsymbol\Gamma$]
\label{lem:angle_fiber}
The coordinates $(\boldsymbol\gamma, \mathbf{w}, \mathbf{B})$ satisfy
\begin{equation}
\label{eq:admissible}
\|\boldsymbol\gamma\|_2 = \|\mathbf{w}\|_2 = 1, \quad \mathbf{a}^\top\mathbf{w} = 0, \quad \mathbf{B}\boldsymbol\gamma = \mathbf 0, \quad
\mathbf{B}^\top\mathbf{a} = \mathbf{B}^\top\mathbf{w} = \mathbf 0, \quad \mathbf{B}^\top\mathbf{B} = \mathbf{I}_p - \boldsymbol\gamma\boldsymbol\gamma^\top .
\end{equation}
Conversely, for every triple $(\boldsymbol\gamma, \mathbf{w}, \mathbf{B})$ that satisfies \eqref{eq:admissible} and every $u \in (0, 1)$, the
matrix $\mathbf{B} + (\sqrt u\,\mathbf{a} + \sqrt{1-u}\,\mathbf{w})\boldsymbol\gamma^\top$ lies in $\St(r-1, p)$ and has coordinates
$(\boldsymbol\gamma, \mathbf{w}, \mathbf{B})$ and between-cluster fraction $u$. Hence the polar factors with the same $\boldsymbol\Gamma$ as $\mathbf{Q}$ form the
curve
\begin{equation}
\label{eq:hotelling_fiber}
\mathbf{Q}_{\boldsymbol\Gamma}(\phi) = \mathbf{B} + (\cos\phi\,\mathbf{a} + \sin\phi\,\mathbf{w})\boldsymbol\gamma^\top, \qquad 0 < \phi < \pi/2,
\end{equation}
along which $u(\mathbf{Q}_{\boldsymbol\Gamma}(\phi)) = \cos^2\phi$.
\end{proposition}

\begin{proof}
We check \eqref{eq:admissible} one identity at a time. The vector $\boldsymbol\gamma$ is a direction, and $\mathbf{w}$ has length one
because $\|\mathbf{q} - \sqrt u\,\mathbf{a}\|_2^2 = 1 - 2u + u = 1 - u$. Next, $\mathbf{a}^\top\mathbf{w} = (\sqrt u - \sqrt u)/\sqrt{1-u} = 0$, and
$\mathbf{B}\boldsymbol\gamma = \mathbf{Q}(\boldsymbol\gamma - \boldsymbol\gamma) = \mathbf 0$. For the products with $\mathbf{B}$,
\[
\mathbf{B}^\top\mathbf{a} = (\mathbf{I}_p - \boldsymbol\gamma\boldsymbol\gamma^\top)\mathbf{Q}^\top\mathbf{a} = \sqrt u\,(\mathbf{I}_p - \boldsymbol\gamma\boldsymbol\gamma^\top)\boldsymbol\gamma = \mathbf 0,
\qquad
\mathbf{B}^\top\mathbf{q} = (\mathbf{I}_p - \boldsymbol\gamma\boldsymbol\gamma^\top)\mathbf{Q}^\top\mathbf{Q}\boldsymbol\gamma = (\mathbf{I}_p - \boldsymbol\gamma\boldsymbol\gamma^\top)\boldsymbol\gamma = \mathbf 0,
\]
and $\mathbf{B}^\top\mathbf{w}$ is a combination of these two, so it is zero too. Finally,
$\mathbf{B}^\top\mathbf{B} = (\mathbf{I}_p - \boldsymbol\gamma\boldsymbol\gamma^\top)\mathbf{Q}^\top\mathbf{Q}(\mathbf{I}_p - \boldsymbol\gamma\boldsymbol\gamma^\top) = \mathbf{I}_p - \boldsymbol\gamma\boldsymbol\gamma^\top$.

For the converse, fix a triple that satisfies \eqref{eq:admissible} and $u \in (0, 1)$, and let
$\mathbf{q}_u = \sqrt u\,\mathbf{a} + \sqrt{1-u}\,\mathbf{w}$ and $\mathbf{Q}_u = \mathbf{B} + \mathbf{q}_u\boldsymbol\gamma^\top$. The vector $\mathbf{q}_u$ has length one, because
$\mathbf{a}$ and $\mathbf{w}$ are orthogonal unit vectors, and $\mathbf{B}^\top\mathbf{q}_u = \mathbf 0$. So
\[
\mathbf{Q}_u^\top\mathbf{Q}_u = \mathbf{B}^\top\mathbf{B} + \boldsymbol\gamma\,\mathbf{q}_u^\top\mathbf{q}_u\,\boldsymbol\gamma^\top = \mathbf{I}_p - \boldsymbol\gamma\boldsymbol\gamma^\top + \boldsymbol\gamma\boldsymbol\gamma^\top = \mathbf{I}_p,
\]
and $\mathbf{Q}_u \in \St(r-1, p)$. Its coordinates are the given ones: $\mathbf{Q}_u^\top\mathbf{a} = \mathbf{B}^\top\mathbf{a} + \boldsymbol\gamma\,\mathbf{q}_u^\top\mathbf{a} = \sqrt u\,\boldsymbol\gamma$, so
$u(\mathbf{Q}_u) = u$ and the direction of $\mathbf{Q}_u^\top\mathbf{a}$ is $\boldsymbol\gamma$; then $\mathbf{Q}_u\boldsymbol\gamma = \mathbf{B}\boldsymbol\gamma + \mathbf{q}_u = \mathbf{q}_u$, whose split
\eqref{eq:q_split} returns $\mathbf{w}$; and $\mathbf{Q}_u(\mathbf{I}_p - \boldsymbol\gamma\boldsymbol\gamma^\top) = \mathbf{B} - \mathbf{B}\boldsymbol\gamma\boldsymbol\gamma^\top = \mathbf{B}$. Writing
$u = \cos^2\phi$ with $\phi \in (0, \pi/2)$ gives \eqref{eq:hotelling_fiber}.
\end{proof}

\begin{proof}[Proof of the independence in Proposition~\ref{thm:angle_independent}]
Under the null, $\mathbf{Q}$ is uniformly distributed on $\St(r-1, p)$ (Theorem~\ref{thm:polar_uniform}). We show that $u$ and
$\boldsymbol\Gamma$ are independent, using maps that change $\boldsymbol\Gamma$ but not $u$ or the distribution of $\mathbf{Q}$. Call a triple
admissible if it satisfies \eqref{eq:admissible}. By Proposition~\ref{lem:angle_fiber}, the set of admissible triples does not
depend on $u$.

\emph{Step 1: the maps.} Let $\mathbf{O}_1$ be an $(r-1) \times (r-1)$ orthogonal matrix with $\mathbf{O}_1\mathbf{a} = \mathbf{a}$, and $\mathbf{O}_2$ a
$p \times p$ orthogonal matrix. The map $\mathbf{Q} \mapsto \mathbf{O}_1\mathbf{Q}\mathbf{O}_2^\top$ does not change the uniform distribution on
$\St(r-1, p)$. It keeps $u$, because $(\mathbf{O}_1\mathbf{Q}\mathbf{O}_2^\top)^\top\mathbf{a} = \mathbf{O}_2\mathbf{Q}^\top\mathbf{a}$ has the same length as $\mathbf{Q}^\top\mathbf{a}$, and it
changes the coordinates to $(\mathbf{O}_2\boldsymbol\gamma, \mathbf{O}_1\mathbf{w}, \mathbf{O}_1\mathbf{B}\mathbf{O}_2^\top)$.

\emph{Step 2: the maps reach every admissible triple.} Take two admissible triples $(\boldsymbol\gamma_1, \mathbf{w}_1, \mathbf{B}_1)$ and
$(\boldsymbol\gamma_2, \mathbf{w}_2, \mathbf{B}_2)$. Choose $\mathbf{O}_2$ with $\mathbf{O}_2\boldsymbol\gamma_1 = \boldsymbol\gamma_2$, and let $\tilde{\mathbf{B}} = \mathbf{B}_1\mathbf{O}_2^\top$. Let
$\mathbf{v}_1, \ldots, \mathbf{v}_{p-1}$ be an orthonormal basis of the vectors orthogonal to $\boldsymbol\gamma_2$. By \eqref{eq:admissible}, the
two lists
\[
\tilde{\mathbf{B}}\mathbf{v}_1, \ldots, \tilde{\mathbf{B}}\mathbf{v}_{p-1}, \mathbf{w}_1
\qquad\text{and}\qquad
\mathbf{B}_2\mathbf{v}_1, \ldots, \mathbf{B}_2\mathbf{v}_{p-1}, \mathbf{w}_2
\]
each consist of $p$ orthonormal vectors orthogonal to $\mathbf{a}$. The vectors orthogonal to $\mathbf{a}$ form a space of dimension
$r - 2 \ge p$, so there is an orthogonal $\mathbf{O}_1$ with $\mathbf{O}_1\mathbf{a} = \mathbf{a}$ that maps the first list to the second. Then
$\mathbf{O}_1\mathbf{w}_1 = \mathbf{w}_2$ and $\mathbf{O}_1\tilde{\mathbf{B}} = \mathbf{B}_2$, since both sides agree on $\mathbf{v}_1, \ldots, \mathbf{v}_{p-1}$ and vanish on
$\boldsymbol\gamma_2$. So $(\mathbf{O}_1, \mathbf{O}_2)$ carries the first triple to the second. For $p = 1$, there are no $\mathbf{v}$'s.

\emph{Step 3: averaging.} Draw $(\mathbf{O}_1, \mathbf{O}_2)$ uniformly among such pairs, independently of $\mathbf{Q}$, and write
$(\mathbf{O}_1, \mathbf{O}_2) \cdot \boldsymbol\Gamma$ for the transformed coordinates. For a fixed admissible triple $\boldsymbol\Gamma_0$, the distribution of
$(\mathbf{O}_1, \mathbf{O}_2) \cdot \boldsymbol\Gamma_0$ does not depend on $\boldsymbol\Gamma_0$: by Step 2, any other triple is a fixed map applied to
$\boldsymbol\Gamma_0$, and composing a uniform draw with a fixed map gives another uniform draw. So for a bounded function $h_2$ of
the triple, $\bar h_2 = \E[h_2((\mathbf{O}_1, \mathbf{O}_2) \cdot \boldsymbol\Gamma_0)]$ is the same for every $\boldsymbol\Gamma_0$. For a bounded function $h_1$
on $(0, 1)$, Step 1 gives the first equality below:
\[
\E[h_1(u)\,h_2(\boldsymbol\Gamma)] = \E[h_1(u)\,h_2((\mathbf{O}_1, \mathbf{O}_2) \cdot \boldsymbol\Gamma)]
= \E\big[h_1(u)\,\E\{h_2((\mathbf{O}_1, \mathbf{O}_2) \cdot \boldsymbol\Gamma) \mid \mathbf{Q}\}\big] = \bar h_2\,\E[h_1(u)].
\]
Taking $h_1$ to be the constant $1$ shows $\bar h_2 = \E[h_2(\boldsymbol\Gamma)]$, so $\E[h_1(u)\,h_2(\boldsymbol\Gamma)] = \E[h_1(u)]\,\E[h_2(\boldsymbol\Gamma)]$: the
expectation factors, and $u$ and $\boldsymbol\Gamma$ are independent.
\end{proof}

The factor $\cos^{p-1}\phi$ in $f$ is what a direct count misses. Given $\boldsymbol\Gamma$, the vector $\mathbf{Q}\boldsymbol\gamma$ lies on the
unit sphere of the $(r - p)$-dimensional space orthogonal to the columns of $\mathbf{B}$. If it were uniform there, the angle
would have density proportional to $\sin^{r-p-2}\phi$, which contradicts Proposition~\ref{prop:T2_share}.

\begin{proof}[Proof of the conditional statement in Proposition~\ref{cor:truncated_F}]
We condition on $\boldsymbol\Gamma$, $\mathbf{H}$, $\mathbf P_2\mathbf{X}$, and the clustering, in four steps. Write $\phi \in (0, \pi/2)$ for the angle with
$u = \cos^2\phi$.

\emph{Step 1: the angle given $(\boldsymbol\Gamma, \mathbf{H}, \mathbf P_2\mathbf{X})$.} By Theorem~\ref{thm:polar_uniform}, $\mathbf{Q}$ is independent of
$(\mathbf{H}, \mathbf P_2\mathbf{X})$, and by the independence just proved, $u$ is independent of $\boldsymbol\Gamma$. So given $(\boldsymbol\Gamma, \mathbf{H}, \mathbf P_2\mathbf{X})$,
$u$ keeps its $\mathrm{Beta}(p/2, (r-1-p)/2)$ distribution, and the angle $\phi$ has the density $f$ of
Section~\ref{sec:onedim_mc}.

\emph{Step 2: the selection event in terms of the angle.} Given $(\boldsymbol\Gamma, \mathbf{H}, \mathbf P_2\mathbf{X})$, Proposition~\ref{lem:angle_fiber}
gives $\mathbf{Q} = \mathbf{Q}_{\boldsymbol\Gamma}(\phi)$, so the data are $\mathbf P_2\mathbf{X} + \mathbf{A}\mathbf{Q}_{\boldsymbol\Gamma}(\phi)\mathbf{H}$, a function of $\phi$ alone. The clustering
therefore returns $\hat{\mathcal C}_1$ and $\hat{\mathcal C}_2$ exactly when $\phi \in \mathcal S_\theta$, with $\mathcal S_\theta$ in \eqref{eq:S_theta}
computed from $(\boldsymbol\Gamma, \mathbf{H}, \mathbf P_2\mathbf{X})$.

\emph{Step 3: Bayes' rule.} Conditioning the density $f$ on the event $\phi \in \mathcal S_\theta$ gives the truncated density
\eqref{eq:truncated_law}, provided $\int_{\mathcal S_\theta} f > 0$. Values of $(\boldsymbol\Gamma, \mathbf{H}, \mathbf P_2\mathbf{X})$ with
$\int_{\mathcal S_\theta} f = 0$ have probability zero given the selection event, as in the proof of
Proposition~\ref{prop:conditional_selection}(a).

\emph{Step 4: uniformity.} The truncated density \eqref{eq:truncated_law} has a continuous distribution function, even
when $\mathcal S_\theta$ is a union of intervals. Since $T^2 = (r-2)\cot^2\phi$ decreases in $\phi$, the event
$T^2 \ge T^2(\mathbf{x})$ is the event $\phi \le \theta$, so $p_{\mathrm{polar,1d}}$ in \eqref{eq:pF} is the distribution function of
\eqref{eq:truncated_law} evaluated at the observed angle $\theta$, which is a draw from \eqref{eq:truncated_law} under the null.
By the probability integral transform it is uniformly distributed, and averaging over $(\boldsymbol\Gamma, \mathbf{H}, \mathbf P_2\mathbf{X})$ given the
selection event keeps it uniform. Finally, the change of variables $F = \frac{r-p-1}{p}\cot^2\phi$ turns
\eqref{eq:truncated_law} into the $F_{p,\,r-p-1}$ density truncated to $\mathcal S_{\boldsymbol\Gamma}$.
\end{proof}

\subsection{Propose-and-check kernels}
\label{appendix:cayley}

The random Givens kernel of Section~\ref{sec:mc} proposes a random rotation of $\mathbf{Q}$ and keeps it only if the
clustering algorithm returns the tested clusters. Two facts make such kernels reversible with respect to the uniform
distribution on $\mathcal S$; we state them in the coordinates of this appendix.

\begin{proposition}[Accepting only feasible proposals]
\label{prop:metropolis_feasible}
Let $\mathcal R(\mathbf{Q}, d\mathbf{Q}')$ be a proposal on $\St(r-1, p)$ that is reversible with respect to the uniform distribution. The
kernel that moves from $\mathbf{Q} \in \mathcal S$ to a proposal $\mathbf{Q}'$ if $\mathbf{Q}' \in \mathcal S$, and stays at $\mathbf{Q}$ otherwise, is
reversible with respect to the uniform distribution on $\mathcal S$.
\end{proposition}

\begin{proof}
For bounded functions $h$, reversibility of $\mathcal R$ means
\[
\iint h(\mathbf{Q}, \mathbf{Q}')\,\mathcal R(\mathbf{Q}, d\mathbf{Q}')\,d\mathbf{Q} = \iint h(\mathbf{Q}', \mathbf{Q})\,\mathcal R(\mathbf{Q}, d\mathbf{Q}')\,d\mathbf{Q}.
\]
Applied to
$h(\mathbf{Q}, \mathbf{Q}')\one_{\mathcal S}(\mathbf{Q})\one_{\mathcal S}(\mathbf{Q}')$, it shows that the accepted moves are symmetric in
$(\mathbf{Q}, \mathbf{Q}')$; the rejected moves leave $\mathbf{Q}$ in place and are symmetric trivially. Their sum is the kernel.
\end{proof}

\begin{proposition}[Symmetric random rotations]
\label{prop:cayley_symmetric}
\label{prop:rotation_symmetric}
Let $\mathbf{C}$ be a random $(r-1) \times (r-1)$ orthogonal matrix, drawn independently of $\mathbf{Q}$, such that $\mathbf{C}^{-1}$ has the same
distribution as $\mathbf{C}$. Then the proposal $\mathbf{Q} \mapsto \mathbf{C}\mathbf{Q}$ is reversible with respect to the uniform
distribution. The distribution of $\mathbf{C}$ may depend on $\mathbf P_2\mathbf{X}$, $\mathbf{H}$, and the clustering. Two examples are
\begin{itemize}
  \item the random Givens kernel, $\mathbf{C} = \mathbf{G}_{ij}(\omega)$ with $i < j$ uniform and $\omega$ symmetric about zero, for
  which $\mathbf{G}_{ij}(\omega)^{-1} = \mathbf{G}_{ij}(-\omega)$;
  \item the Cayley random walk, $\mathbf{C} = (\mathbf{I}_{r-1} - \tfrac\epsilon2\mathbf{W})^{-1}(\mathbf{I}_{r-1} + \tfrac\epsilon2\mathbf{W})$ with $\mathbf{W}$ a random
  skew-symmetric matrix such that $-\mathbf{W}$ has the same distribution as $\mathbf{W}$, for which $\mathbf{C}^{-1}$ is this map at
  $-\mathbf{W}$.
\end{itemize}
\end{proposition}

\begin{proof}
For a bounded function $h$, the substitution $\mathbf{Q}' = \mathbf{C}\mathbf{Q}$, under which the uniform distribution is unchanged,
and then the equality in distribution of $\mathbf{C}^{-1}$ and $\mathbf{C}$ give
\[
\E\Big[\int h(\mathbf{Q}, \mathbf{C}\mathbf{Q})\,d\mathbf{Q}\Big] = \E\Big[\int h(\mathbf{C}^{-1}\mathbf{Q}', \mathbf{Q}')\,d\mathbf{Q}'\Big]
= \E\Big[\int h(\mathbf{C}\mathbf{Q}', \mathbf{Q}')\,d\mathbf{Q}'\Big],
\]
which is the reversibility of the proposal. For the Cayley map, the eigenvalues of a real skew-symmetric matrix are
purely imaginary, so $\mathbf{I}_{r-1} - \tfrac\epsilon2\mathbf{W}$ is invertible, and because $\mathbf{I}_{r-1} \pm \tfrac\epsilon2\mathbf{W}$ commute,
$\mathbf{C}^\top = (\mathbf{I}_{r-1} - \tfrac\epsilon2\mathbf{W})(\mathbf{I}_{r-1} + \tfrac\epsilon2\mathbf{W})^{-1} = (\mathbf{I}_{r-1} + \tfrac\epsilon2\mathbf{W})^{-1}(\mathbf{I}_{r-1} - \tfrac\epsilon2\mathbf{W}) = \mathbf{C}^{-1}$.
\end{proof}

\begin{proof}[Proof of Proposition~\ref{prop:givens_reversible}]
For a fixed step size $\epsilon$, the random Givens kernel accepts only feasible proposals from the proposal of
Proposition~\ref{prop:rotation_symmetric}, so it is reversible by Proposition~\ref{prop:metropolis_feasible}. A power of
a reversible kernel is reversible, and the kernel that draws $\epsilon$ from a fixed distribution and then takes $L$
steps is an average of reversible kernels with weights free of $\mathbf{Q}$, so it is also reversible.
\end{proof}

Our Cayley random walk draws $\mathbf{W} = \mathbf{N} - \mathbf{N}^\top$ with $\mathbf{N}$ a matrix of independent standard normal entries, and
draws $\epsilon$ as the random Givens kernel does. Neither proposition covers a step size tuned on trial moves from
$\mathbf{Q}_{\text{obs}}$ (Section~\ref{sec:bc}).

\subsection{Proof of the exactness of the Besag--Clifford $p$-value}
\label{appendix:bc_proof}

\begin{proof}[Proof of Proposition~\ref{prop:bc_validity}]
Fix $\mathbf P_2\mathbf{X}$, $\mathbf{H}$, and the event that the clustering returns $\hat{\mathcal C}_1$ and $\hat{\mathcal C}_2$. Because $\mathcal K$, $M$,
and $L$ depend on the data only through these, $\mathcal K$ is a fixed kernel and $M$ a fixed integer, and $J$, $\xi$, and the
random inputs of every step are drawn independently of $\mathbf{Q}_{\text{obs}}$. Write $\mathcal K(\mathbf{Q}, d\mathbf{Q}')$ for the distribution of
one step of the path from $\mathbf{Q}$, and $c = \int \one_{\mathcal S}(\mathbf{Q})\,d\mathbf{Q}$. By
Proposition~\ref{prop:conditional_selection}, $\mathbf{Q}_{\text{obs}}$ has distribution $c^{-1}\one_{\mathcal S}(\mathbf{Q})\,d\mathbf{Q}$, and
reversibility means
\[
\one_{\mathcal S}(\mathbf{Q})\,d\mathbf{Q}\;\mathcal K(\mathbf{Q}, d\mathbf{Q}') = \one_{\mathcal S}(\mathbf{Q}')\,d\mathbf{Q}'\;\mathcal K(\mathbf{Q}', d\mathbf{Q}).
\]
A Givens step has no density, so these are statements about distributions, not densities.

\emph{The path.} Applying reversibility to $(\mathbf{Q}_k, \mathbf{Q}_{k-1})$, $(\mathbf{Q}_{k-1}, \mathbf{Q}_{k-2})$, and so on, gives for
every $k$
\begin{equation}
\label{eq:reverse_prefix}
\one_{\mathcal S}(\mathbf{Q}_k)\,d\mathbf{Q}_k\prod_{j=1}^{k}\mathcal K(\mathbf{Q}_j, d\mathbf{Q}_{j-1})
= \one_{\mathcal S}(\mathbf{Q}_0)\,d\mathbf{Q}_0\prod_{j=0}^{k-1}\mathcal K(\mathbf{Q}_j, d\mathbf{Q}_{j+1}).
\end{equation}
Given $J = k$, Algorithm~\ref{alg:bc} starts at $\mathbf{Q}_k = \mathbf{Q}_{\text{obs}}$ and runs the chain backward to $\mathbf{Q}_0$ and
forward to $\mathbf{Q}_M$, so the path $(\mathbf{Q}_0, \ldots, \mathbf{Q}_M)$ has distribution
\[
c^{-1}\one_{\mathcal S}(\mathbf{Q}_k)\,d\mathbf{Q}_k\Big\{\prod_{j=1}^{k}\mathcal K(\mathbf{Q}_j, d\mathbf{Q}_{j-1})\Big\}\Big\{\prod_{j=k}^{M-1}\mathcal K(\mathbf{Q}_j, d\mathbf{Q}_{j+1})\Big\}
= c^{-1}\one_{\mathcal S}(\mathbf{Q}_0)\,d\mathbf{Q}_0\prod_{j=0}^{M-1}\mathcal K(\mathbf{Q}_j, d\mathbf{Q}_{j+1})
\]
by \eqref{eq:reverse_prefix}. The right side does not depend on $k$: it is the distribution of a chain started from the
uniform distribution on $\mathcal S$. So $J$, which is uniform on $\{0, \ldots, M\}$, is independent of the path. The points of
the path need not be exchangeable; only the position of the observed point is uniform
\citep[Section~3.2]{Howes2023}.

\emph{The rank.} Fix a path, and let its distinct values of $u$, in decreasing order, occur $N_1, \ldots, N_D$ times,
with $\bar N_d = \sum_{d' < d} N_{d'}$. Given the path, the observed point falls in the $d$th group with probability
$N_d/(M+1)$, and then $\hat p_{\mathrm{polar}} = (\bar N_d + N_d\xi)/(M+1)$ is uniform on the interval
$\mathcal I_d = \big(\bar N_d/(M+1), (\bar N_d + N_d)/(M+1)\big)$. So, given the path, $\hat p_{\mathrm{polar}}$ has density
\[
\sum_{d=1}^{D}\frac{N_d}{M+1}\cdot\frac{M+1}{N_d}\,\one\{x \in \mathcal I_d\} = \sum_{d=1}^{D}\one\{x \in \mathcal I_d\} = 1,
\qquad 0 < x < 1,
\]
because the intervals $\mathcal I_1, \ldots, \mathcal I_D$ cover $(0, 1)$ without overlap. So $\hat p_{\mathrm{polar}}$ is uniform
given the path, even when all values are tied. Averaging over the path completes the proof.

\emph{The kernels.} The random Givens kernel is reversible by Proposition~\ref{prop:givens_reversible} and the
hit-and-run kernel by Proposition~\ref{prop:K_reversible}; in both cases $L$ steps are reversible too.
\end{proof}

\section{Additional material on the Monte Carlo test}
\label{appendix:mc}

This appendix collects material on the Monte Carlo test of Sections~\ref{sec:mc} and~\ref{sec:bc} that the main text
only mentions. Appendix~\ref{appendix:alg_onedim} gives the algorithm for the one-dimensional test,
Appendix~\ref{appendix:mc-budget} explains why a naive Markov chain estimator is not a valid
$p$-value, Appendix~\ref{appendix:bc_tuning} shows how the path length, the number of steps between points of the path, and the step size affect the test,
and Appendix~\ref{appendix:kernels} compares kernels.

\subsection{The algorithm for the one-dimensional test}
\label{appendix:alg_onedim}

Algorithm~\ref{alg:bc1d} computes $p_{\mathrm{polar,1d}}$ in \eqref{eq:pF} for any clustering algorithm, as described in
Section~\ref{sec:onedim_mc}. It follows Algorithm~\ref{alg:bc}, with the angle $\phi$ in place of the polar factor
and a Metropolis step for the density \eqref{eq:truncated_law} in place of the random Givens kernel.

\begin{algorithm}[t]
\caption{The one-dimensional selective test $\hat p_{\mathrm{polar,1d}}$ for any clustering algorithm}
\label{alg:bc1d}
\begin{algorithmic}[1]
\Require data $\mathbf{x} \in \R^{n \times p}$, clustering algorithm $\mathcal C$, tested clusters
  $\hat{\mathcal C}_1, \hat{\mathcal C}_2 \in \mathcal C(\mathbf{x})$, path length $M$, number of steps $L$
\State Compute $\mathbf P_0$, $\mathbf P_1$, $\mathbf P_2$, $\mathbf{H}$, and $\mathbf{Q}_{\text{obs}}$ as in Algorithm~\ref{alg:bc}
\State $\boldsymbol\gamma \gets \mathrm{dir}(\mathbf{Q}_{\text{obs}}^\top\boldsymbol\nu)$, $\mathbf{q} \gets \mathbf{Q}_{\text{obs}}\boldsymbol\gamma$,
  and $\theta \gets \arccos\|\mathbf P_0\mathbf{q}\|_2$
  \Comment{Section~\ref{sec:onedim}}
\State Draw $J$ uniformly from $\{0, 1, \ldots, M\}$ and set $\phi_J \gets \theta$
\For{$j = J, J+1, \ldots, M-1$}
  \State $\phi_{j+1} \gets \Call{RandomStep}{\phi_j}$ \Comment{forward}
\EndFor
\For{$j = J, J-1, \ldots, 1$}
  \State $\phi_{j-1} \gets \Call{RandomStep}{\phi_j}$ \Comment{backward}
\EndFor
\State Draw $\xi \sim \text{Uniform}(0,1)$
\Ensure $\hat p_{\mathrm{polar,1d}} = \frac{1}{M+1}\big(\sum_{j=0}^{M}\one\{\phi_j < \phi_J\} + \xi\sum_{j=0}^{M}\one\{\phi_j = \phi_J\}\big)$
\Statex
\Function{RandomStep}{$\phi$} \Comment{target density \eqref{eq:truncated_law}}
  \State Draw $\log\epsilon$ uniformly on $[\log 10^{-3}, 0]$
  \For{$\ell = 1, \ldots, L$}
    \State Draw $\omega \sim \N\big(0, (\pi\epsilon)^2\big)$ and $U \sim \text{Uniform}(0,1)$, and set $\phi' \gets \phi + \omega$
    \If{$\phi' \in (0, \pi/2)$ and $U \le f(\phi')/f(\phi)$} \Comment{$f(\phi) \propto \cos^{p-1}\phi\,\sin^{r-p-2}\phi$}
      \If{$\hat{\mathcal C}_1, \hat{\mathcal C}_2 \in \mathcal C\big(\mathbf P_2\mathbf{x} + \mathbf{Q}_{\boldsymbol\Gamma}(\phi')\,\mathbf{H}\big)$}
        \Comment{path \eqref{eq:QM_path}}
        \State $\phi \gets \phi'$
      \EndIf
    \EndIf
  \EndFor
  \State \Return $\phi$
\EndFunction
\end{algorithmic}
\end{algorithm}

\subsection{A naive Markov chain estimator}
\label{appendix:mc-budget}
\label{rem:mcmc}

This subsection records the naive Markov chain estimator mentioned in Section~\ref{sec:mc} and
why it fails. Its target
distribution, the uniform distribution on $\mathcal S_{\mathbf{Q}}$, is the right conditional reference, but drawing useful
Monte Carlo samples from that target is hard unless the proposal is designed around
the geometry of the selection event. The Besag--Clifford test of Section~\ref{sec:bc} needs none.

The naive estimator is built from $B$ post-burn-in draws
$\mathbf{Q}^{(1)},\dots,\mathbf{Q}^{(B)}$ with the uniform distribution on $\mathcal S_{\mathbf{Q}}$ as stationary distribution,
\begin{equation*}
\widehat p_B \;=\; \frac{1+\sum_{b=1}^B \one\{u(\mathbf{Q}^{(b)}) \ge u(\mathbf{Q}_{\text{obs}})\}}{B+1}.
\end{equation*}
If the draws were close to independent conditional samples, this would approximate
the selective tail probability. In our problem they are not. The selection event
leaves a thin feasible set around the observed configuration, and a generic
random walk that accepts only feasible moves travels through it slowly. The dimension of the Stiefel manifold is
\begin{equation*}
\dim\big(\St(r-1, p)\big) \;=\; (r-1)\,p - \frac{p(p+1)}{2},
\end{equation*}
so larger tested pairs and larger feature dimension both make the exploration
harder. On the $24$ null cells with $n \le 200$, $p \le 20$, and isotropic, AR(1), or
compound-symmetric $\boldsymbol\Sigma$ ($600$ replicates each), the naive estimator with a
Cayley random-walk kernel, $1{,}000$ burn-in steps, and $2{,}500$ post-burn-in draws thinned by $5$ has median selective
Type~I error $0.077$ and reaches $0.185$ at $n = 200$, $p = 10$; at $n = 200$ and
$p \in \{50, 100\}$ it lies between $0.12$ and $0.20$, while the Besag--Clifford test
is calibrated in the same cells. The bottleneck is not rejection of proposals;
step-size adaptation keeps the acceptance rate moderate:
\begin{center}
\begin{tabular}{lccc}
\toprule
 & $p = 2$ & $p = 10$ & $p = 50$ \\
\midrule
Cayley acceptance rate & $\approx 0.40$ & $\approx 0.36$ & $\approx 0.31$ \\
\bottomrule
\end{tabular}
\end{center}
The naive estimator is valid only when the chain has mixed, which the Type I errors above show it has not. The
Besag--Clifford $p$-value of Section~\ref{sec:bc} does not need mixing: it is exact for any reversible kernel and any path
\looseness=-1 length (Proposition~\ref{prop:bc_validity}), and better kernels and longer paths only add power.

\subsection{Path length, number of steps, and step size}
\label{appendix:bc_tuning}

\begin{figure}[t]
\centering
\includegraphics[width=\linewidth]{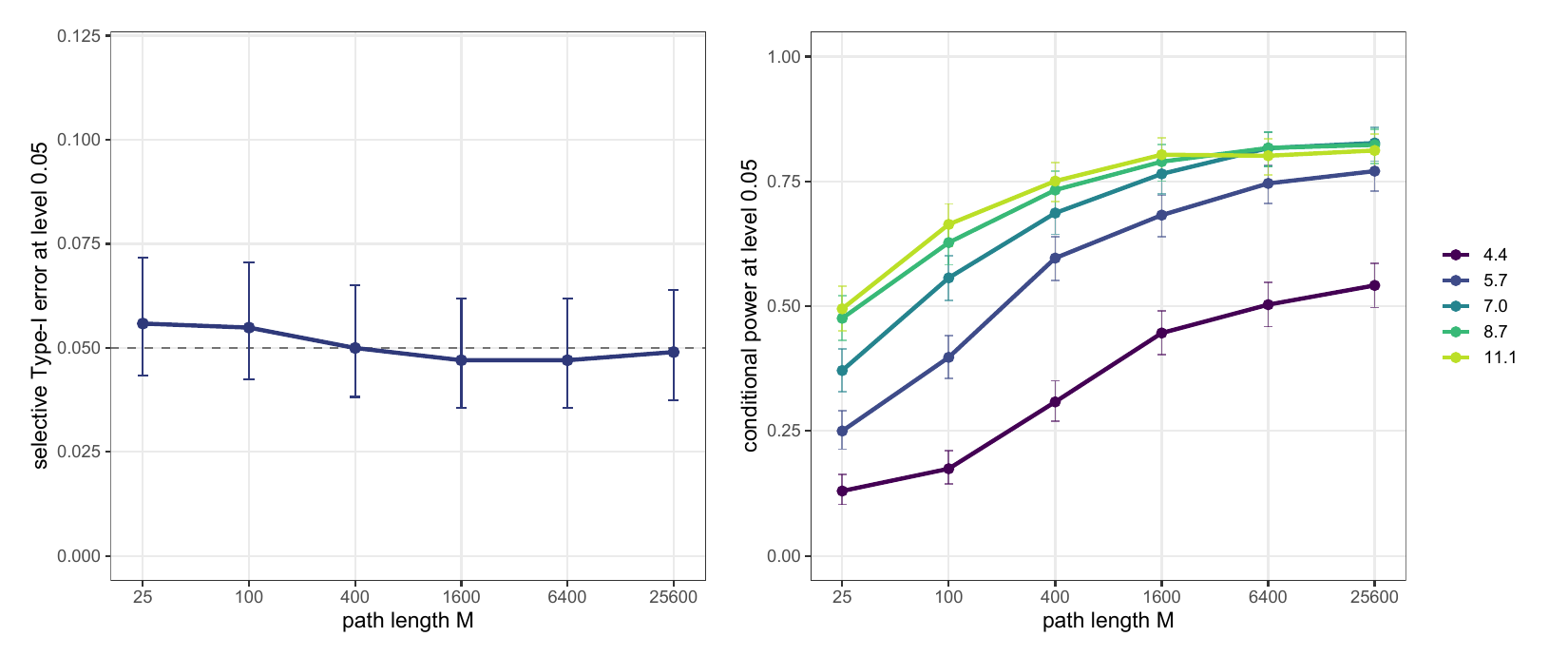}
\caption{Path length changes power but not the Type I error ($n = 45$, $p = 10$, $K = 3$). Left: selective Type I
error against the path length $M$ on null data, each replicate evaluated at every $M$. Right: conditional power against $M$
for five separations $\delta$ (one curve each).}
\label{fig:validity_vs_M}
\end{figure}

Figure~\ref{fig:validity_vs_M} illustrates Section~\ref{sec:bc}. On null data the
Type~I error of $\hat p_{\mathrm{polar}}$ stays at the nominal level from $M = 25$ to $M = 25{,}600$
(between $0.047$ and $0.056$ over $1{,}021$ replicates, each evaluated at every $M$, with $L = 2$ steps between points),
while on data with a signal the power grows with $M$ and levels off ($473$ to $493$ recovered replicates per
separation). The identity kernel is an
extreme case: it is reversible, and it returns $\hat p_{\mathrm{polar}} = \xi$, which is exactly uniform
and has no power.

The requirement that the kernel, $M$, and $L$ depend on the data only through $\mathbf P_2\mathbf{X}$, $\mathbf{H}$, and the clustering cannot
be dropped. Take the two-point space $\{0,1\}$ with the uniform distribution, $u(x) = x$, and
the kernel that always moves to the other point, which is reversible. Taking $M = 1$ and
$L = 1$ when $\mathbf{Q}_{\text{obs}} = 1$ but $L = 2$ when $\mathbf{Q}_{\text{obs}} = 0$ gives
$\Prob(\hat p_{\mathrm{polar}} \le 1/2) = 3/4$, although each fixed $\mathcal K^L$ is reversible. Tuning a
step size on trial moves from $\mathbf{Q}_{\text{obs}}$ is a choice of this kind. A step size, or a number of steps, drawn
afresh for each block of steps from a fixed distribution is allowed, because the kernel of one
block is then a mixture of reversible kernels with weights that do not depend on the current point of the path.

\subsection{A pilot comparison of kernels}
\label{appendix:kernels}

We compared four kernels in the construction of Section~\ref{sec:bc} on the design of
Figure~\ref{fig:motivation} ($n = 100$, $p = 2$, compound-symmetric $\boldsymbol\Sigma$ with
\looseness=-1 correlation $0.6$, $k$-means with $K = 3$), with $M = 1{,}000$ and $L = 2$:
\begin{itemize}
  \item the random Givens kernel of Section~\ref{sec:mc};
  \item the Cayley random walk of Appendix~\ref{appendix:cayley};
  \item a Givens step that draws the angle by a slice sampler with shrinkage \citep{Neal2003}, which
  moves at every step but runs the clustering several times per step;
  \item the hit-and-run step on the exact arcs of Section~\ref{sec:kmeans}, which conditions on the
  Lloyd trajectory.
\end{itemize}
On $588$ null pairs, the Type~I error at level $0.05$ was $0.029$ for the random Givens kernel,
$0.070$ for the Cayley random walk, $0.049$ for the slice version, and $0.067$ for the exact
arcs ($297$ pairs). All four kernels are exact by Proposition~\ref{prop:bc_validity}, and on a
toy problem where $\mathbf{Q}_{\text{obs}}$ is drawn uniformly from a known set, both the Givens and the Cayley
kernels gave rejection rates of $0.050$ (Givens) and $0.053$ (Cayley) over $2{,}000$ replicates, so we attribute the
spread above to Monte Carlo error. With separation $\delta = 4$
($111$ pairs), the power was $0.973$ for the random Givens kernel ($7$ seconds per $p$-value),
$0.955$ for the Cayley random walk ($10$ seconds), and $1.000$ for the slice version ($38$ seconds,
about $10{,}000$ runs of the clustering against $2{,}000$ for the other two).

\begin{figure}[t]
\centering
\includegraphics[width=\textwidth]{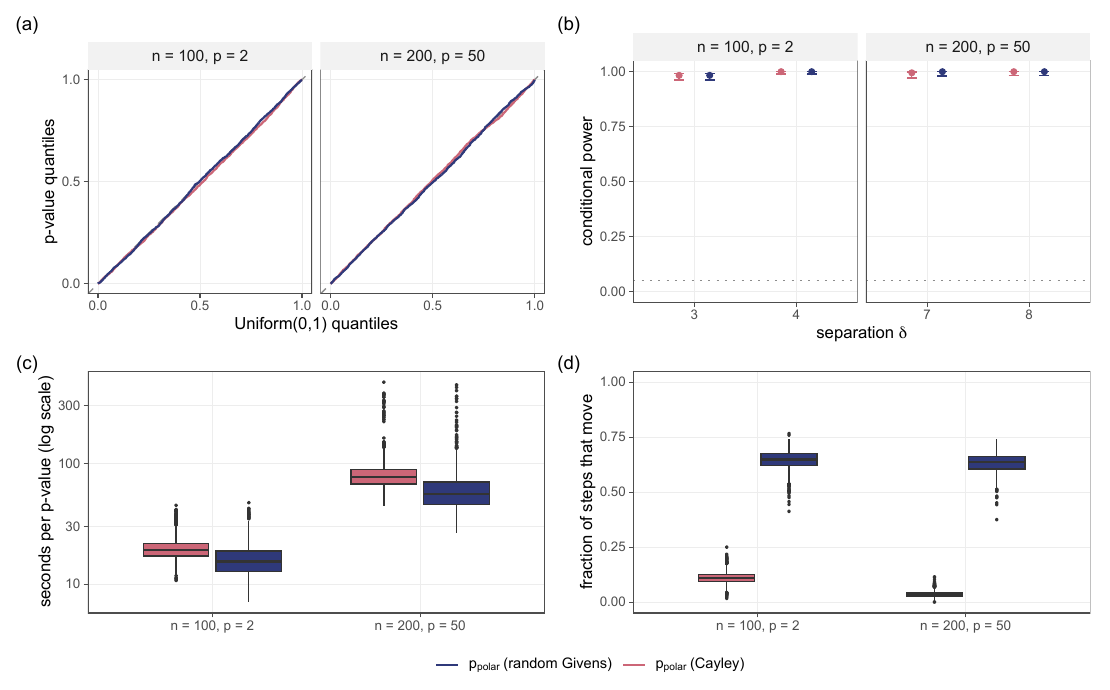}
\caption{The random Givens kernel and the Cayley random walk with $M = 5{,}000$ and $L = 2$, on null data
with compound-symmetric $\boldsymbol\Sigma$. \textbf{(a)} Quantile-quantile plots of the $p$-values
against the Uniform$(0,1)$ distribution. \textbf{(b)} Seconds of computing time per $p$-value.
\textbf{(c)} The fraction of steps at which the chain moves.}
\label{fig:kernel_compare}
\end{figure}

A larger run compared the random Givens kernel and the Cayley random walk with $M = 5{,}000$ and
$L = 2$ in two settings with compound-symmetric $\boldsymbol\Sigma$: the design above, and $n = 200$ with
$p = 50$, each with about $2{,}100$ null pairs. Both kernels control the selective Type~I error
(Figure~\ref{fig:kernel_compare}(a)), and both reject $H_0$ for $98\%$ to $100\%$ of the pairs of true
clusters at the separations we ran. The random Givens kernel is $20\%$ to $30\%$ faster
(Figure~\ref{fig:kernel_compare}(b)) and moves at about $65\%$ of its steps, against $3\%$ to $11\%$ for
the Cayley random walk (Figure~\ref{fig:kernel_compare}(c)).

\section{Additional simulation results}
\label{appendix:grid-extra}

This appendix reports every cell of the $k$-means simulation study of Section~\ref{sec:sim}, which
Figure~\ref{fig:kmeans} summarizes. It also gives the results for Leiden community detection, the
rows of Section~\ref{sec:robust} that are not Gaussian, and the computing time of each test. The figures of this appendix
also show a reference that is not available in practice, $p_{\mathrm{polar}}(\boldsymbol\Sigma)$, which replaces Hotelling's
$T^2$ in $p_{\mathrm{polar}}$ by $(\bar{\mathbf{x}}_{\hat{\mathcal C}_1} - \bar{\mathbf{x}}_{\hat{\mathcal C}_2})^\top\boldsymbol\Sigma^{-1}
(\bar{\mathbf{x}}_{\hat{\mathcal C}_1} - \bar{\mathbf{x}}_{\hat{\mathcal C}_2})$ and so requires the true $\boldsymbol\Sigma$. Its conditional
power is close to that of $p_{\mathrm{polar}}$, except when $p$ is close to the number of tested rows $r$, where
estimating $\boldsymbol\Sigma$ costs power.

\begin{figure}[p]
\centering
\includegraphics[width=\textwidth]{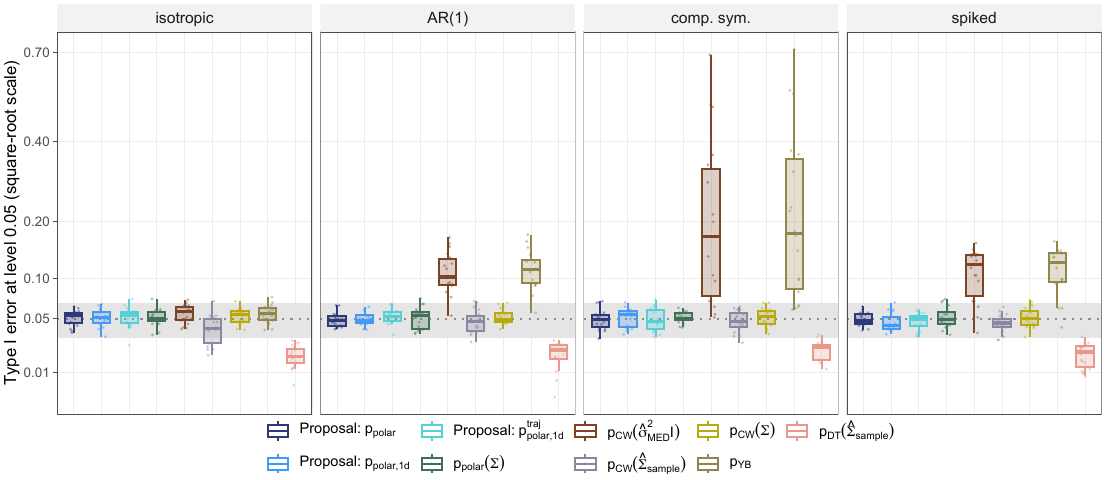}
\caption{Selective Type I error at $\alpha = 0.05$ after $k$-means, by covariance structure, on a square-root scale. Each
box summarizes one test over the cells $(n, p)$, which are drawn as points; the naive test is omitted. The gray band
is the $95\%$ Monte Carlo interval for a test of exact level $0.05$.}
\label{fig:kmeans_typeI_cells}
\end{figure}

\begin{figure}[p]
\centering
\includegraphics[width=\textwidth]{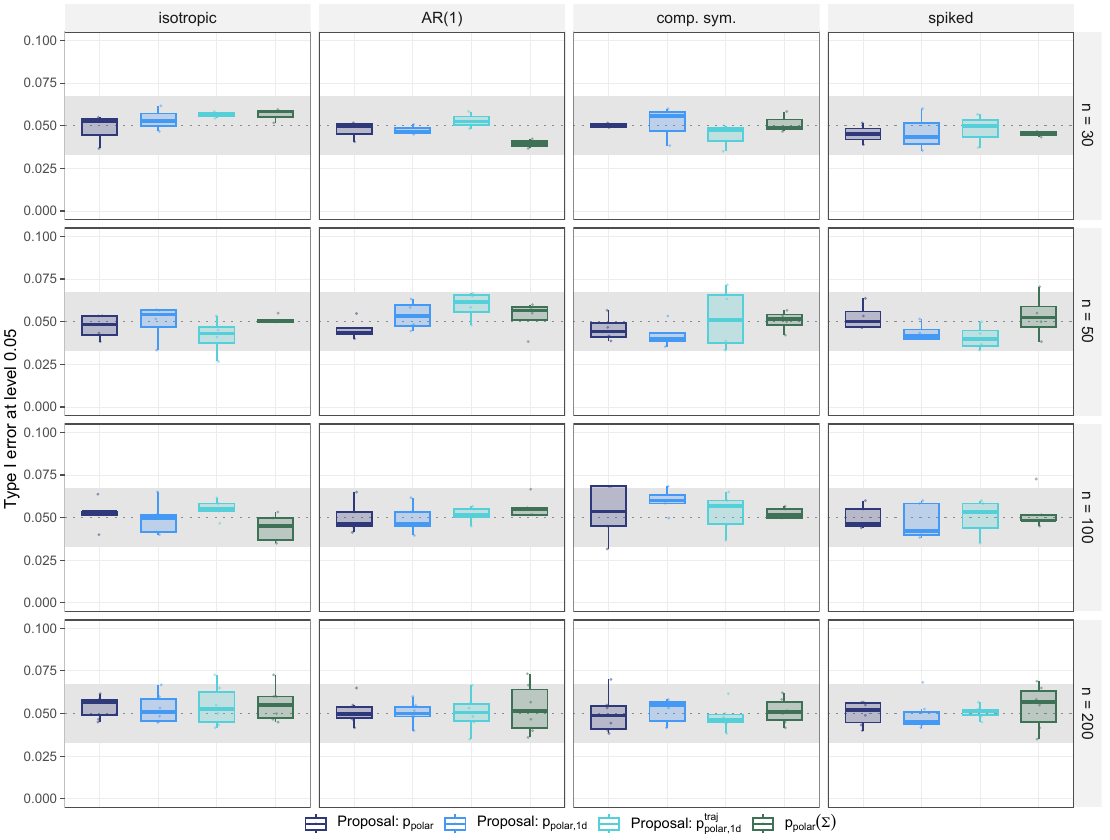}
\caption{Selective Type I error at $\alpha = 0.05$ after $k$-means for the tests that we propose, stratified by $n$
(rows) and $\boldsymbol\Sigma$ (columns). Each box summarizes one test over the values of $p$ with $p \le n/2$, which are
drawn as points. The gray band is as in Figure~\ref{fig:kmeans_typeI_cells}.}
\label{fig:kmeans_typeI_exact}
\end{figure}

\paragraph{Selective Type I error of $k$-means.}
Table~\ref{tab:kmeans_typeI} gives the selective Type~I error at $\alpha = 0.05$ as the median and the
maximum over the cells $(n, p)$, by covariance. Figure~\ref{fig:kmeans_typeI_cells} summarizes the cells for every test,
and Figure~\ref{fig:kmeans_typeI_exact} shows the tests that we propose on a magnified scale. With
$600$ data sets per cell, the $95\%$ Monte Carlo interval for a test with selective Type~I error
$0.05$ is $[0.033, 0.067]$. Two comparators appear only here. The unknown-scale test
$p_{\mathrm{YB}}$ of \citet{YunFoygelBarber2023} divides by the within-cluster sum of squares of the two tested
clusters, so it needs no estimate of $\sigma$; it also assumes $\boldsymbol\Sigma = \sigma^2\mathbf{I}_p$
and, like $p_{\mathrm{CW}}(\hat\sigma^2_{\mathrm{MED}}\mathbf{I})$, holds the level only under isotropic
noise. We computed it, with \texttt{KmeansInference}, on the same null data sets as the other tests. Data thinning \citep{NeufeldDharamshiGaoWitten2024},
$p_{\mathrm{DT}}(\hat{\boldsymbol\Sigma}_{\mathrm{sample}})$, splits $\mathbf{X}$ into two parts using
the sample covariance of $\mathbf{X}$ about its overall mean, clusters one part, and applies the
classical Hotelling test to the other.
Its two parts are independent only when the true $\boldsymbol\Sigma$ is used, but with the sample
covariance it is conservative in all of our cells.

\begin{table}[t]
\centering
\footnotesize
\setlength{\tabcolsep}{4pt}
\begin{tabular}{lcccc}
\toprule
Method & isotropic & AR(1) & comp.\ sym. & spiked \\
\midrule
$p_{\mathrm{naive}}$ & 1.000 [1.000] & 1.000 [1.000] & 1.000 [1.000] & 1.000 [1.000] \\
$p_{\mathrm{polar}}$ & 0.053 [0.064] & 0.048 [0.065] & 0.049 [0.070] & 0.048 [0.064] \\
$p_{\mathrm{polar,1d}}$ & 0.052 [0.067] & 0.048 [0.063] & 0.054 [0.068] & 0.043 [0.068] \\
$p_{\mathrm{polar,1d}}^{\mathrm{traj}}$ & 0.054 [0.073] & 0.053 [0.067] & 0.048 [0.072] & 0.050 [0.060] \\
$p_{\mathrm{polar}}(\boldsymbol\Sigma)$ & 0.050 [0.073] & 0.053 [0.073] & 0.051 [0.062] & 0.049 [0.073] \\
$p_{\mathrm{CW}}(\hat\sigma^2_{\mathrm{MED}}\mathbf{I})$ & 0.058 [0.072] & 0.103 [0.170] & 0.171 [0.693] & 0.122 [0.159] \\
$p_{\mathrm{CW}}(\hat{\boldsymbol\Sigma}_{\mathrm{sample}})$ & 0.041 [0.070] & 0.048 [0.070] & 0.048 [0.063] & 0.046 [0.063] \\
$p_{\mathrm{CW}}(\boldsymbol\Sigma)$ & 0.055 [0.070] & 0.049 [0.068] & 0.053 [0.068] & 0.051 [0.071] \\
$p_{\mathrm{YB}}$ & 0.056 [0.075] & 0.114 [0.175] & 0.177 [0.714] & 0.125 [0.162] \\
$p_{\mathrm{DT}}(\hat{\boldsymbol\Sigma}_{\mathrm{sample}})$ & 0.019 [0.030] & 0.023 [0.030] & 0.025 [0.035] & 0.022 [0.033] \\
\bottomrule
\end{tabular}

\caption{Selective Type I error at $\alpha = 0.05$ after $k$-means, by covariance: median [maximum] over
the cells $(n, p)$ of Section~\ref{sec:typeI}.}
\label{tab:kmeans_typeI}
\end{table}

\begin{figure}[p]
\centering
\includegraphics[width=\textwidth]{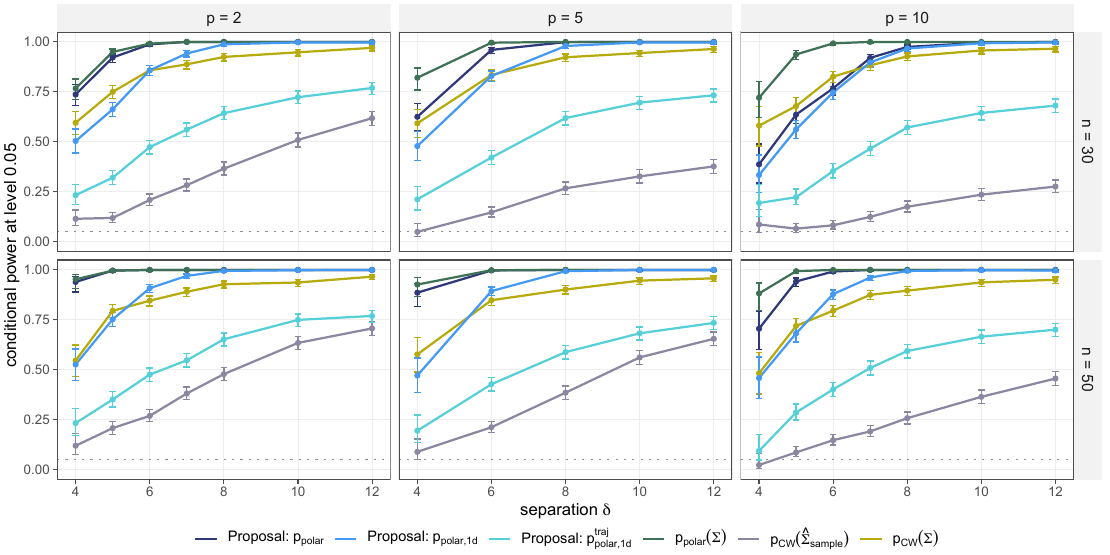}
\caption{Conditional power \eqref{eq:cond_power} at $\alpha = 0.05$ after $k$-means with
$\boldsymbol\Sigma = \mathbf{I}_p$, stratified by $n$ (rows) and $p$ (columns), with $95\%$ Wilson intervals.}
\label{fig:kmeans_power_all}
\end{figure}

\begin{figure}[p]
\centering
\includegraphics[width=\textwidth]{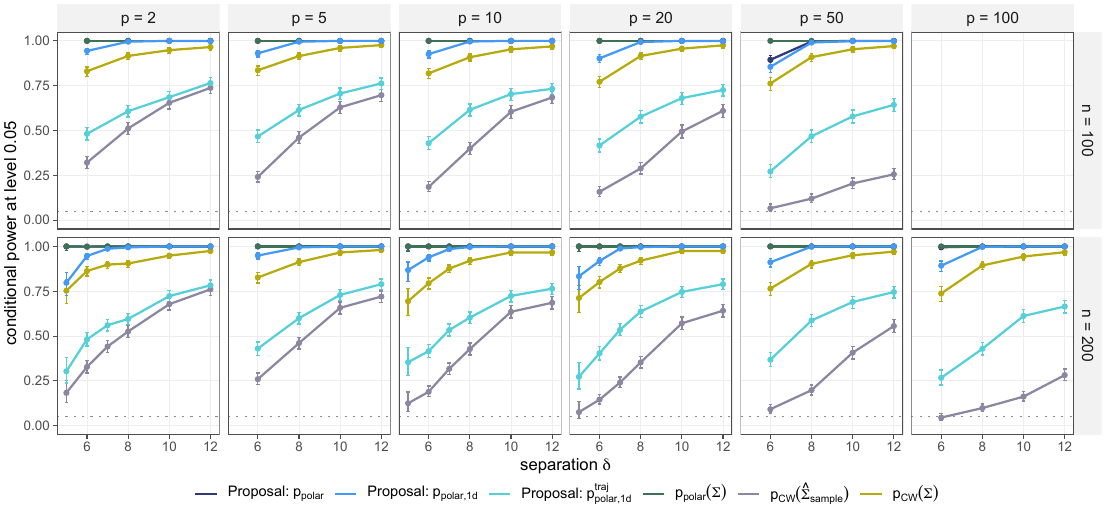}
\caption{As Figure~\ref{fig:kmeans_power_all}, at $n = 100$ and $200$.}
\label{fig:kmeans_power_large_n}
\end{figure}

\begin{figure}[p]
\centering
\includegraphics[width=\textwidth]{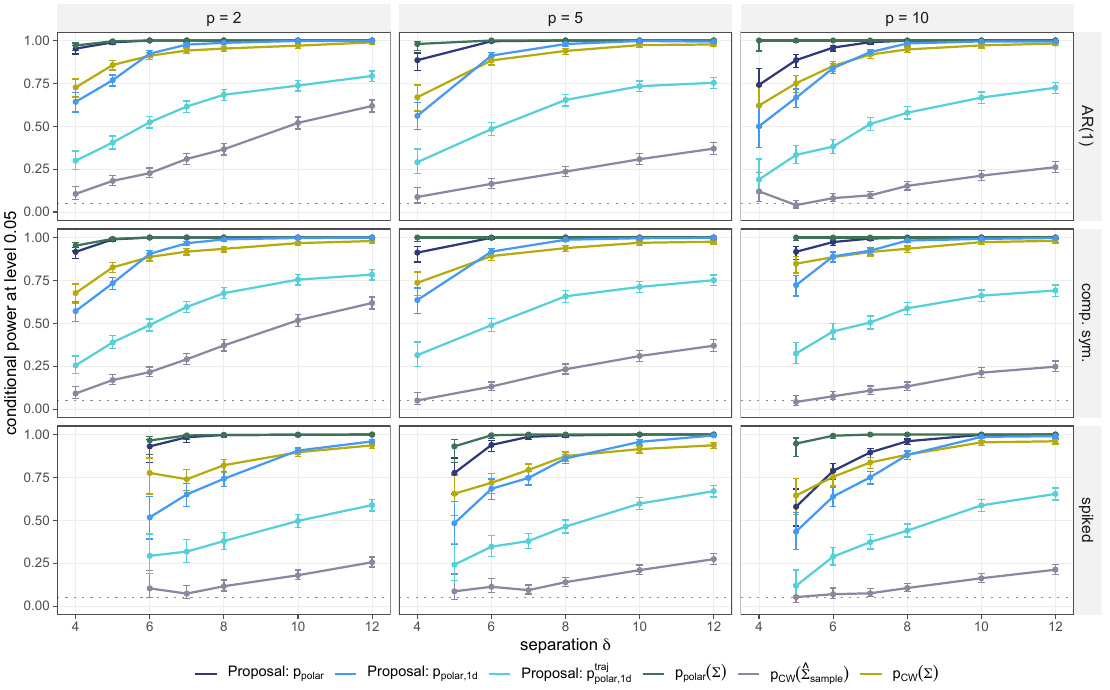}
\caption{As Figure~\ref{fig:kmeans_power_all}, at $n = 30$ under the AR(1), compound-symmetric, and
spiked covariances (rows), for the tests that control the selective Type~I error under them.}
\label{fig:kmeans_power_noniso}
\end{figure}

\paragraph{Conditional power of $k$-means.}
Figure~\ref{fig:kmeans_power_all} repeats Figure~\ref{fig:kmeans}(b) at $n = 30$ and $50$ and
Figure~\ref{fig:kmeans_power_large_n} at $n = 100$ and $200$. Figure~\ref{fig:kmeans_power_noniso}
shows the other covariances at $n = 30$, for the tests that control the selective Type~I error under
them. Conditional power is shown where at least $50$ data sets returned two true clusters.

\begin{figure}[t]
\centering
\includegraphics[width=\textwidth]{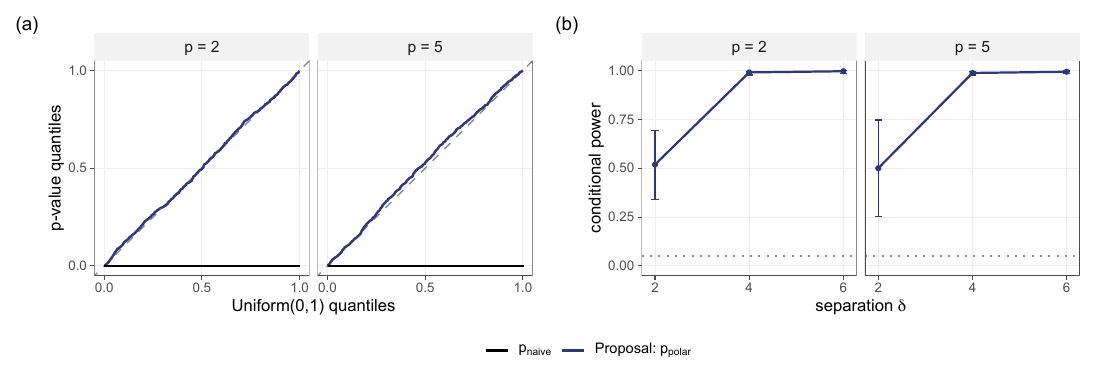}
\caption{Leiden community detection in the design of Figure~\ref{fig:motivation}. \textbf{(a)}
Quantile-quantile plots of the $p$-values with one true group ($\delta = 0$), stratified by $p$.
\textbf{(b)} The conditional power \eqref{eq:cond_power} of $p_{\mathrm{polar}}$ at $\alpha = 0.05$
against $\delta$, with $95\%$ Wilson intervals.}
\label{fig:leiden}
\end{figure}

\paragraph{Leiden community detection.}
Leiden community detection \citep{TraagWaltmanVanEck2019} on a graph that joins each row to its $15$
nearest neighbors has no closed-form selection event, so no existing selective test applies; we report $p_{\mathrm{polar}}$. We used the design of Figure~\ref{fig:motivation}: $n = 100$, compound-symmetric
$\boldsymbol\Sigma$ with correlation $0.6$, and two true groups with means $\pm(\delta/2)\mathbf{u}$, where
$\mathbf{u} = (1, -1, 0, \ldots, 0)^\top/\sqrt 2$, with $p \in \{2, 5\}$ and $\delta \in \{0, 2, 4, 6\}$.
The resolution of the Leiden algorithm was fixed for each $p$ before the study so that it returns about
three clusters, and every pair of clusters with at least $p + 2$ members was tested.
Figure~\ref{fig:leiden}(a) shows that the null $p$-values of $p_{\mathrm{polar}}$ are uniform, with
selective Type~I error $0.043$ at $p = 2$ and $0.051$ at $p = 5$, and Figure~\ref{fig:leiden}(b) shows
that its conditional power is $0.99$ at $\delta = 4$ for both dimensions.

\begin{figure}[t]
\centering
\includegraphics[width=\textwidth]{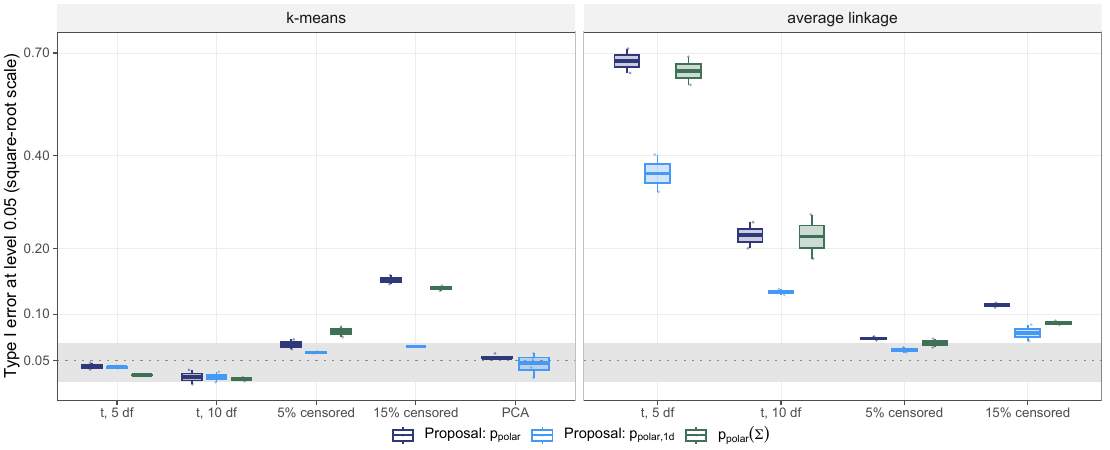}
\caption{Selective Type I error at $\alpha = 0.05$ with rows that are not Gaussian (Section~\ref{sec:robust}),
after $k$-means (left) and average-linkage hierarchical clustering (right), on a square-root scale. Each box summarizes
one test over the values of $p$ (and of $q$ for principal components), which are drawn as points. The gray band is as in
Figure~\ref{fig:kmeans_typeI_cells}. There is no reference $p_{\mathrm{polar}}(\boldsymbol\Sigma)$ for principal components.}
\label{fig:robust}
\end{figure}

\paragraph{Rows that are not Gaussian.}
Figure~\ref{fig:robust} reports the study of Section~\ref{sec:robust}. All data sets have $n = 200$,
compound-symmetric $\boldsymbol\Sigma$ with correlation $0.5$, and no signal. The $t$ rows are
multivariate $t$ with $5$ or $10$ degrees of freedom. The censored rows are Gaussian rows shifted so that
a fraction $s \in \{0.05, 0.15\}$ of each coordinate falls below zero and then censored at zero. For
principal components, Gaussian rows with $p \in \{50, 200\}$ are reduced to their first $q \in \{5, 10\}$
principal components before $k$-means, and the tests use the component scores.

\begin{figure}[t]
\centering
\includegraphics[width=0.75\textwidth]{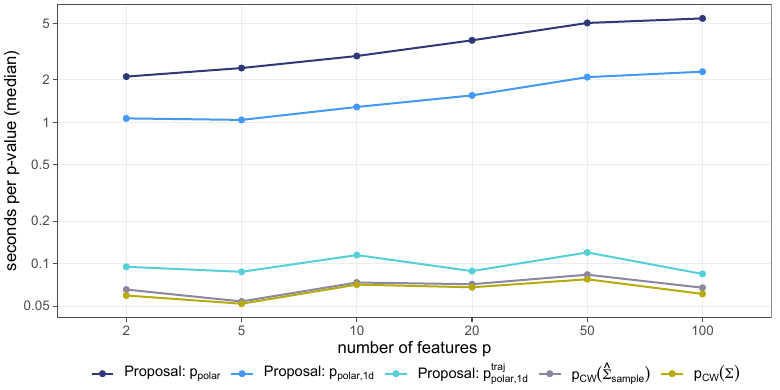}
\caption{Median seconds for one $p$-value after $k$-means at $n = 200$, on one core, by $p$. The
vertical axis is on a log scale.}
\label{fig:kmeans_timing}
\end{figure}

\paragraph{Computing time.}
Figure~\ref{fig:kmeans_timing} gives the median time for one $p$-value after $k$-means at $n = 200$.
We reran the simulation code of this section on six data sets for each value of $p$, three under the
global null and three from the power study, with $M = 5{,}000$ and $L = 2$. Each $p$-value was computed
on one core of an Apple M4 Max processor, with macOS 15.4, R 4.5.2, and the Apple Accelerate linear
algebra library restricted to one thread. One value of $p_{\mathrm{polar}}$ takes $2$ to $6$ seconds,
and one value of $p_{\mathrm{polar,1d}}$ takes $1$ to $2.5$ seconds. Each of the $ML = 10{,}000$ steps
of the path reruns $k$-means once, so the time grows with the cost of one run of the clustering
algorithm. The closed-form $p_{\mathrm{polar,1d}}^{\mathrm{traj}}$ and the tests of
\citet{ChenWitten2023} take about $0.1$ seconds.

\end{document}